\documentclass[emulateapj]{aastex6}

\usepackage{graphicx}
\usepackage{amssymb}
\usepackage{amsmath}
\usepackage{enumitem}
\usepackage{natbib}
\usepackage{booktabs}
\usepackage{multirow}
\usepackage{nicefrac}
\usepackage{mathtools}

\shorttitle{GMC Clump Catalog: Method and Initial Results}
\shortauthors{Zetterlund et al.}
\slugcomment{\today}     

\begin{document}
\title{Northern Galactic Molecular Cloud Clumps in Hi-GAL: Clump and Star Formation within Clouds} 

\author{Erika Zetterlund, Jason Glenn}
\affil{CASA, Department of Astrophysical and Planetary Sciences, University of Colorado 389-UCB, Boulder, CO 80309, USA}
\email{erika.zetterlund@colorado.edu}
\and
\author{Erik Rosolowsky}
\affil{Department of Physics, University of Alberta, Edmonton, Alberta, Canada}

\begin{abstract}

We investigate how the properties of Galactic giant molecular clouds (GMCs) and their denser substructures (clumps) correlate with the local star formation rate. We trace clouds using the $^{12}$CO(3-2) transition, as observed by the CO High Resolution Survey (COHRS). We identify their constituent clumps using thermal dust emission, as observed by the \textit{Herschel} infrared GALactic plane survey (Hi-GAL). We estimate star formation rates in these clouds using 70 $\mu$m emission. In total, we match 3,674 clumps to 473 clouds in position-position-velocity space spanning the Galactic longitude range $10^\circ<\ell<56^\circ$. We find that more massive clouds produce more clumps and more massive clumps. These clumps have average number densities an order of magnitude greater than their host clouds. We find a mean clump mass fraction of $0.20^{+0.13}_{-0.10}$. This mass fraction weakly varies with mass and mass surface density of clouds, and shows no clear dependence on the virial parameter and line width of the clouds. The average clump mass fraction is only weakly dependent upon Galactocentric radius. Although the scatter in our measured properties is significant, the star formation rate for clouds is independent of clump mass fraction. However, there is a positive correlation between the depletion times for clouds and clump mass fraction. We find a star formation efficiency per free fall time of $\epsilon_{\mathrm{ff}}=0.15\%$ for GMCs but $\epsilon_{\mathrm{ff}}=0.37\%$ for clumps.
\end{abstract}


\section{Introduction}

All stars form from molecular gas. However, not all molecular gas forms stars. Rather, the formation of stars is most closely tied to the presence of cold, dense molecular gas \citep{Gao04}. This dense gas is typically found in giant molecular clouds (GMCs), although smaller clouds do exist and form stars \citep[e.g.,][]{Enoch08}. The amount of dense molecular gas and the star formation rate appear to be linearly correlated on both global and local scales \citep{Wu07}. However, when using dense molecular gas to predict star formation, one must be careful, as relations which work globally may not show any correlation on a cloud-by-cloud basis \citep{Evans14}.


Numerical simulations use recipes based on this linear relationship between star formation and gas to incorporate star formation into their studies.  Various star formation prescriptions are evaluated based on their ability to reproduce the star formation law.
Once some physical criteria are met, a set fraction of the gas mass is converted to stars, typically 1\% to 10\% per local dynamical time. The criteria for triggering star formation typically include a density threshold, the presence of converging flows, and a cooling time which is shorter than the dynamical time. This ensures that the gas is in GMCs, which are composed of cold, dense gas. While the density thresholds correspond to, at best, the neutral medium \citep[e.g., 5 cm$^{-3}$ in][]{Guedes11}, GMCs may be formed by colliding flows of warm neutral medium \citep[e.g.,][]{Vazquez11}, which is the reasoning for having a converging flows criterion. This route is taken out of necessity, due to the resolution of galaxy-scale simulations.

However, this does not consider the complexities of a cloud's internal physics. Stars do not form randomly throughout GMCs. Rather, they form where the clouds are densest. We model this substructure with the concept of clumps embedded in clouds.Furthermore, these clumps contain cores, which are interpreted as precursors to individual stellar systems. Clouds have characteristic radii of $R = 1-7.5$ pc; clumps have $R = 0.15-1.5$ pc , cores show radii of $R = 0.015-0.1$ pc. Meanwhile, as size decreases H$_2$ densities increase, with typical densities being $n = 50-500$ cm$^{-3}$ for clouds, $n=10^3-10^4$ cm$^{-3}$ for clumps, and $n=10^4-10^5$ cm$^{-3}$ for cores \citep{Bergin2007}.  There are not clear boundaries between these classes of objects, and the classification is primarily useful in describing hierarchy of substructure \citep{BGPS7}. When optimized for resolution, entire-galaxy simulations are capable of resolving to cloud scales \citep[e.g.,][]{Dobbs11,Fujimoto14}. However, clump and core scales are not currently attainable.

Regardless of its necessity, a universal star formation efficiency that is activated at a universal density threshold is a very simplistic model for such an important process. Such a model ignores the conversion of relatively diffuse molecular gas in clouds to denser substructures. The fraction of gas mass converted to clumps and then to cores surely influences the number and mass of stars formed. These conversions are complex. For instance, \citet{Usero15} and \citet{Bigiel16} showed that the dense gas fraction of the disk increases, but the star formation efficiency of dense gas decreases, in areas of high surface density and high molecular gas fraction. Furthermore, clouds are turbulent, and this turbulence may be important in driving star formation \citep[e.g.,][]{Padoan97,Enoch08}, or hindering it \citep[particularly in the central molecular zone,][]{Rathborne14,Federrath16,Barnes17}. Because of these simplifications, it is imperative that these star formation recipes be tested against reality.

Such a test requires observations of clouds and their substructures using the tracers best suited to each density level. 
Whole clouds are traditionally identified using emission from the lower rotational transitions in CO \citep[e.g.,][]{Scoville75,Solomon87}. However, these transitions are generally optically thick, impeding our ability to detect denser substructures. This can be somewhat mitigated by observing higher-J transitions of CO or by observing the isotopologues of CO, most commonly $^{13}$CO, but at the cost of weaker lines. 

Dense cloud substructures are better traced by the solid dust grains found amongst the gas. Dust makes up approximately 1\% of the ISM by mass \citep{Draine03} and there is evidence for grain growth within molecular clouds. This dust emits thermally in the millimeter and submillimeter regimes. Importantly, dust emission at these wavelengths is optically thin, allowing us to see through the entire Galactic disk, despite lacking velocity information.

Dust is also an effective tracer of recent star formation, if viewed at shorter wavelengths. Newly formed stars are still embedded in the dense gas from which they formed, making them difficult to observe directly. Instead, we observe the effects of the new stars on their environment. The heat from these stars can increase the temperature of the gas and dust from 20 K (a typical temperature for molecular cloud clumps) to around 120 K, corresponding to a peak wavelength of 24 $\mu$m. Naturally then, 24 $\mu$m is a popular waveband in which to view ongoing star formation \citep[e.g.,][]{Rieke09}. At longer wavelengths, the emission revealing star formation becomes contaminated with cirrus emission. Nevertheless, 70 $\mu$m is another popular waveband for such work \citep[e.g.,][]{Calzetti10}. In fact, it is preferred by some, as contamination from cirrus is easier to remove than the contamination from evolved stars found in 24 $\mu$m emission \citep{Svoboda16}.

Since molecular clouds set the initial conditions of star formation, their properties should regulate how of star formation within them.  Theories of star formation \citep[e.g.,][]{McKee07, Krumholz09} outline that the star formation rate depends on local gas conditions within the clouds.  Some empirical models \citep[e.g.,][]{Lada12} argue that star formation proceeds in a small, constant mass fraction of molecular clouds.  However, studying the conditions of molecular gas in a range of galactic environments \citep{Longmore13, Utomo18} show that there is intrinsic variation in the efficiency with which molecular gas can form stars.  It remains an open question as to how the galactic environment regulates molecular gas properties and whether all molecular gas participates equally in the star formation process.  Such investigations rely on small scale (1-100 pc) observations of molecular clouds to determine their properties and connect to their internal star formation properties.  Observations of molecular clouds in extragalactic systems are limited by angular resolution and studies in the Milky Way have been limited by line of sight blending.  Recently, \citet{Colombo18} have released a catalog of molecular clouds using the CO High Resolution Survey \citep[COHRS;][]{Dempsey13}, made with $16''$ resolution.  Using a new cataloging approach, their work yielded a catalog of 85\,000 molecular clouds in the Inner Milky Way with excellent linear resolution ($\lesssim 1$ pc).  Such high resolution cloud identifications allow for the determination of cloud properties and carry out matching to other studies at comparable resolution.

 In this work, we study a well resolved set of molecular clouds identified in \citet{Colombo18} explore the empirical relationships between molecular cloud properties, cloud substructure (``clumps''), and the corresponding star formation rate.  Our goal is to identify which of the properties within the molecular gas are most linked to substructure formation and the subsequent formation of stars.
We build on the previous efforts of \citet{Zetterlund18}, which produced a catalog of molecular cloud clumps using 500 $\mu$m thermal dust maps from the \textit{Herschel} infrared GALactic plane survey (Hi-GAL) and the distance determination techniques of \citep{BGPS8}. The application of those methods to Hi-GAL is described in \citet{Zetterlund17}. We combine this molecular cloud clump catalog with the molecular cloud catalog of \citet{Colombo18}. We also derive star formation rates from 70 $\mu$m Hi-GAL data.  Using the combination of these data sets, we explore whether higher star formation rates or shorter molecular gas depletion times are correlated with molecular cloud properties or clump properties.

\section{Data}
\subsection{Hi-GAL clump catalog}

The \textit{Herschel} infrared GALactic plane survey (Hi-GAL) \citep{Molinari2010} observed the entire $360^\circ$ of the Galactic Plane in a $2^\circ$ strip. Hi-GAL used the SPIRE \citep{Griffin2010} and PACS \citep{Poglitsch2010} detectors on the \textit{Herschel Space Observatory}, obtaining broadband imaging at 70, 160, 250, 350, and 500 $\mu$m. We use the SPIRE 500 $\mu$m maps, which show thermal dust emission, mostly from the dense molecular gas. The thermal dust emission has a low optical depth, which allows us to trace dense molecular gas throughout the Galactic disk. In addition, we use PACS 70-$\mu$m maps in order to measure star formation rates as inferred by heated dust.

\begin{figure}[!htbp]
\begin{center}
	\includegraphics[width=0.49\textwidth]{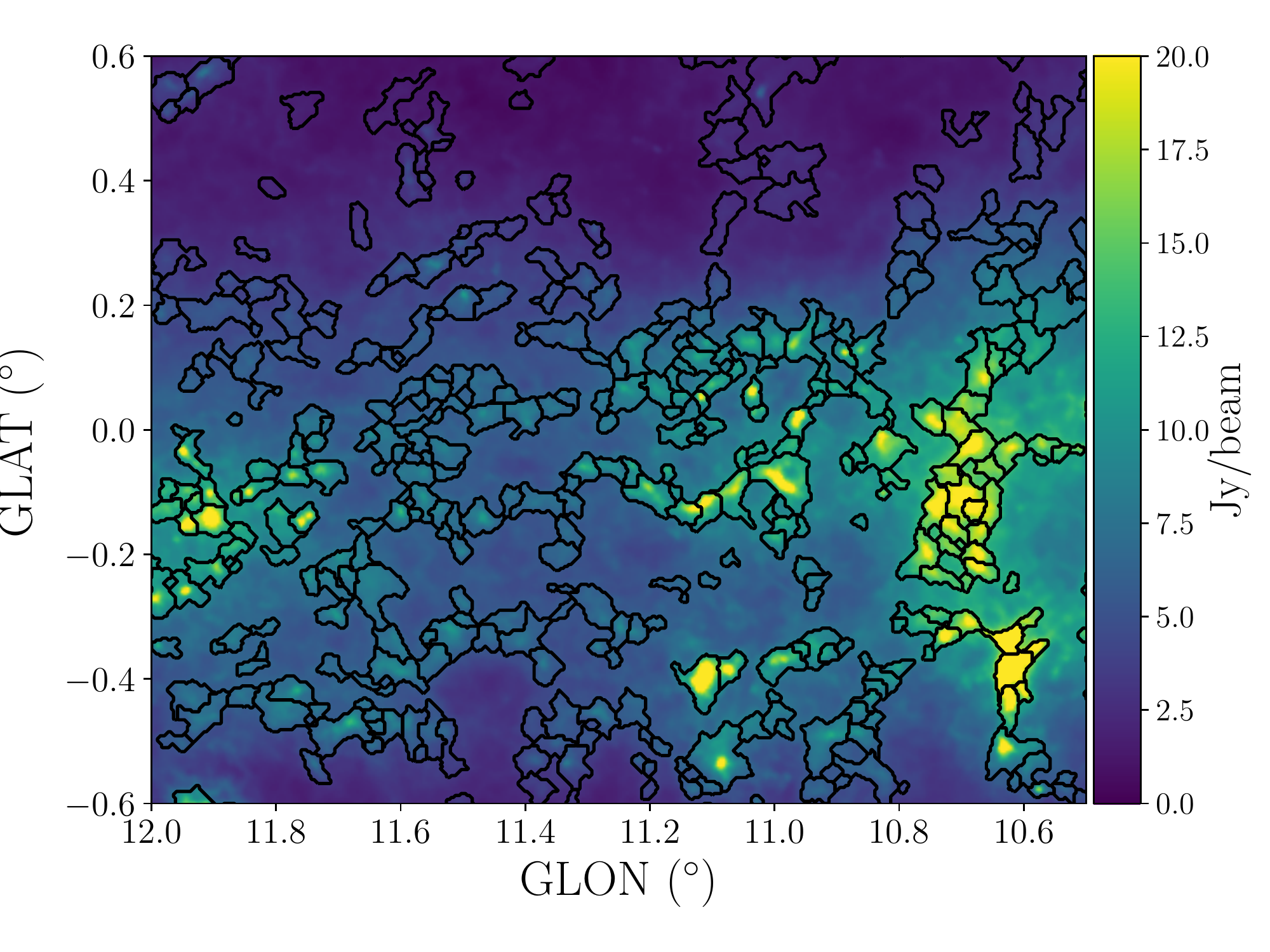}
	\includegraphics[width=0.49\textwidth]{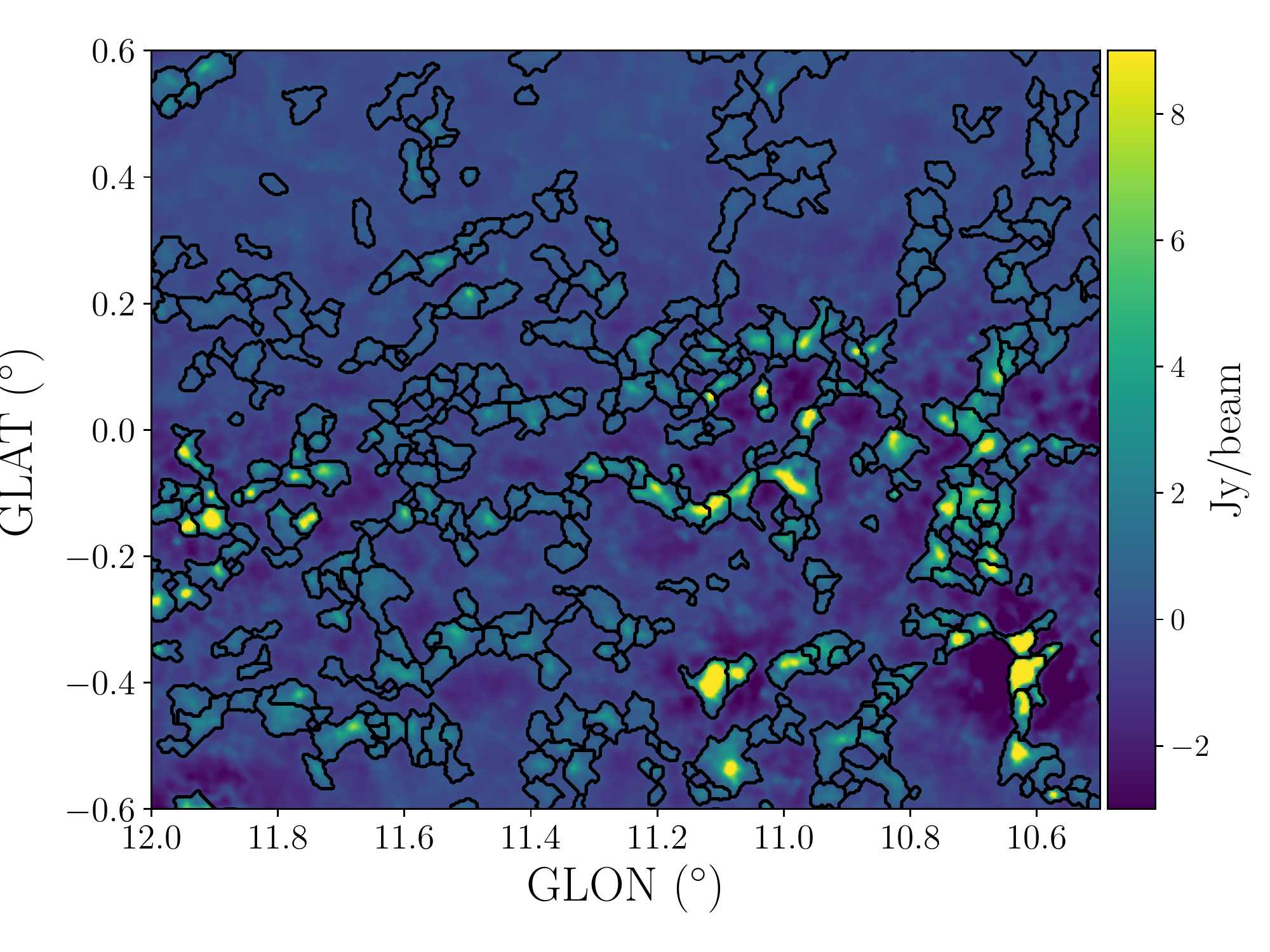}
\caption{A sample Hi-GAL map with molecular cloud clump borders overlaid. The map is shown both unfiltered (left) and high-pass filtered in order to remove large-scale structures (right).}
\label{fig:Hsample}
\end{center}
\end{figure}

Molecular cloud clumps were extracted from the 500 $\mu$m maps in \citet{Zetterlund18}. Each Hi-GAL map within the Galactic longitudes $10^\circ < \ell < 56^\circ$ was high-pass filtered to remove large-scale flux contributed from the infrared cirrus. Then, source identification was done with {\sc Bolocat} \citep{BGPS2}, a seeded watershed algorithm, on the processed maps. {\sc Bolocat} begins by identifying the pixels in a map whose emission is considered significant. It then smooths out any holes in those areas, and divides them based on the differences between peaks and saddle points in the emission map. The thresholds for significant emission and object division are based on the noise level of the map. Hi-GAL maps are confusion-limited, and their emission levels vary drastically with Galactic longitude. Therefore, instead of determining the noise level based on the instrumental noise levels, we define our noise levels on a map-by-map basis, as determined by the distribution of pixel values. To do this, we take the all of the positive pixel values from a filtered map and fit their distribution to an exponential function. The various thresholds are set proportional to the scale factor of that fit. This method reliably identifies objects consistent with visual inspection, over all Galactic longitudes, and thus it was selected. A sample map is seen in Figure \ref{fig:Hsample}. For more details, see \citet{Zetterlund17}.

Many line-of-sight clump velocities were known through {pointed observations toward BGPS clumps of HCO$^{+}(3-2)$ and N$_2$H$^+(3-2)$ emission made by \citet{Shirley13}. This catalog established velocities for 10\% of clumps, and targets without a dense gas velocity measurement had their velocities established through the morphological matching approach used in \citet{BGPS12}.  This approach matches the emission in the dust continuum to the $^{13}$CO(1-0) emission in the BU-Galactic Ring Survey \citep{Jackson06}. Morphological matching determined velocities for an additional 40\% of clumps. The remaining clumps had a significant second velocity component in the $^{13}$CO(1-0) data.

Using these velocities, distances were determined using the method developed for BGPS by \citet{BGPS8} and \citet{BGPS12}. This method probabilistically breaks the kinematic distance ambiguity using a Bayesian framework. A distance probability density function (DPDF) is calculated for each clump. The posterior DPDF is found by,
\begin{equation}
	\text{DPDF} = \mathcal{L}(d_\odot | \ell,b,v_\text{LSR}) \prod_i P_i (d_\odot | \ell,b),
\label{eq:dpdf}
\end{equation}
where $\mathcal{L}(d_\odot | \ell,b,v_\text{LSR})$ is the kinematic distance likelihood, and the $P_i (d_\odot | \ell,b)$ are various Bayesian priors that will help determine which distance has the higher probability of being true. Parallax excluded, the most powerful prior morphologically matches 8 $\mu$m absorption features to the dust continuum emission, most often placing objects exhibiting significant absorption at their near kinematic distances. Using Bayesian statistics allows for a continuous distance probability function, as opposed to a simple near-far discrimination. Once DPDFs are calculated, distance estimates can be drawn in Monte Carlo fashion, and provide robust uncertainties for cloud clump properties.

Our algorithmic definitions for a molecular cloud clump and its property extraction rely on a complex set of algorithms with free parameters that have been chosen to produce a single optimal extraction.  However, these choices represent a major systematic uncertainty in our entire analysis. To check the effects of our choices on our results, we perform the same a parallel analysis using the clump catalog from \citet{Molinari16b}. This catalog is also derived from Hi-GAL maps, but uses an entirely different clump finding algorithm. By comparing the results for these parallel analyses, we can evaluate the systematic effects of the clump catalog algorithm.

\subsection{COHRS cloud catalog}
\label{ss:cohrsdata}

The CO High Resolution Survey (COHRS) \citep{Dempsey13} is a large-scale CO survey that imaged the inner Galactic Plane in $^{12}$CO($J = 3-2$) using the James Clerk Maxwell Telescope (JCMT). The survey mapped the Galactic Plane over the longitude range $10^\circ.25 < \ell < 55^\circ.25$, with width varying from $0^\circ.5 - 1^\circ$ of latitude, for a velocity range --30\,km\,s$^{-1}<v_{LSR}<155$\,km\,s$^{-1}$. COHRS has a spatial resolution of 16'' and a velocity resolution of 1\,km\,s$^{-1}$.
The $^{12}$CO($J = 3-2$) transition was chosen as being less optically thick than the lower-J transitions of $^{12}$CO. Nevertheless, even this transition will be optically thick towards dense cores. However, the relatively diffuse material found between cores --- which is the main concern when identifying clouds --- will not have this issue. While a $^{13}$CO transition could have been used in order to avoid the optical thickness of dense cores in $^{12}$CO, it would be at the sacrifice of missing much of the diffuse material between cores. This $^{12}$CO line traces warm molecular gas (10-50\,K) and moderate densities, and thus is sensitive to giant molecular clouds (GMCs) with active star formation.

Molecular gas clouds in COHRS were identified using the Spectral Clustering for Interstellar Molecular Emission Segmentation (SCIMES) algorithm \citep{Colombo15}. This algorithm extracts a series of nested hierarchical sources in position-position-velocity (PPV) space, called a dendrogram, based on emission levels. The dendrograms consist of leaves, which correspond to local maxima and have no substructures; branches, which contain leaves and/or other branches; and a trunk, which is the superstructure which contains all other structures. These dendrograms, calculated using {\sc astrodendro}, are used to identify clouds using a graph theory framework. Dendrograms are converted into fully-connected graphs (i.e. every leaf is connected to every other leaf), which are then weighted based on properties of the isosurfaces connecting each pair of leaves. For each pair of leaves, the connecting isosurface is the surface with a constant emission level in PPV space which encloses both of the leaves. For minimally connected leaves, this isosurface is the boundary of the trunk. The weighted graphs are then used to identify coherent clouds \citep[see ][ for illustrations of this process]{Colombo15}. The similarity criteria utilized by SCIMES are based on the luminosities of the emission within the isosurface connecting two leaves, and the volumes of those isosurfaces. A pair of leaves are considered more similar if their connecting isosurface contains greater emission levels within a smaller volume. Leaves are separated into clusters which define clouds by globally maximizing intracluster similarity and minimizing intercluster similarity. Unlike algorithms such as {\sc Bolocat}, SCIMES does not separate GMCs into their substructures, thus allowing us to compare properties of entire clouds (from COHRS) to those of their embedded clumps (from Hi-GAL). For details on the application of SCIMES to the COHRS dataset, see \citet{Colombo18}. A sample of the COHRS clouds identified by SCIMES is shown in Figure \ref{fig:dendro}. The right shows a visual representation of the full dendrogram, whereas the left shows the ultimate mask selection.

\begin{figure}[!htbp]
\begin{center}
	\includegraphics[width=0.49\textwidth]{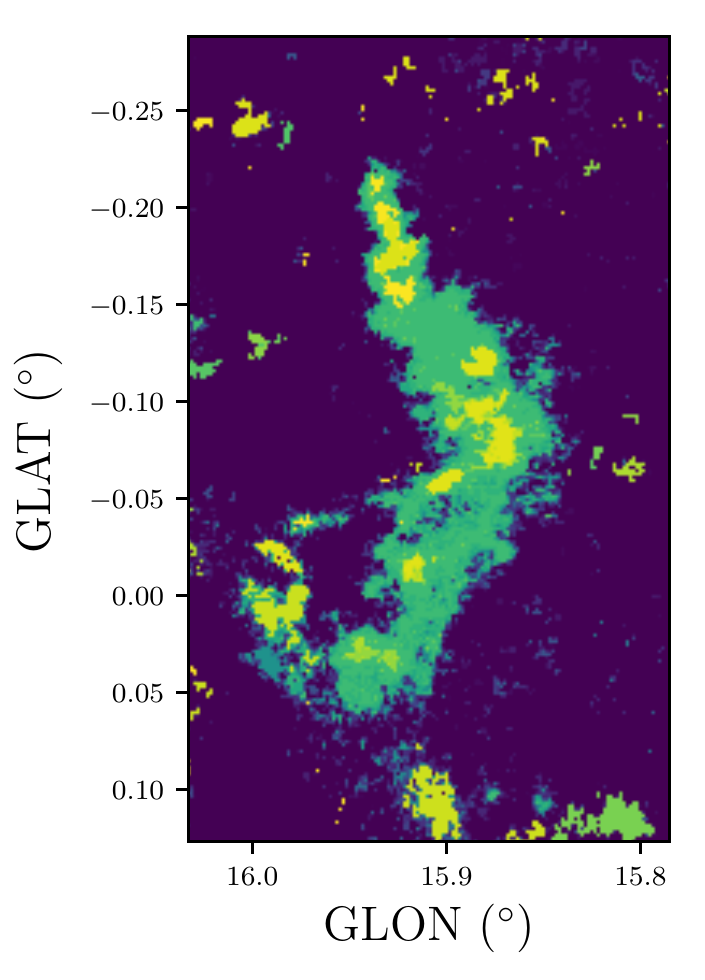}
	\includegraphics[width=0.49\textwidth]{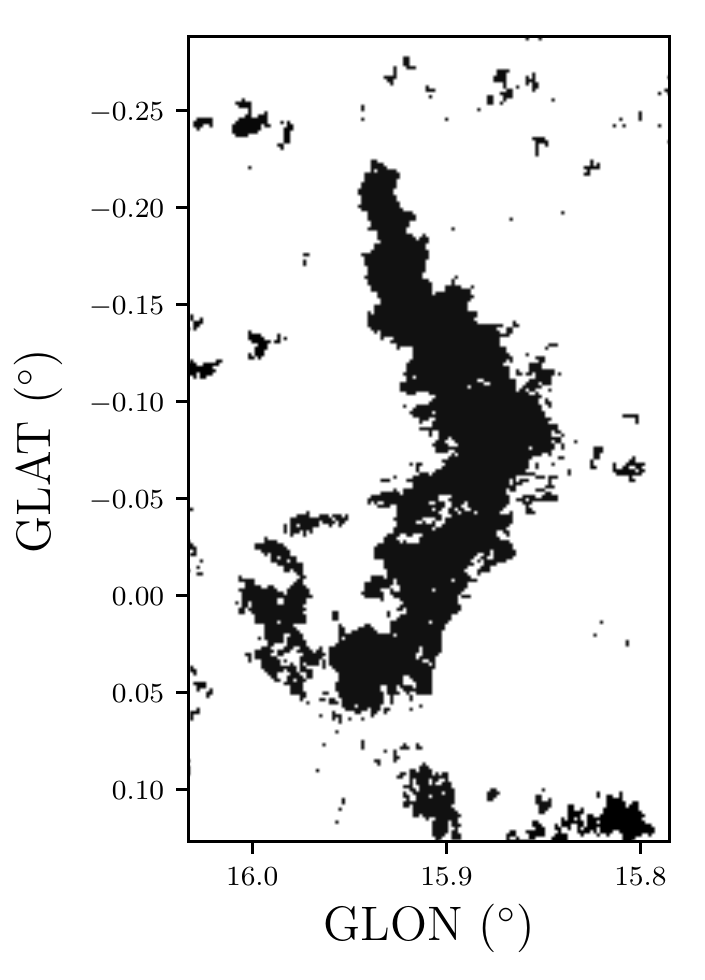}
\caption{A sample image of the dendrogram structure underlying the COHRS clouds. {\it Left}: The full dendrogram structure for a sample region, represented as colored masks, with masks for smaller structures taking priority to masks for the larger structures in which they are nested. Colors are assigned based on dendrogram structure number, and therefore more yellow pixels correspond to objects higher in the dendrogram hierarchy. {\it Right}: A binary view of the resulting COHRS cloud masks. In addition to the clouds on the right collapsing the dendrogram substructures, there are dendrogram levels lower than those chosen as COHRS objects, which were considered insignificant by the graph theory algorithm that made the separations into clusters.}
\label{fig:dendro}
\end{center}
\end{figure}

Distances are assigned to COHRS clouds using the DPDFs from \citet{Zetterlund18}. Each Hi-GAL distance is assigned to a pixel in PPV space using their positions on the sky and line-of-sight velocities. The mean of any distance pixels falling within a cloud is assigned to that cloud. Clouds containing no distance pixels take on the distance of the nearest cloud within the dendrogram structure. Where this is not possible, as with isolated clouds, the nearest distance in PPV space is used \citep{Colombo18}.

\section{Method}
\subsection{Matching Hi-GAL clumps to COHRS clouds}
\label{ss:match}

We match Hi-GAL clumps to their host COHRS clouds using the COHRS masks and Hi-GAL peak positions and velocities. For clumps with known velocities, matching is straightforward. Clumps which meet the local flux threshold are associated with whichever cloud's PPV mask they fall within. We adopt a local flux threshold in response to the variance with Galactic longitude of the flux sensitivity of the maps \citep[as discussed in][]{Zetterlund17}. We set this threshold to be equivalent to three beams' worth of emission at the clump identification threshold used in {\sc Bolocat}. We chose this threshold as the value that minimizes dubious matches, while retaining as many likely matches as possible. So long as a cloud has at least one clump that lies within the cloud's PPV mask and meets the flux criterion, all of the cloud's velocity-matched clumps are kept, including those below the integrated flux density threshold. We also allow for secondary clumps to be associated to the cloud if their velocity is within $1\sigma$ of the cloud, but not strictly inside its mask. Each of these secondary match criteria account for fewer than 10\% of matched clumps, and were included because they recover otherwise unmatched clumps which are clearly contained within clouds on visual inspection. We have determined that it is more likely that their inclusion avoids an underestimation of mass contained within clumps than causes an overestimation.

However, not all Hi-GAL clumps have known velocities. While we do not allow clouds to be matched solely to velocity-less clumps, we make additional matches based on the dendrograms from which the clouds were identified. To do this, we make a list of all the dendrogram structures at the peak clump $(\ell,b)$ position, collapsing the dendrogram object masks along the velocity axis. We then trace each dendrogram structure back to its parent cloud. If that dendrogram structure contains 4 or fewer leaves, contains 10\% or fewer of the total number of leaves contained in the full cloud, and there is only one cloud which meets the prior two criteria, we associate the clump with the cloud. Since leaves in SCIMES are often smaller than a beam, it is not useful to require the clump be aligned with a dendrogram leaf. Since SCIMES only allows for binary mergers in its dendrograms, 4 leaves allows for the object to be two hierarchical levels down from a single leaf. However, not all clouds contain a large number of leaves, and therefore we use the $\le$10\% of total leaves criterion to ensure that the clump is aligned with a peak in the emission and not merely a small cloud. We found our cut-offs to simultaneously optimize for confidence in the match and number of matches found. Tighter requirements made for fewer matches, and for missing many matches that were obvious by eye. Loosening the requirements resulted in fewer matches at lower confidence, as more than one cloud along the line of sight could meet the criteria, leading to the clump not being matched to either cloud. The clump is also matched if there is exactly one cloud along the line of sight, regardless of leaf quantities.

Within a turbulent ISM that has structure at all detectable scales, defining clouds, clumps, and cores in a physically meaningful way has some degree of subjectivity.  The utility of a catalog approach is the ability to measure physical properties of structures that can be identified in the ISM.  Our cloud catalog filters for regions of CO emission based on compactness of the structures in a position-position-velocity data set and the clump catalog is built around a high-pass filter with a characteristic angular scale.  The requirement that clumps form clear substructures of clouds means that the clump measurement is by definition picking distinct substructures with the cloud regions. To assess the systematics in our clump catalog approach, we carry out a parallel matching process using the catalog of \citet{Molinari16b}. We find $>90\%$ of the clouds are matched to clumps in both catalogs, which enables using this catalog as a direct test of the influence of the clump catalog on our results.

\subsection{Clump properties}
\label{ss:clumpprops}

Physical radii are given as $R = \theta_R d_\odot$, where $\theta_R$ is angular radius and $d_\odot$ is heliocentric distance. $\theta_R$ is calculated as the geometric mean of the deconvolved major and minor axes of the flux density distribution of the clump,
\begin{equation}
\theta_R = \eta [(\sigma_{\text{maj}}^2 - \sigma_{\text{beam}}^2)(\sigma_{\text{min}}^2 - \sigma_{\text{beam}}^2)]^{1/4},
\label{eq:dc}
\end{equation}
where, for Hi-GAL PLW (500 $\mu$m), $\sigma_{\text{beam}} = \theta_{\text{FWHM}}/\sqrt{8\ln{2}} = 15''$, $\theta_{\text{FWHM}} = 35''$, and $\eta = 2.4$ is a factor relating the rms size of the emission distribution to the true size of the source, which is adopted from \citet{BGPS2}.

Our clump mass estimates use high-pass filtered flux densities at 500 $\mu$m. The conversion from flux density to mass is 
\begin{equation}
	M = \frac{\Gamma d_\odot^2}{\kappa_{500} B_{500}(T)} S_{500},
\label{eq:mass}
\end{equation}
where $\Gamma=100$ is the gas-to-dust mass ratio, $\kappa=5.04$ cm$^2$g$^{-1}$ is the opacity at 500 $\mu$m \citep{OH94}, $B_{500}(T)$ is the Planck function at 500 $\mu$m, $d_\odot$ is the heliocentric distance, and $S_{500}$ is the integrated flux density from 500 $\mu$m maps.  We do not carry out a multiband analysis of our clump properties to determine the dust temperature. Instead, we rely on the temperature determinations from \citet{Elia17} who have carried out modified blackbody fits to the catalog of \citet{Molinari16b}. We match the catalogs, assigning a dust temperature as being the same as for the closest clump from \citet{Elia17}. In the catalog matching, $>90\%$ of our sources have temperature measurement from a clump within $30''$, and only two sources have a separations larger than $3'$. The mean dust temperature found for these clumps is $\langle T \rangle = 14~\mathrm{K}$ and the distribution of adopted dust temperatures is shown in Figure \ref{fig:tdust_hist}.  Comparing our extracted temperatures to Figure 5 in \citet{Elia17} shows that the dust clumps follow a similar distribution, spanning the ranges seen in both prestellar and protostellar clumps in their analysis.

\begin{figure}
    \centering
    \includegraphics[width=0.45\textwidth]{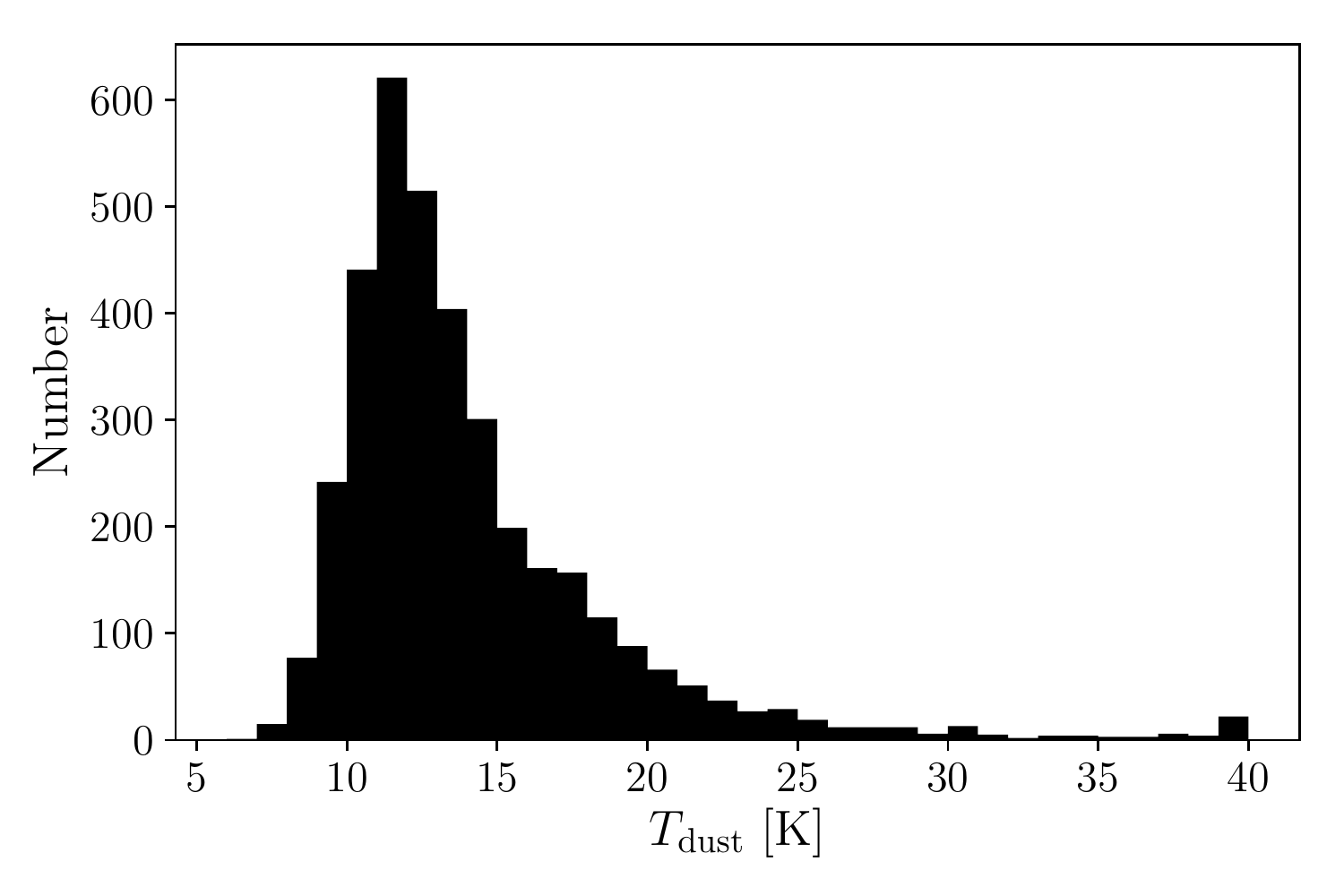}
    \caption{Histogram of adopted dust temperature obtained by spatially matching our clump catalog to the catalog used in \citet{Elia17}. The mean dust temperature is $\langle T\rangle = 14~\mathrm{K}$}
    \label{fig:tdust_hist}
\end{figure}

The primary sources of uncertainty in our clump masses come from two sources: photometric errors and the absolute scale of dust opacities in the ISM.  Photometric errors are relatively low and arise from the confusion in the Hi-GAL maps. We calculate photometric errors by assuming the per-pixel confusion noise level can be estimated as the flux density level used for clump identification with {\sc Bolocat} \citep[see][]{Zetterlund17}.  Dust opacities in the ISM are known to a factor of two \citep[e.g.,][]{OH94, Draine07} and this drives our systematic uncertainty.

Number densities for clumps are calculated assuming that the volume of a clump can be approximated as a sphere with radius $R$. Using this volume, along with the mass, $M$, from Equation \ref{eq:mass}, we obtain number densities for clumps using the equation,
\begin{equation}
	n = \frac{3 M}{4\pi \mu m_\mathrm{H} R^3},
\label{eq:nclump}
\end{equation}
where $\mu=2.4$ is the mean molecular mass for the molecular ISM, and $m_\mathrm{H}$ is the mass of a hydrogen atom. Note that this calculates the average number density for the clump, but we expect clumps to be centrally concentrated, so these number densities are sensitive to the extent of their borders defined by {\sc Bolocat}.

\subsection{Cloud properties}
\label{ss:cloudprops}

We define a cloud's clump formation efficiency as the fraction of the cloud's gas which was converted into dense clumps that may form stars. We estimate this quantity as our observed clump mass fraction,
\begin{equation}
	\text{Clump Mass Fraction} =  \frac{M_{\text{clump}}}{M_\text{GMC}} = 
		\frac{\sum\limits_{i} M_{\text{clump},i}}{M_\text{GMC}}.
\label{eq:cfe}
\end{equation}
That is, the sum of the clump masses contained in a cloud divided by the cloud mass, where $i$ indexes the individual clumps.

Cloud properties derived from $^{12}$CO(3-2) emission were calculated by \citet{Colombo18} and are summarized here. The effective radius of a cloud is calculated from its Gaussian fit as 
\begin{equation}
	R_\mathrm{eff} = \eta \sqrt{\sigma_\mathrm{maj} \sigma_\mathrm{min}},
\label{eq:cr}
\end{equation}
where $\sigma_\mathrm{maj}$ and $\sigma_\mathrm{min}$ are the major and minor axes calculated as the intensity-weighted second moments along the two spatial dimensions of all of the pixels within the object mask. $\eta = 1.91$ is adopted as the scaling factor, following \citet{Rosolowsky06} and \citet{Solomon87}. This $\eta$ value is different than the one cited in Section \ref{ss:clumpprops} for determining clump radii. The $\eta=2.4$ scaling factor used for clumps was calculated analytically for a spherical density profile of $\rho \propto r^{-1}$. The $\eta=1.91$ value used for clouds includes an empirical correction calculated by \citet{Solomon87}, who studied GMCs using CO data.

We calculate the mass of a cloud using its $^{12}$CO(3-2) luminosity. This so-called luminosity mass is calculated as
\begin{equation}
	M_\mathrm{lum} = \frac{\alpha_\mathrm{CO}}{R_{31}} L_\mathrm{CO},
\label{eq:cml}
\end{equation}
where $L_\mathrm{CO}$ is the $^{12}$CO(3-2) luminosity. $R_{31} = 0.5$ is the CO(3-2)/CO(1-0) flux ratio, calculated using the $^{12}$CO(3-2)/$^{12}$CO(1-0) UMASS-Stony Brook and COHRS data \citep{sanders86, Colombo18}. This ratio is less certain towards the low-mass end, which is fortunately not where our clouds generally lie. The Galactic $^{12}$CO(1-0)-to-H$_2$ conversion factor is $\alpha_\mathrm{CO} = 4.35$ M$_\odot$ pc$^{-2}$ (K km s$^{-1}$)$^{-1}$, known to a precision of $\pm0.1$ dex through the Milky Way, based on the aggregate of a variety of techniques \citep[e.g.,][]{Bolatto13}. \citet{Lee18} also demonstrate that a constant CO-to-H$_2$ conversion factor is a good assumption on a cloud-by-cloud basis for systems with solar and supersolar metallicity, with a variations of less than factor of 2 between individual clouds.  Combining the cloud-to-cloud scatter with the spread in line ratio at high luminosity, we note that CO-emission-based masses for clouds can be uncertain by 0.5 dex. In the analysis that follows, cloud mass, $M_\mathrm{GMC}$, is taken to be the cloud's mass derived from $^{12}$CO(3-2) line luminosity, as defined above.
From this, we also calculate the surface mass density as
\begin{equation}
	\Sigma (\mathrm{M_\odot \ pc^{-2}}) = \frac{M_\mathrm{lum}}{\pi R_\mathrm{eff}^2},
\label{eq:surfden}
\end{equation}
and the number density as
\begin{equation}
	n = \frac{3 M_\mathrm{lum}}{4\pi \mu m_\mathrm{H} R_\mathrm{eff}^3},
\label{eq:ncloud}
\end{equation}
where $\mu=2.4$ is the mean molecular mass, and $m_\mathrm{H}$ is the mass of a hydrogen atom.  The well resolved clouds are not spherical (Figure \ref{fig:dendro}), so changing densities of clouds may also reflect changing morphologies.

In addition, the virial mass is calculated using the cloud's radius, $R_\mathrm{eff}$, and velocity dispersion, $\sigma_v$, as
\begin{equation}
	M_\mathrm{vir} = \frac{5\sigma_v^2R_\mathrm{eff}}{G},
\label{eq:cmv}
\end{equation}
where $G$ is the gravitational constant, and $\sigma_v$ is calculated as the intensity-weighted second moment along the velocity axis. The two mass estimates are used to determine the virial parameter,
\begin{equation}
	\alpha_\mathrm{vir} = \frac{M_\mathrm{vir}}{M_\mathrm{lum}} = \frac{2E_{\mathrm{kin}}}{|E_\mathrm{grav}|},
\label{eq:vir}
\end{equation}
which compares the kinetic energy $E_{\mathrm{kin}}$ to the gravitational binding energy $E_{\mathrm{grav}}$.  In the absence of other effects such as rapid time variation of the mass distribution, magnetism, and pressure, $\alpha_\mathrm{vir} > 2$ indicates that the cloud is stable against collapse, whereas $\alpha_\mathrm{vir} \ll 2$ suggests that the random motions within the cloud are not great enough to support the cloud, which may be collapsing. We regard the virial parameter as an approximate comparison of the magnitude of the gravitational binding energy to the internal kinetic energies of the clouds rather than a strict diagnostic of the cloud's dynamical state.

In the subsequent analysis, we partition the sample into clouds those dominated self-gravity $\alpha_\mathrm{vir}<2$ and those dominated by higher internal kinetic energies.  The goal of this partitioning is to determine whether the large-scale cloud potential is connected the formation of substructure or the internal star formation rate.  In the study of \citet{Colombo18}, there is 1 dex scatter in the virial parameter at a given mass, but on average, the mean virial parameters decreases as cloud mass increases to that higher mass clouds are typically more strongly self-gravitating. Even so there is significant scatter so we divide the sample to look for variation, returning to a synthesis of this analysis in Section \ref{s:vir}.

The methods of \citet{Colombo18} correct for the finite sensitivity of COHRS using the extrapolation techniques developed by \citet{Rosolowsky06}.  These corrections are made by first determining the basic moments of the cloud ($\sigma_\mathrm{maj}$, $\sigma_\mathrm{min}$, $\sigma_v$, $F_\mathrm{CO}$) at various identification thresholds. The values at these thresholds are then linearly fit (quadratic polynomial fit for $F_\mathrm{CO}$) in order to determine the moments at infinite sensitivity (i.e. 0 K). Once these extrapolated moments are calculated, the beam width is deconvolved from the size of the cloud ($\sigma_\mathrm{maj}$, $\sigma_\mathrm{min}$, $\sigma_v$) by subtracting in quadrature. Note that these corrections have a more significant effect on smaller sources. As we are concerned with the largest clouds in the COHRS catalog, the effects on our data are minimal.

Star formation rates for clouds are calculated using Hi-GAL 70 $\mu$m images. This is the short wavelength limit for Hi-GAL and corresponds to dust that is heated by new stars. We relate star-formation rate to 70 $\mu$m luminosity using the calibration of \citet{Calzetti10}:
\begin{equation}
	\mathrm{SFR}(\mathrm{M}_\odot~\mathrm{yr}^{-1})
			= \frac{L(70) (\mathrm{erg}~\mathrm{s}^{-1})}{1.7\times10^{43}}.
\end{equation}
The scatter about this relation is $\sim0.2$ dex, or about 60\%. However, this calibration was developed using individual regions with typical scale 100s of pc and it will be less reliable for individual GMCs. At this wavelength, the emission traces light from stars of ages $0-100$ Myr, but the mean emission-contributing stellar age is 5 Myr.

While our clouds are defined in PPV space, we are using a continuum tracer to measure star formation. We control for this and estimate our uncertainties due to this mismatch as follows. To extract the flux densities required for conversion to SFRs, we integrate the flux density found within the velocity-collapsed COHRS mask. We account for background flux and cirrus emission by creating a ``rind'' around the mask with a width of 1 to 1.5 resolution elements, mimicking the process of aperture photometry but for irregular ISM structures. We fit a polynomial with the form $f(x) = c_0 + c_1x + c_2y+c_3xy$ to this rind and subtract the resulting background flux density from the source flux density. Our main source of systematic uncertainty is due to the superposition of multiple clouds along a single line-of-sight, although background subtraction mitigates much of this. We estimate our fractional SFR uncertainty as the fraction of CO flux along the line of sight which is not associated with the cloud. This accounts for uncertainties in our background subtraction, and asks the question, if all of the gas emitting in $^{12}$CO(3-2) is equally forming stars, how much 70 $\mu$m emission did we mistakenly attribute to this cloud?

\section{Results of Clump and Cloud Matching}

We matched a total of 4,704 Hi-GAL clumps to 1,182 COHRS clouds. Of those 4,704 clumps, 3,157 were matched using their velocities (of which 423 did not meet the flux threshold, and 121 had velocities slightly outside the COHRS mask), and 1,124 were matched using the COHRS dendrograms. We focus this work on massive, well-resolved clouds for which we can carry out good matching to clump substructure, so we limit the main analysis to the clouds at size scales of $\sigma_\mathrm{major}>3'$ (see Section \ref{ss:cfedist}). Using this cut, we retain a subset of the above containing 3,674 clumps matched to 473 clouds. Two sample clouds with their matched clumps are shown in Figure \ref{fig:matching}. Clouds (black contours) at large angular size scales typically have multiple clumps matched to them, using both velocity information (blue contours) and dendrogram structure (green contours) for those clumps without velocity information.

\begin{figure}[!htbp]
\begin{center}
	\includegraphics[height=0.35\textheight]{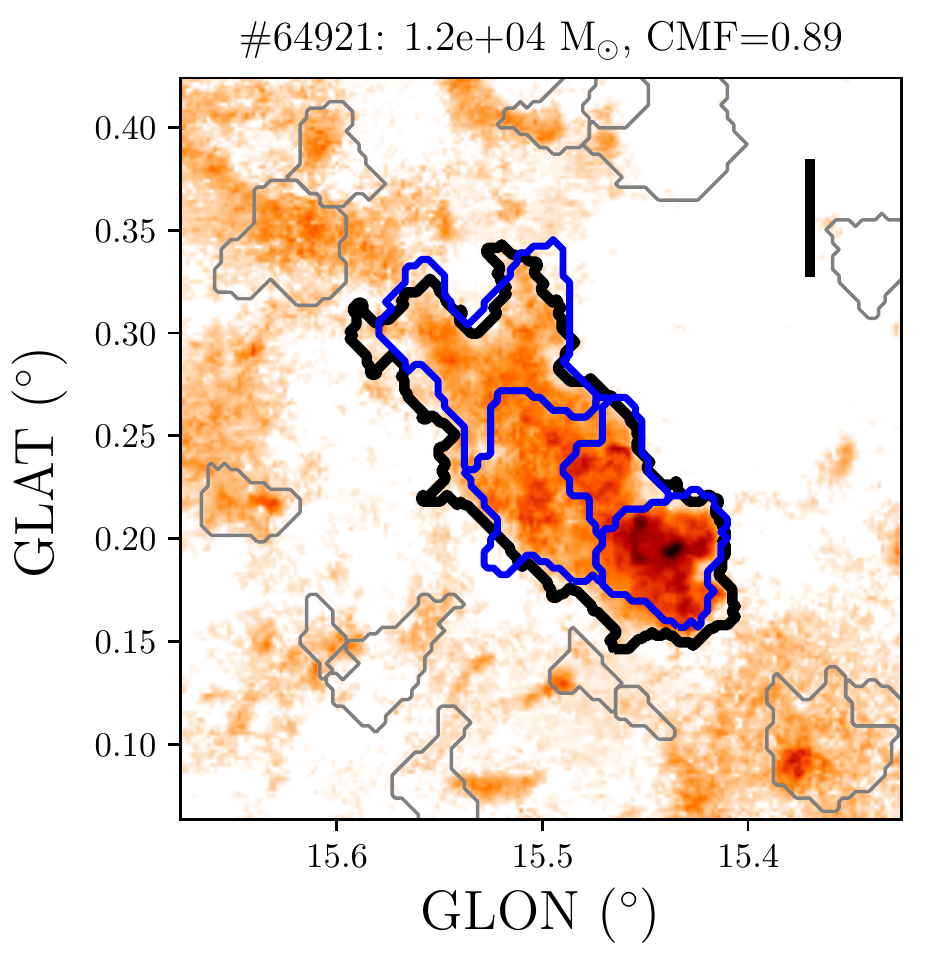}
	\includegraphics[height=0.35\textheight]{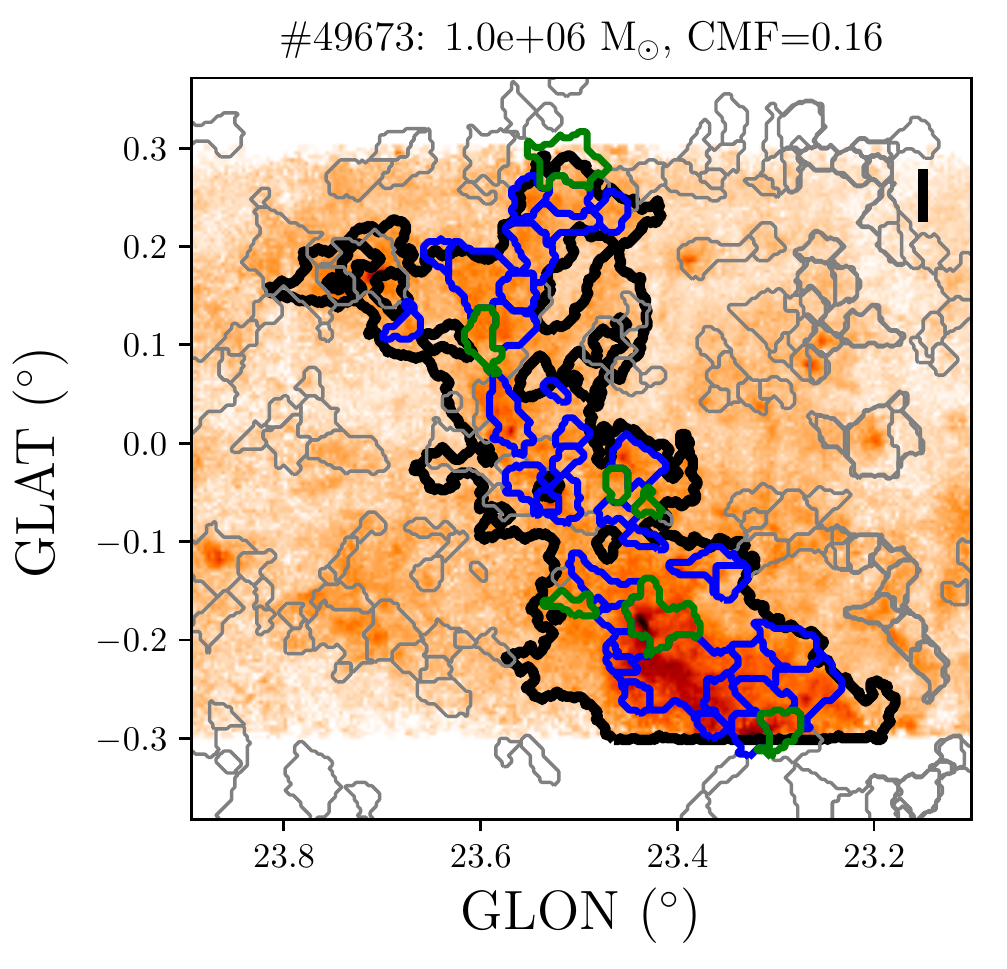}
\caption{Sample clump-to-cloud matching, at two different size scales. $^{12}$CO($J = 3-2$) emission is shown in orange, integrated over the velocity extent of the cloud. The thick black contour shows the COHRS cloud. Blue and green contours show Hi-GAL clumps matched to the cloud via velocity and dendrogram structure, respectively. Thin grey contours show other Hi-GAL clumps, which were not matched to the cloud contoured in the image. Emission (orange) outside of the black contour occurs in the same velocity range as the cloud shown, but is not associated with it, according to SCIMES. The black bars in the upper right corners are 5 pc in length, for scale. The COHRS catalog number, cloud mass, and clump mass fraction are found above each image. Note that the left hand cloud resides in the high-end tail of the clump mass fraction distribution.}
\label{fig:matching}
\end{center}
\end{figure}

Although the COHRS catalog is extensive, containing 85,020 objects of widely varying sizes, only 724 of those are clouds that have angular sizes of $\sigma_\mathrm{major}>3'$. We have matched clumps to 65\% of these clouds. If we additionally require the clouds to have $M>10^4~\mathrm{M}_\odot$, we associate clumps with 75\% of clouds, and 95\% for $M>10^5~\mathrm{M}_\odot$. Of the 19,886 clumps in our Hi-GAL catalog, 7,362 lie within the narrower latitude coverage of COHRS. Of these, 64\% were matched to COHRS clouds (before instituting the angular size cut). Furthermore, 4,046 clumps both lie within the COHRS coverage and have known velocities. Our match rate for these known-velocity clumps is 78\%.

The selection based on angular size restricts this analysis to the most massive clouds. In comparing our sample to the parent catalog of molecular clouds sample, the angular size cut limits us to clouds with $M>3\times 10^4~M_\odot$, but $\sim 90$\% (319 of 360) of the clouds above this mass in the COHRS sample fulfill the angular size criterion.  Hence, the results that follow are summarizing the properties of the massive cloud population.  The angular size cut also limits the analysis to clouds with $R>12~\mathrm{pc}$ but again, we see $>90\%$ (454 of 488) of the clouds in the COHRS population fulfill this criterion.    Of the cloud population that fulfills the angular size limit, the clouds that match to Hi-GAL clumps are slightly higher in mass with $>90\%$ completeness for $M>5\times 10^4~M_\odot$ (205 of 228) and $R>20~\mathrm{pc}$ (160 of 181).  Hence, the angular size cut limits us to massive, large clouds in physical units, but we are recovering most of the molecular cloud population that is found in the COHRS catalog.  The COHRS catalog itself also suffers from distance selection biases for these massive clouds \citep[see discussion in ][]{Colombo18} but our angular size cut does not significantly change those biases.

\begin{figure}[!htbp]
\begin{center}
	\includegraphics[height=0.6\textheight]{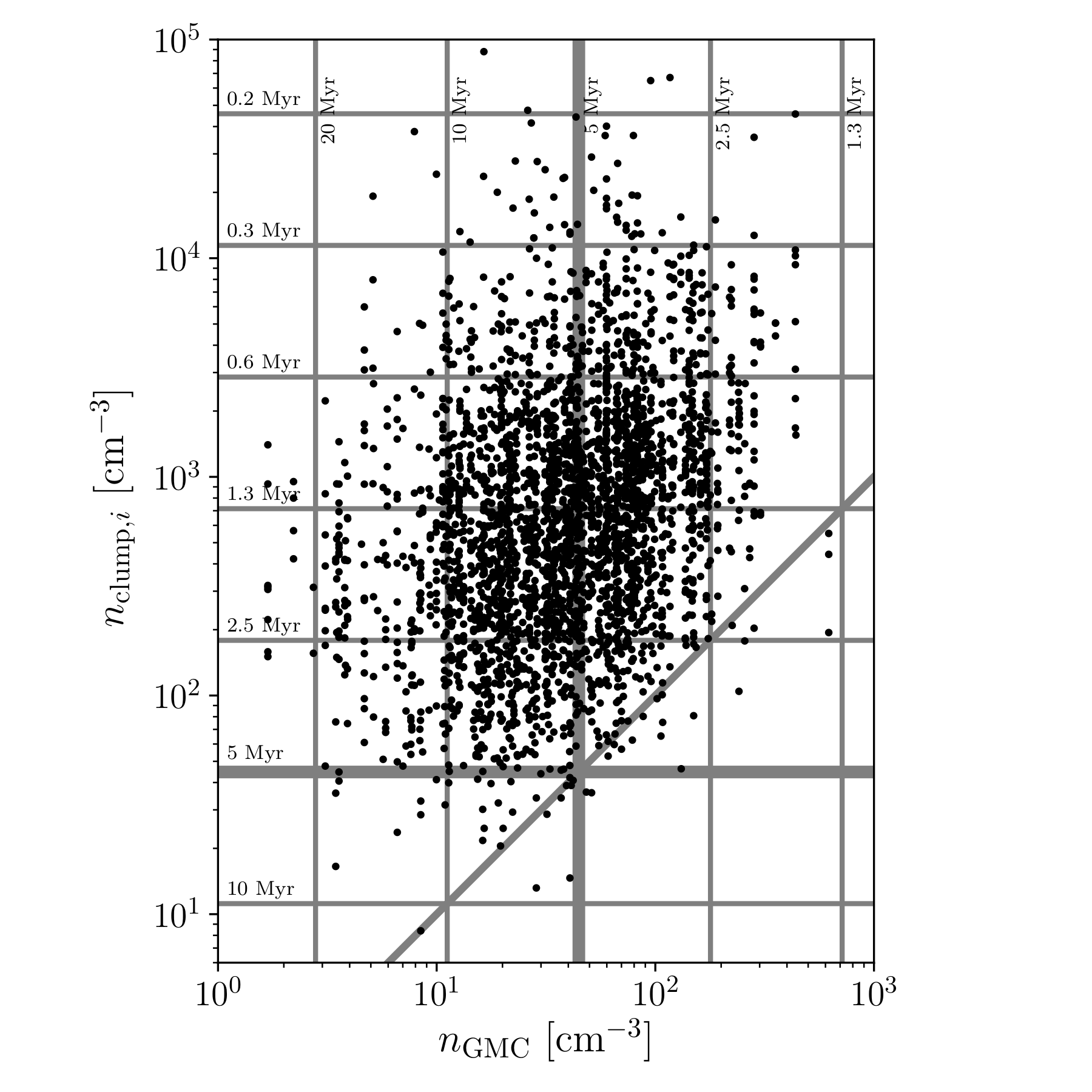}
\caption{Number densities of individual clumps, plotted against their host clouds. Uncertainties (including systematic) are roughly 0.5 dex for clouds \citep{Colombo18} and 0.3 dex for clumps. The locus of equality is shown in grey. Because of the large uncertainties, the handful of points below this line do not pose a problem. Horizontal and vertical lines correspond to free-fall times for clumps and clouds, respectively. The thick lines show $t_\mathrm{ff}=5~\mathrm{Myr}$, with additional lines varying from their neighbors by a factor of 2. Note that we do not expect all GMCs to be collapsing.}
\label{fig:nn}
\end{center}
\end{figure}

In order to check that we are indeed probing denser regions with our Hi-GAL clump catalog than with our COHRS cloud catalog, we plot the number densities of individual clumps against those of their host clouds in Figure \ref{fig:nn}. Mean cloud and clump densities are 58 cm$^{-3}$ and 1700 cm$^{-3}$, respectively. Both of these values are on the sparser end of their nominal density ranges of $n_\mathrm{cloud}=50-500~\mathrm{cm}^{-3}$ and $n_\mathrm{clump}=10^3-10^4~\mathrm{cm}^{-3}$. Despite this, nearly all of the points lie above the identity line, meaning that the clumps are denser than the cloud in which they are embedded. A handful of points lie below the identity line. These are likely a product of the substantial uncertainties on the density values. Clump densities appear to be dependent on cloud densities only in that clumps must be denser than clouds. No further correlation is seen. Typical values for $\log_{10}(n_{\mathrm{clump,}i}/n_\mathrm{GMC})$ are $1.2\pm0.6$. Density distributions for both structure types are well-described by lognormal functions. The parameters for these fits are $\mu = 3.6$, $\sigma = 1.0$ for clouds; and $\mu = 6.4$, $\sigma = 1.2$ for clumps, with densities measured in cm$^{-3}$. Free-fall times are shown as a grid of grey lines. The mean free-fall times for clouds and clumps are 6.1 Myr and 1.6 Myr respectively.

\begin{figure}[!htbp]
\begin{center}
	\includegraphics[width=0.49\textwidth]{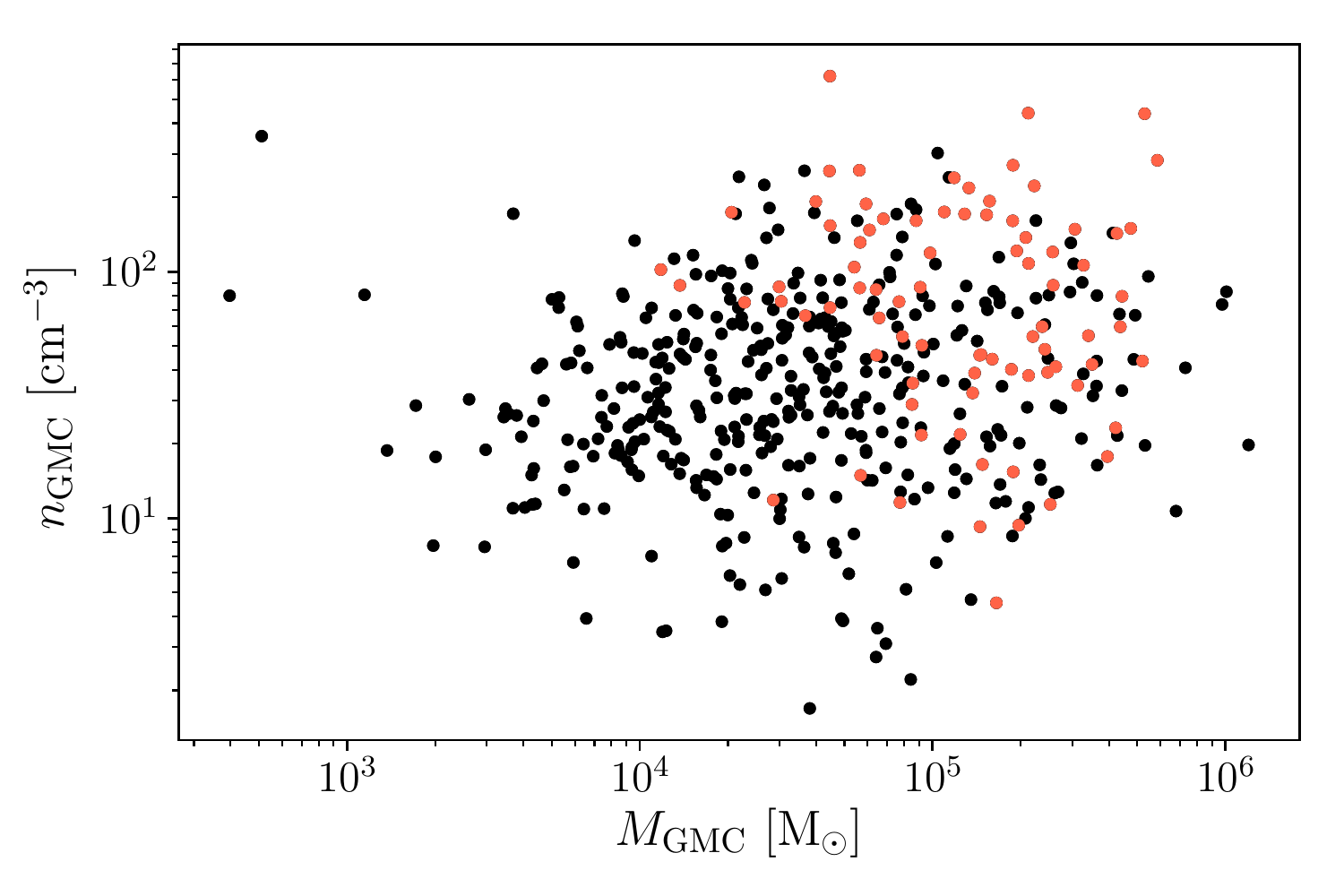}
	\includegraphics[width=0.49\textwidth]{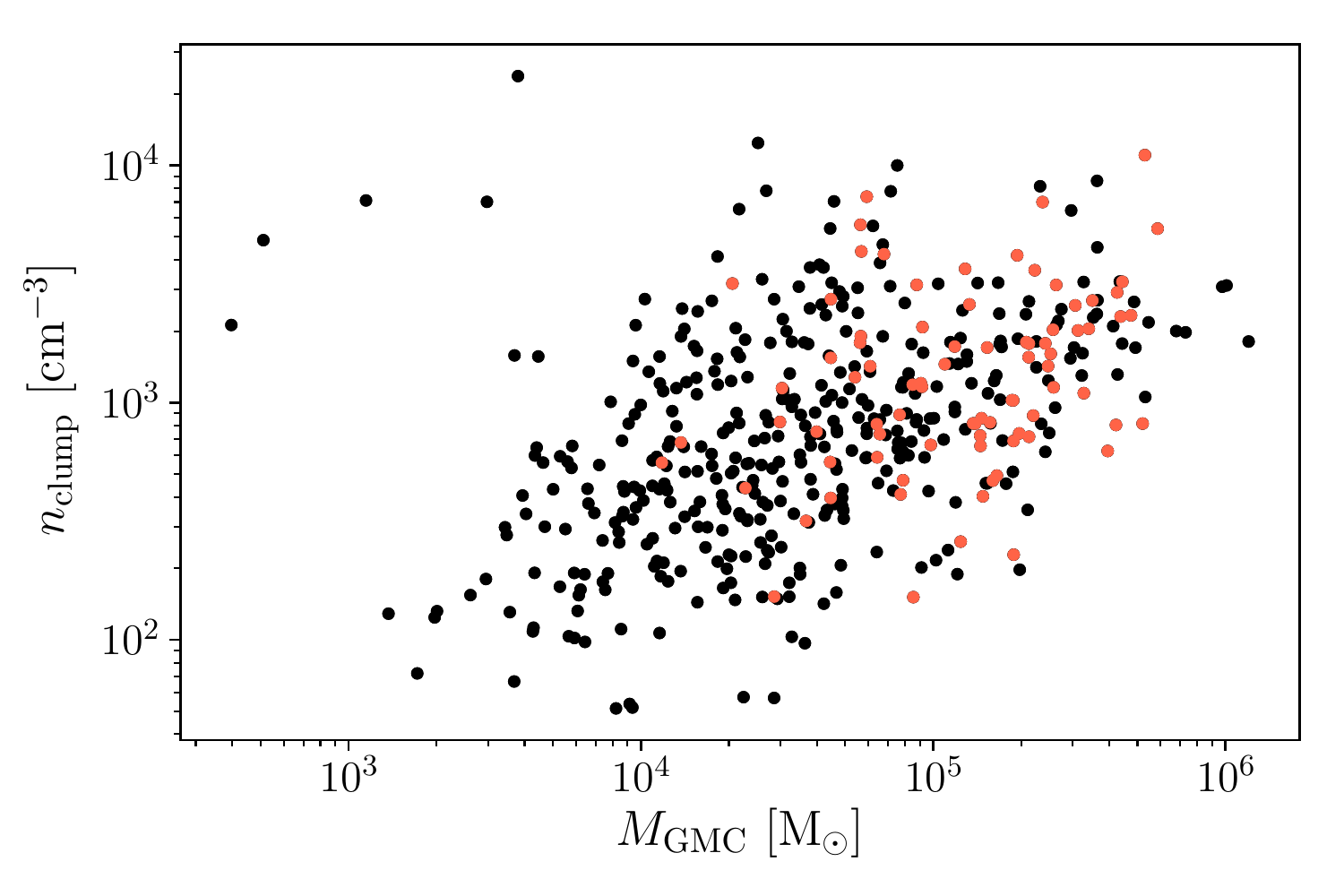}
\caption{{\it Left}: Cloud number densities plotted against their masses. {\it Right}: Average clump number densities plotted against the masses of their host clouds. Red points indicate clouds with virial parameters $\alpha_\mathrm{vir}<2$.}
\label{fig:n_Mgmc}
\end{center}
\end{figure}

In Figure \ref{fig:n_Mgmc} we investigate the effects of cloud mass on cloud and clump number densities. We see that, while cloud number density is unaffected by cloud mass, more massive clouds tend to host denser clumps.  In this figure, and in subsequent analyses, we highlight the subset of clouds with virial parameters $\alpha_{\mathrm{vir}}<2$ to explore whether those clouds that are most dominated by self-gravity on the large scale occupy different parts of the parameter spaces than those with $\alpha_\mathrm{vir} > 2$.  In this case, we that $\alpha_{\mathrm{vir}} < 2$ cloud are preferentially higher mass but do not show obvious offsets in either density measurement.

\section{Clump Formation Efficiency}
\label{s:cfe}

We calculate mass fractions using Equation \refeq{eq:cfe}.  We expect these mass fractions to be reliable estimates of the clump formation efficiencies (CFEs). It is possible that some of our so-called clumps are not actually gravitationally bound, but instead are extended along the line-of-sight, artificially increasing the mass fraction by including gas mass which is less concentrated than a clump. In addition, there were likely clumps which did not meet our matching criteria, but should have been associated with a cloud in our catalog, thus artificially lowering our mass fractions. We expect the overall effect of these caveats to be a slight underestimation in the clump formation efficiency of clouds.

\subsection{CFE Distribution}
\label{ss:cfedist}

Figure \ref{fig:cfes} (left) shows these mass fractions, plotted against cloud mass. The mass fractions span several orders of magnitude, with some points over unity. The Hi-GAL maps were high-pass filtered at a scale of $3'$ before clumps were identified. Due to this, at angular scales smaller than $3'$, we are prone to matching clouds and clumps at similar size scales, and is some cases even match clumps to clouds where the clump is larger than its cloud. This leads to mass fractions greater than one.

\begin{figure}[!htbp]
\begin{center}
	\includegraphics[width=0.49\textwidth]{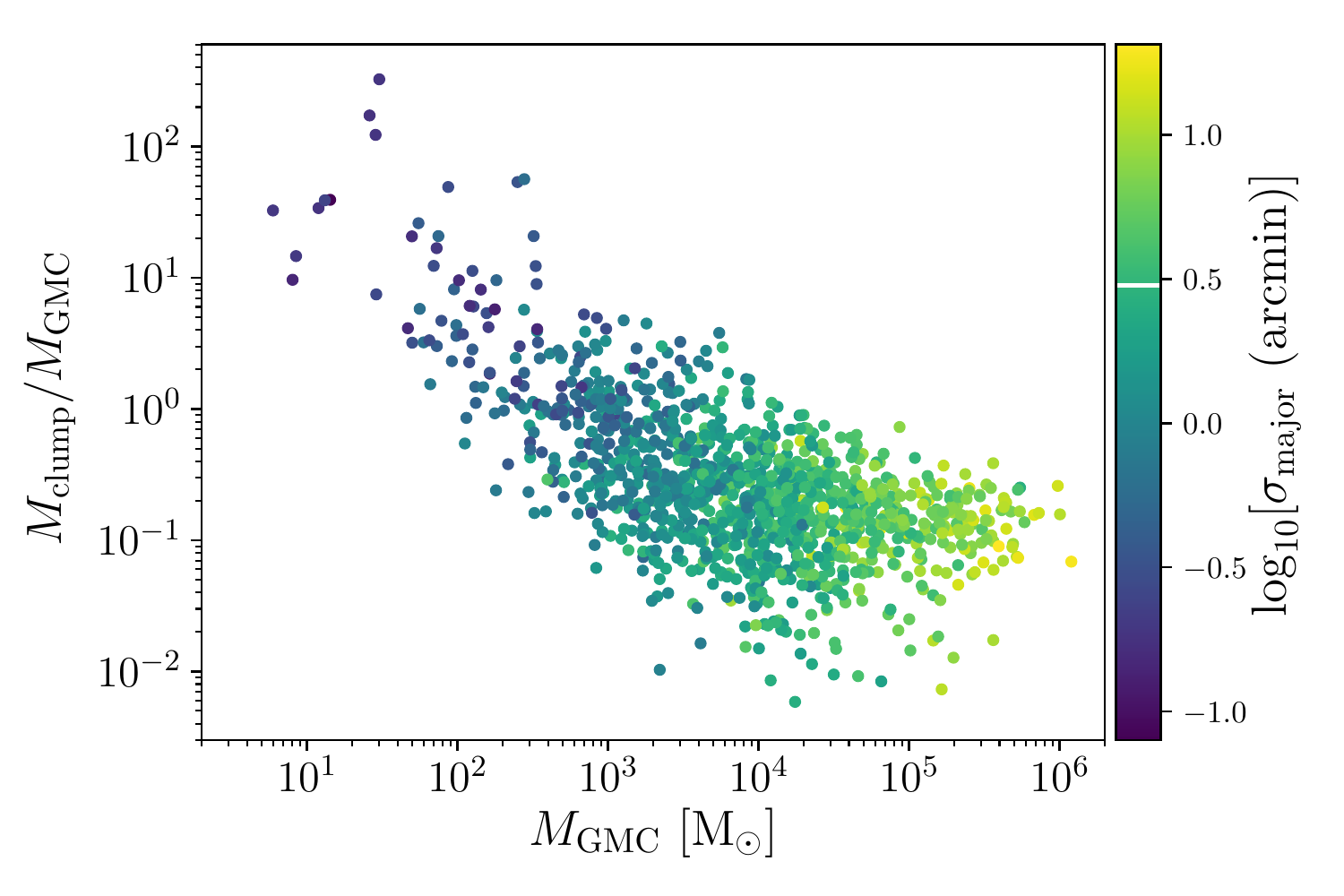}
	\includegraphics[width=0.49\textwidth]{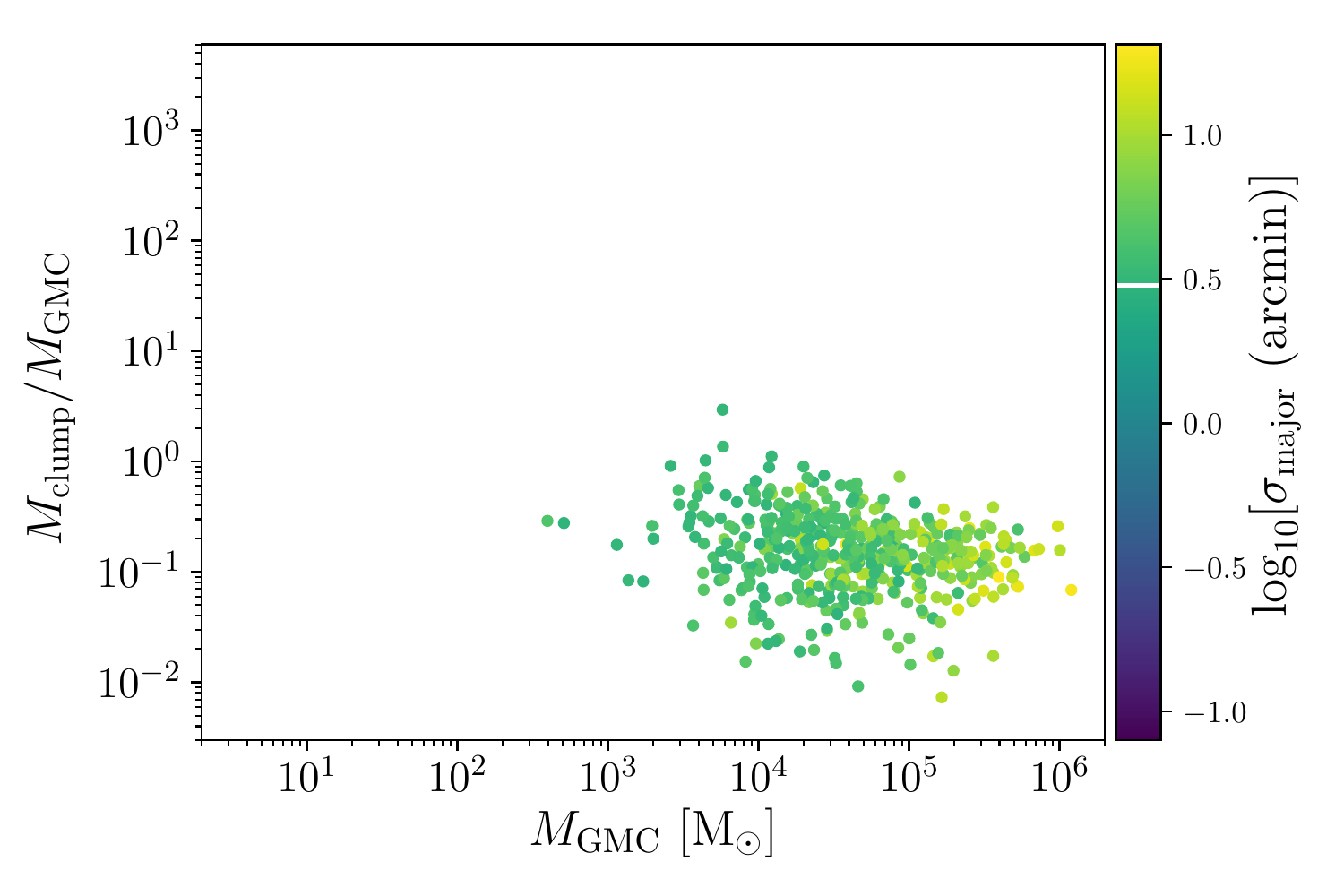}
\caption{Clump mass fraction as a function of cloud mass. Points are colored by cloud major axis. The full sample is found in the left panel. At size scales smaller than $3'$ (marked with a white dash on the colorbar), Hi-GAL detects sources at similar angular sizes to COHRS, and thus similar mass levels in the clumps as the clouds, leading to clump mass fractions over unity. These clumps are excluded from our analysis, leaving only those clouds found in the right panel. Note that this cut fails to exclude a few clouds with clump mass fractions over unity, but these only make up $\sim1\%$ of our sample.}
\label{fig:cfes}
\end{center}
\end{figure}

\begin{figure}[!htbp]
\begin{center}
	\includegraphics[width=0.6\textwidth]{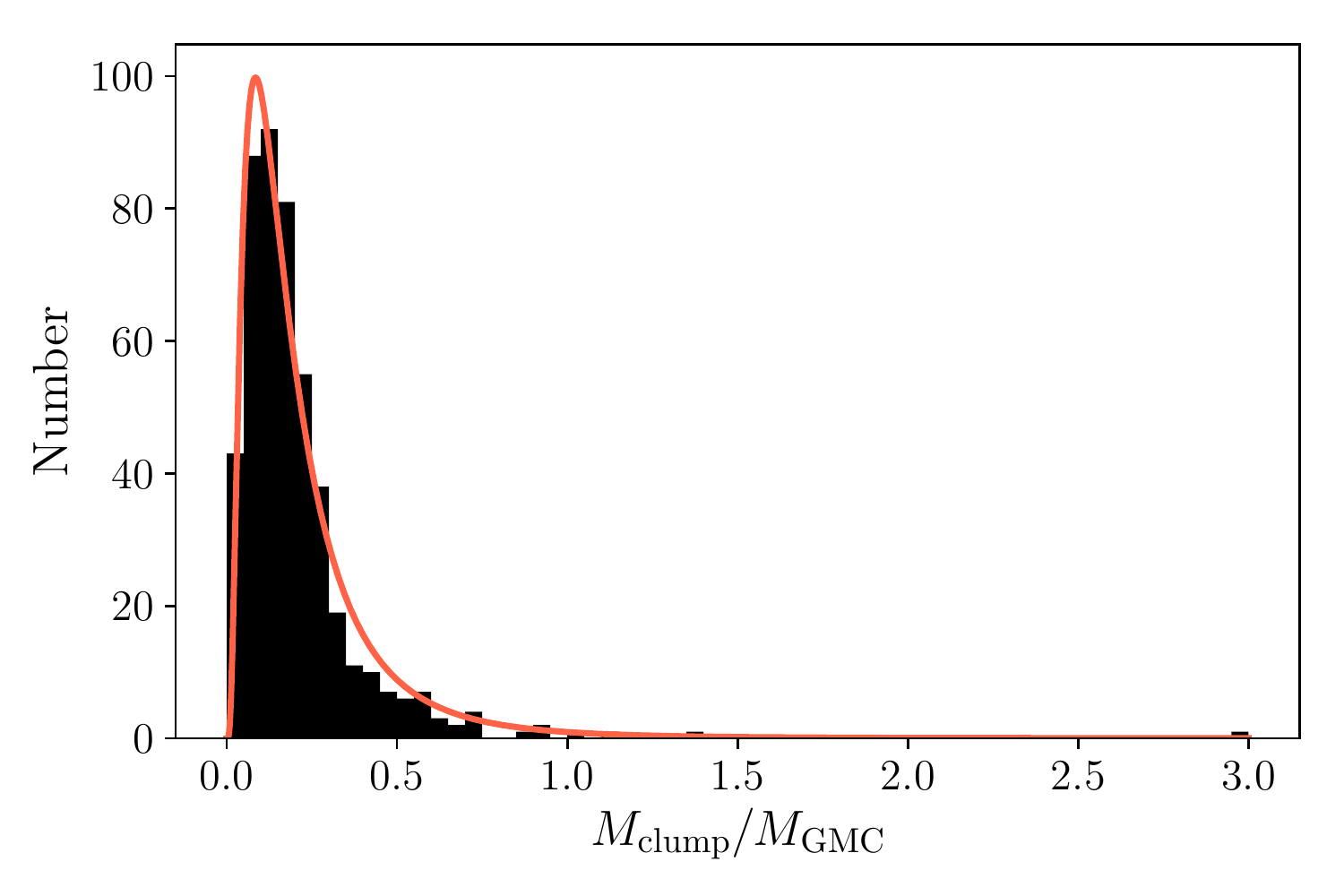}
\caption{The distribution of clump mass fractions for the subset of clouds which have $\sigma_\mathrm{major}>3'$.  The red curve represents the lognormal fit to the data, with parameters $\mu=-1.8$ and $\sigma=0.8$.}
\label{fig:cfehists}
\end{center}
\end{figure}

Because of this, we consider only the GMCs with $\sigma_\mathrm{major}>3'$. This ensures that all clumps are smaller than the cloud which contains them. While we still find a few clouds whose mass in clumps is greater than their cloud mass found with COHRS, these comprise only about 1\% of the sample. If instead we accomplish this by checking the angular size of each clump against that of its host cloud and requiring clouds to be larger than their clumps, the distribution of mass fractions does not change significantly. We see no changes to any of the correlations (or lack thereof) found throughout this paper. The only noticeable changes are the inclusion of less massive clouds and clouds with lower star formation rates. As we are not attempting to make a complete sample, with this check, we can proceed knowing that our angular size cut is not biasing the sample's distribution of clump mass fractions. Also notable is that the lower left-hand quadrant of this plot is empty. This is due to the mass sensitivity of Hi-GAL. The lower the cloud mass, the fainter its substructures, and thus the less likely we are to detect its clumps. Figure \ref{fig:cfes} (right) shows the distribution of mass fractions after the $\sigma_\mathrm{major}>3'$ cut on cloud size was made.

The mass fractions seen in Figure \ref{fig:cfes} (right) are shown in histogram form in Figure \ref{fig:cfehists}. The  mass fraction above the $3'$ GMC size cut is $0.20^{+0.13}_{-0.10}$, where the uncertainties indicate the range found in 68\% of the GMCs. This is consistent with the mean clump mass fraction found by \citet{Battisti14} using BGPS and GRS data, as well as the findings of \citet{Eden12}, who also utilized BGPS and GRS data, specifically in the Scutum tangent region. As the BGPS catalog uses the same clump definition and algorithm as we do here, these findings are readily comparable. \citet{Battisti14} found a mean clump mass fraction of $0.11^{+0.12}_{-0.06}$. \citet{Eden12} divided their sample into foreground, background, and tangent regions, finding CFEs of $13.00\pm1.60$, $8.52\pm0.73$, and $8.38\pm0.30$ per cent, respectively. While our mean clump mass fraction is greater than those of the above publications, our findings agree within $1\sigma$. Clump mass fractions are consistent with a single lognormal distribution. This fit describes our observed tail of higher clump formation efficiencies. This suggests that the clouds in the tail of this distribution are a true part of a single well-sampled population, and not evidence for a separate population at a different evolutionary stage.
A study of CFEs in the Scutum tangent region by \citet{Eden12} found large variations in CFEs, including some extreme values, but similarly concluded that the clouds make up a single population which is unaffected by environment and has a distribution consistent with lognormal. 
Furthermore, this distribution is in agreement with the dense gas fraction map calculated produced by \citet{Zetterlund18} using Hi-GAL clumps and the azimuthally averaged H$_2$ model of \citet{Wolfire03}, when adjusted for temperature measurements. Whereas we measure temperatures for individual clumps, \citet{Zetterlund18} assumed a uniform clump temperature of 20 K. If we, likewise, assume a temperature of 20 K, we find a mean clump mass fraction of $0.08^{+0.05}_{-0.04}$ and a maximum value of 0.46.

\citet{Enoch07} studied the molecular clouds Serpens, Perseus, and Ophiuchus, and found that, within visual extinction contours of $A_V=2$, less than 5\% of the clouds' mass was found in cores (2.7\%, 3.8\%, and 1.2\%, respectively). For clouds defined with $A_V=6$ contours, the core mass fraction was still less than 10\% in all three clouds (4\%, 7\%, and 2\%, respectively). The mean core mass fractions for these three clouds are 2.6\% and 4.3\% for extinction contours of $A_V=2$ and $A_V=6$, respectively. If we consider our clump mass fractions of $0.20^{+0.13}_{-0.10}$, and consider that cores are nested in clumps, this equates to a clump-to-core mass efficiency of approximately 20\%. However, it should be noted that these clouds are only a few thousand solar masses, and therefore on the low-mass end of our cloud sample and this clump-to-core mass efficiency should not be taken as universal or representative of GMCs.

As a check, to assess if our mass fractions could reasonably produce the observed star formation rate in the Milky Way given the cloud mass we see in COHRS, we estimate the clump formation efficiency required for a Galaxy-wide SFR of $1~\mathrm{M}_\odot ~\mathrm{yr}^{-1}$ \citep[e.g.,][]{Robitaille10}, as 
\begin{equation}
	\sum\limits_{i} M_{\text{GMC},i}\, \mathrm{CFE}\, \varepsilon  t_\mathrm{ff}^{-1} 
	= (1~\mathrm{M}_\odot ~\mathrm{yr}^{-1}) f_\mathrm{obs}.
\end{equation}
We use $\sum\limits_{i} M_{\text{GMC},i} = 7.2 \times 10^{7}~\mathrm{M}_\odot$, which is the sum of all GMCs ($M>10^3~\mathrm{M}_\odot$) in the COHRS catalog. We assume a clump-to-star mass efficiency of $\varepsilon=10\%$ as an order-of-magnitude estimate consistent with a core-collapse efficiency of 25\% \citep{Enoch08}. As our timescale, we use the mean free-fall time of a GMC in the COHRS catalog, $t_\mathrm{ff} = \sqrt{3\pi/32G\rho} = 3.2 \times 10^{6}~\mathrm{yr}$. COHRS observed longitudes $10^\circ.25 < \ell < 55^\circ.25$, which encompasses a fraction of approximately $f_\mathrm{obs} = 1/3$, of the Galactic Plane from the Solar System's vantage point. These values combine to give a necessary clump formation efficiency, CFE, of 0.16, which is consistent with our observed value.  The free-fall time is the lower limit on the time frame, with the upper limit being the lifetime of the cloud, $\tau=10t_\mathrm{ff}$ \citep{Kennicutt12}. Using the average lifetime of a cloud puts an upper limit on the necessary clump formation efficiency at $\mathrm{CFE} = 1.6$. While this exercise is only valid to an order of magnitude, it suggests that our clump mass fractions are not unreasonable, given knowledge of star formation across the Galaxy. 

\subsection{The Influence of Clump Cataloging}

We assess the impact of our cataloging method on our basic results. Figure \ref{fig:cfecomp} shows a cloud-by-cloud comparison of the clump mass fractions from the \citet{Zetterlund18} and \citet{Molinari16b} catalogs. Our catalog produces clump mass fractions which are $\sim 10\%$ larger than those found when using the catalog from \citet{Molinari16b}. Whereas {\sc Bolocat} objects do not have a specified shape, {\sc CuTEx\footnote{https://herschel.ssdc.asi.it/index.php?page=cutex.html}}, the clump finding algorithm used by \citet{Molinari16b}, identifies objects which can be approximated as a Gaussian function. This results in their catalog breaking down what we consider a single clump, into multiple pieces. They therefore discard some of the more diffuse gas which we still consider part of a clump, resulting in the slightly lower average clump mass fraction that we see when using their catalog. This overall trend repeats in the analysis that follows with only offsets differences between the results based on the cataloging method and negligible changes in derived scalings. Compared to the large systematics that arise in translating the observed emission into physical properties, the clump cataloging method used contribute negligibly to our uncertainty budget.

\begin{figure}[!htbp]
\begin{center}
	\includegraphics[width=0.6\textwidth]{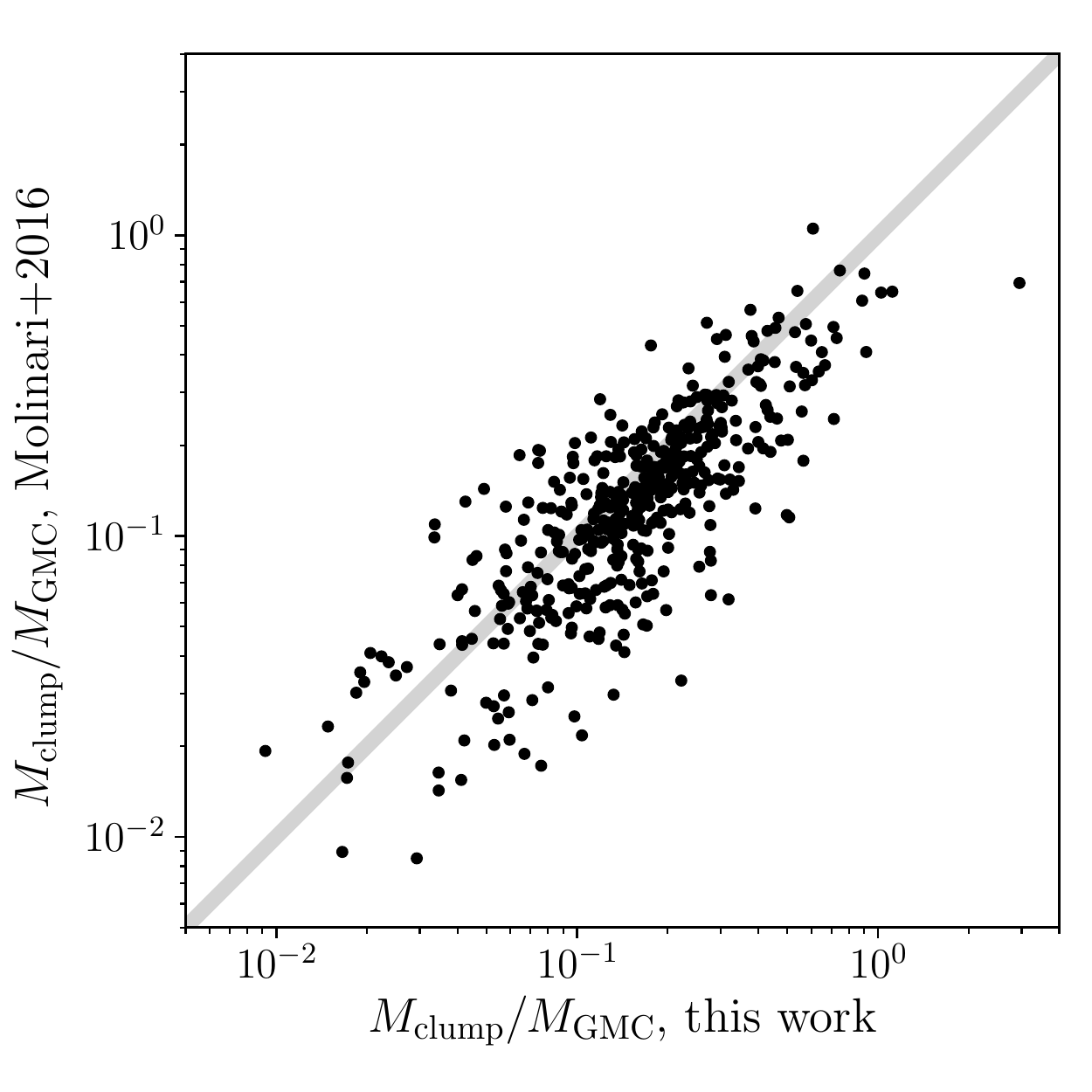}
\caption{A comparison of clump mass fractions found when matching our clump catalog and the clump catalog of \citet{Molinari16b} to the COHRS cloud catalog. Our catalog produces clump mass fractions which are $9\pm47$ per cent larger than those found when using the \citet{Molinari16b} catalog.}
\label{fig:cfecomp}
\end{center}
\end{figure}

\subsection{Relations with Physical Properties}
\label{ss:cfeprops}

\begin{figure}[!htbp]
\begin{center}
	\includegraphics[width=0.49\textwidth]{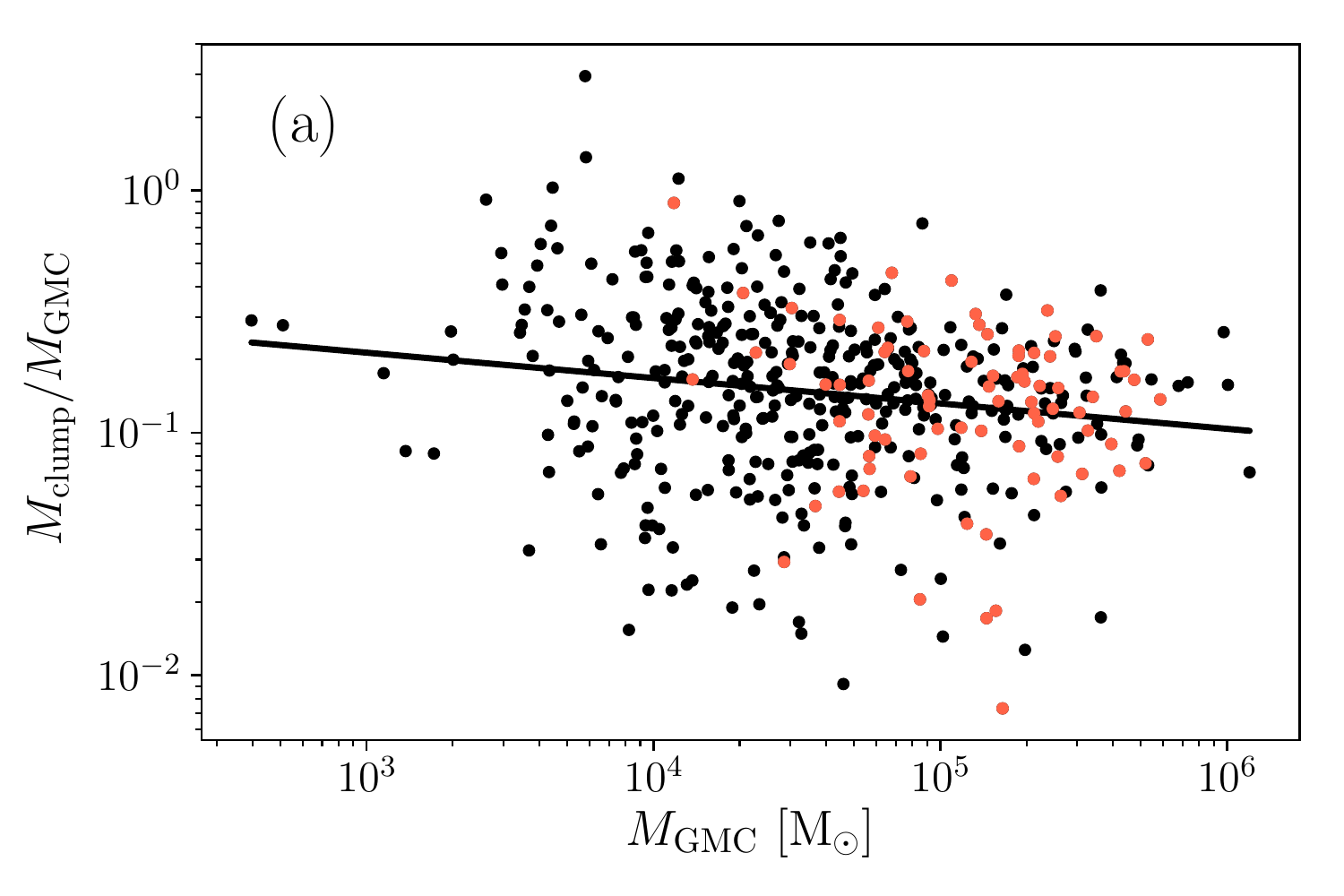}
	\includegraphics[width=0.49\textwidth]{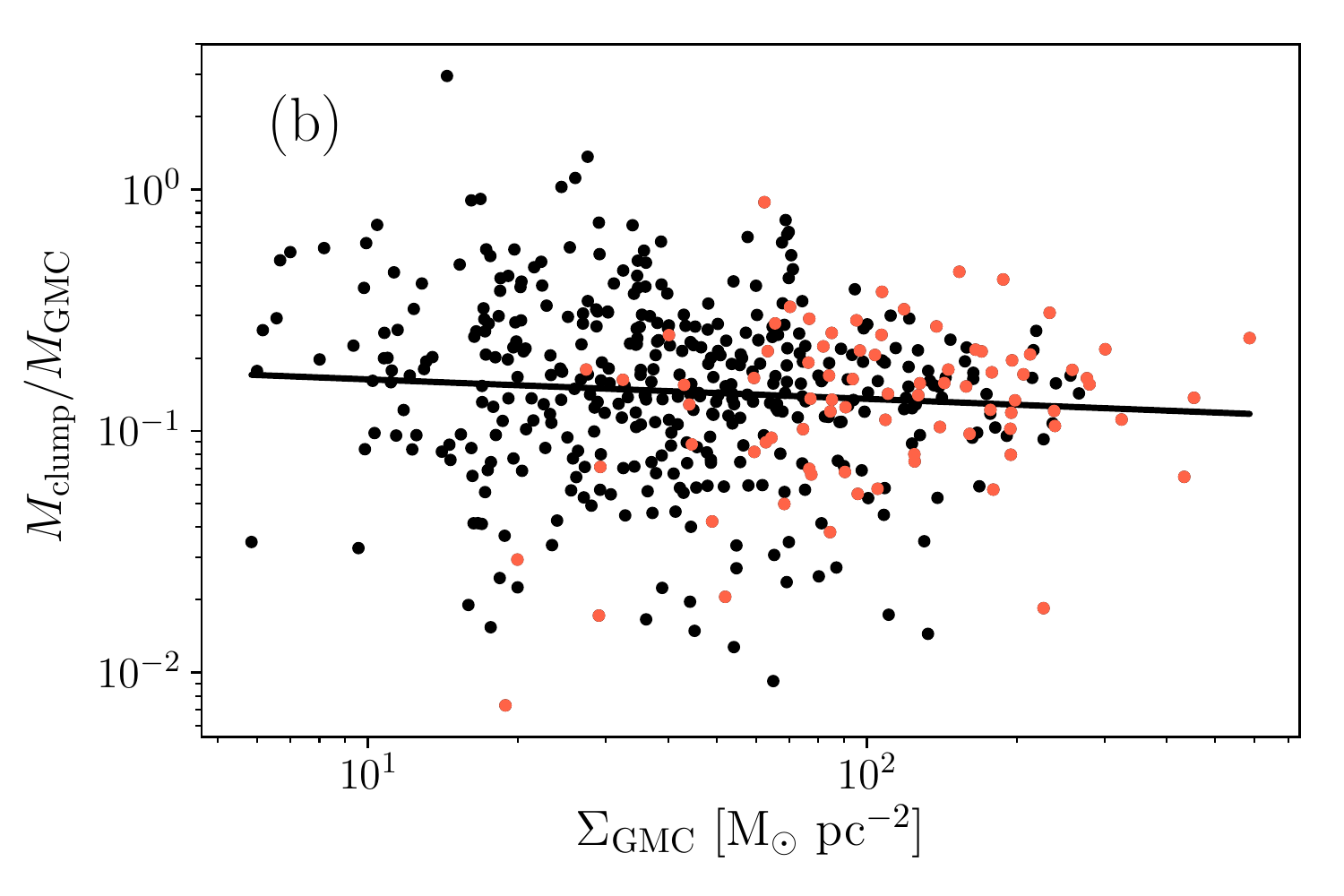}
	\includegraphics[width=0.49\textwidth]{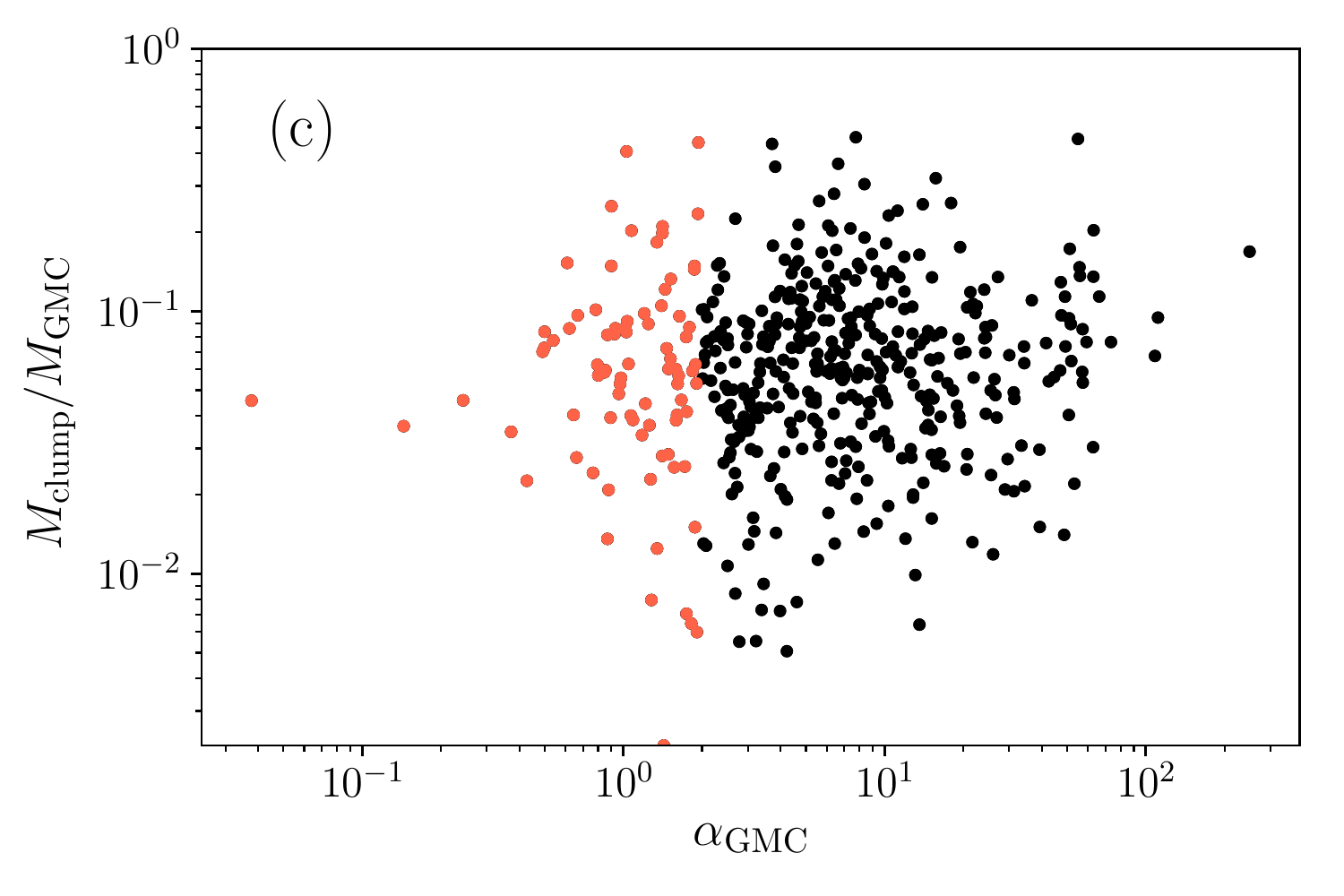}
	\includegraphics[width=0.49\textwidth]{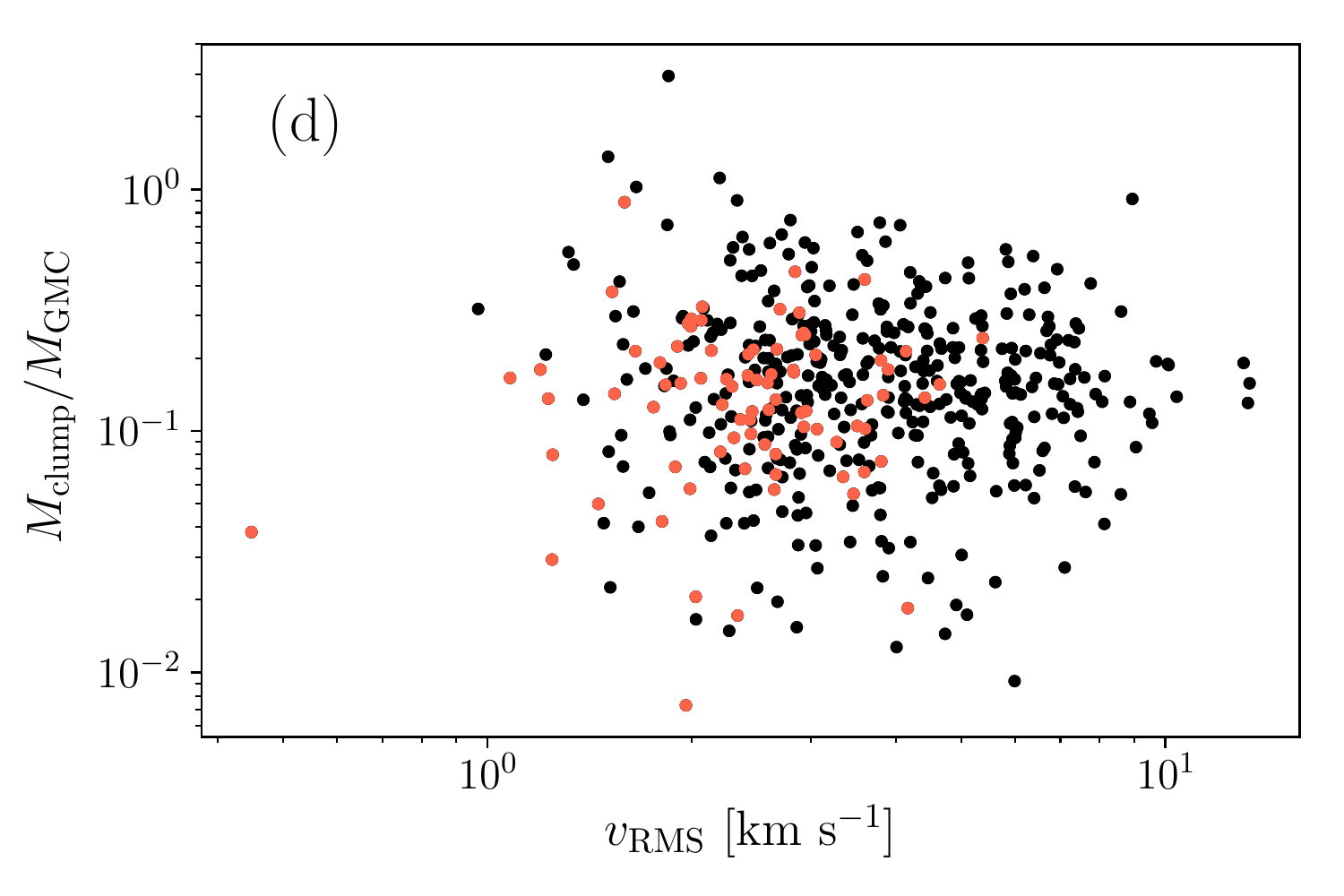}
\caption{Clump mass fraction of the subset of clouds which have $\sigma_\mathrm{major}>3'$, as a function of the following cloud properties: (a) mass, (b) surface mass density, (c) virial parameter, and (d) line width. Red points indicate clouds with virial parameters $\alpha_\mathrm{vir}<2$. The power-law fits in (a) and (b) have slopes of $-0.14\pm0.03$ and $-0.13\pm0.05$, respectively.}
\label{fig:cfeprops}
\end{center}
\end{figure}

To determine what properties of a GMC influence its clump formation efficiency, we plot clump mass fractions of clouds against various physical properties (Figure \ref{fig:cfeprops}). In examining cloud mass, surface mass density, virial parameter, and line width we see no more than slight correlations between clump mass fraction and these cloud properties. In the case of cloud mass and cloud mass surface density, we find shallow power-law slopes of $-0.14\pm0.03$ and $-0.13\pm0.05$, respectively, with large dispersions. A cloud with more mass will, on average, have a slightly lower clump mass fraction. A cloud's stability against collapse as a whole (see Section \ref{s:vir}) does not influence the fraction of a cloud's mass that will collapse into denser clumps. 

 In these plots, red points correspond to non-virialized clouds ($\alpha_\mathrm{GMC}<2$). That is, those unstable to gravitational collapse. This comprises $\sim 25\%$ of our sample. Panel (c) shows that the non-virialized clouds sample from the full distribution of clump mass fractions, as do virialized clouds. As seen in panels (a) and (b), these potentially collapsing clouds are more likely to have higher masses and surface mass densities. However, we know that lower mass clouds ($M\sim10^3~\mathrm{M}_\odot$) can collapse to form stars \citet{Enoch08}. These clouds are not absent from the COHRS catalog, but we are less likely to match Hi-GAL clumps to them. This may be due to the small clouds' clumps being too small for Hi-GAL to detect above the confusion noise. When we relax the $\sigma_\mathrm{major}>3'$ size cut to GMCs and instead simply require the GMC to be larger than its constituent clumps (see Section \ref{ss:cfedist}), we recover some of these lower mass non-virialized clouds. For a more thorough analysis of the distribution of virial parameters for the COHRS catalog as a whole, see Section \ref{s:vir}.


\begin{figure}[!htbp]
\begin{center}
	\includegraphics[width=0.49\textwidth]{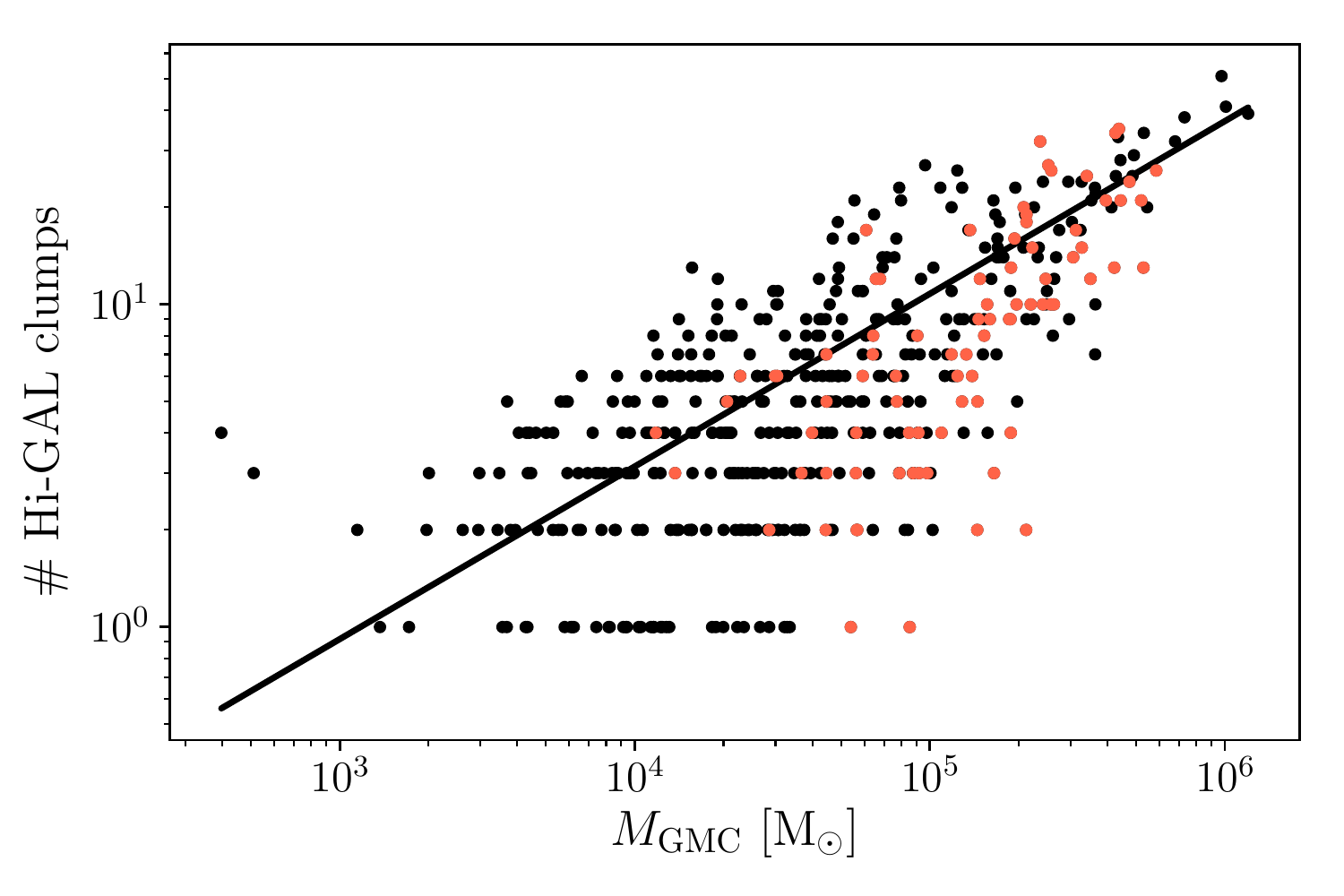}
	\includegraphics[width=0.49\textwidth]{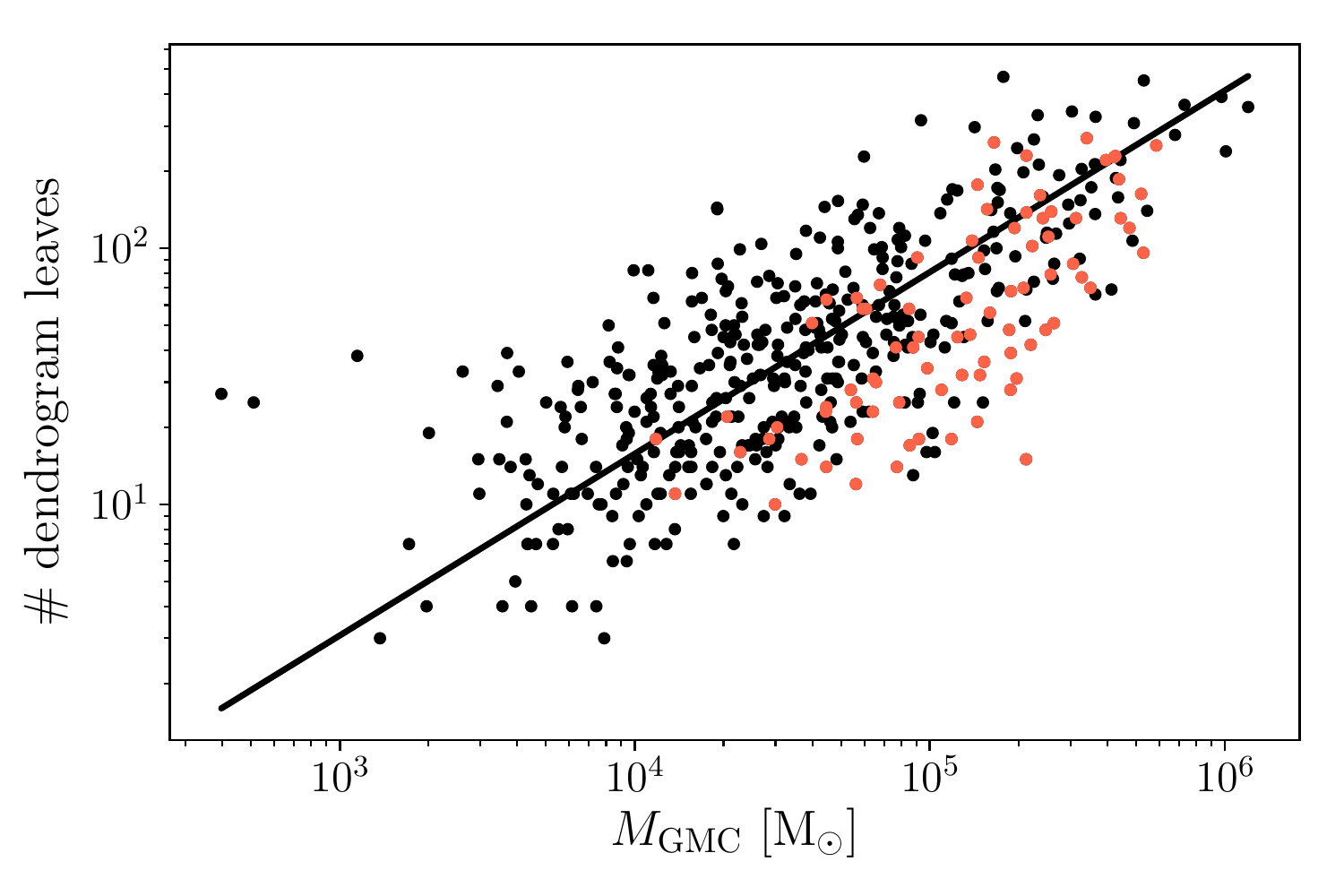}
\caption{The number of clumps (left) and dendrogram leaves (right) in each cloud, as a function of cloud mass. The black lines represent power-law fits, with slopes of $\beta=0.54\pm0.03$ and $\beta=0.71\pm0.04$, respectively, and $1\sigma$ dispersions around the regressions of 0.3 dex.  Red points indicate clouds with virial parameters $\alpha_\mathrm{vir}<2$.}
\label{fig:nclumps}
\end{center}
\end{figure}

\begin{figure}[!htbp]
\begin{center}
    \includegraphics[width=0.6\textwidth]{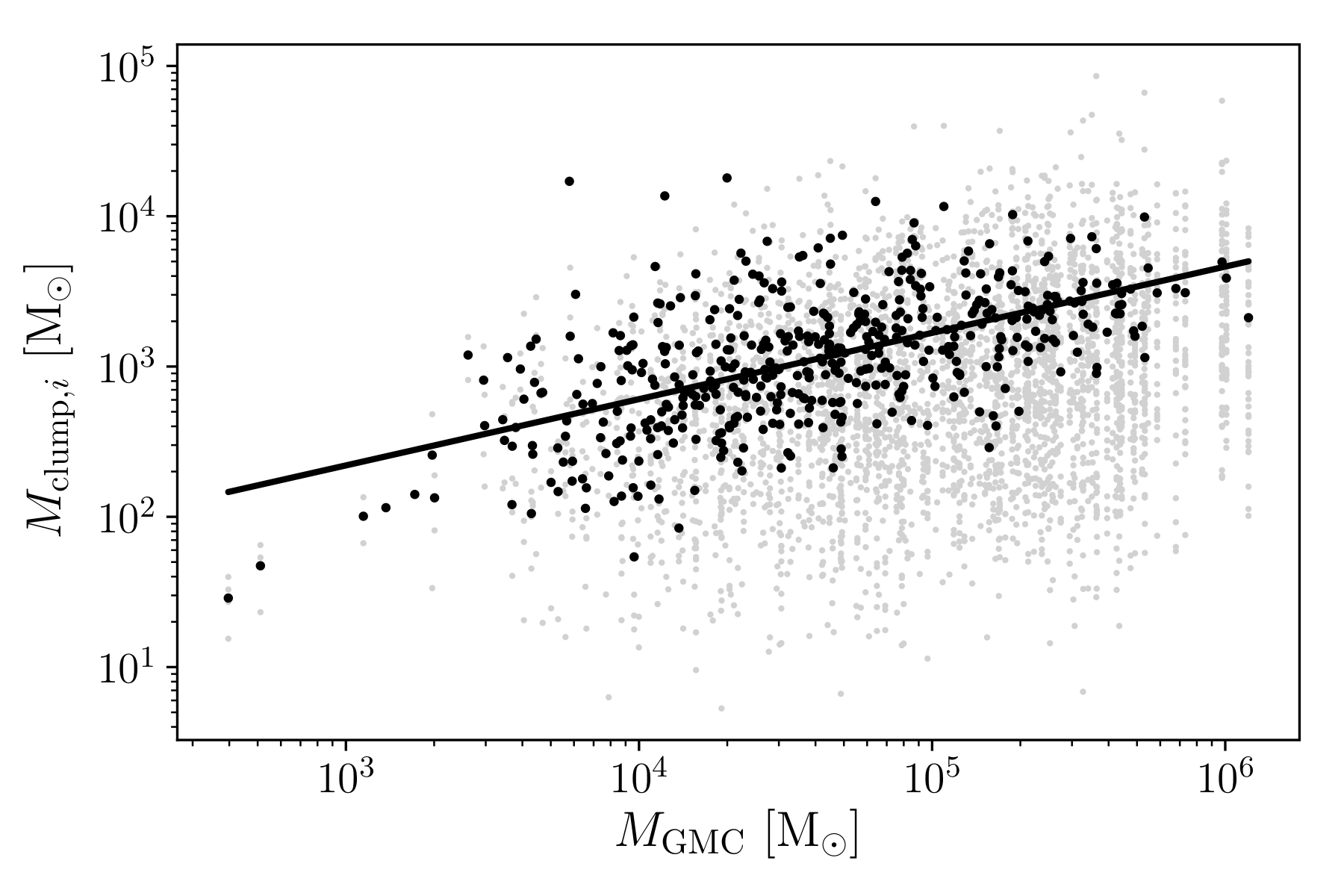}
\caption{The mean clump mass in each cloud (black), as a function of cloud mass. Individual clump masses are shown in grey. The black line represents a power-law fit with a slope of $0.48\pm0.03$ and a $1\sigma$ dispersion around the regression of 0.4 dex.}
\label{fig:Mclump}
\end{center}
\end{figure}

Clump mass fraction is only weakly correlated with cloud mass or cloud surface mass density. There are two possibilities for how this relative lack of dependence could come to be: (1) more massive clouds could have more clumps, or (2) they could have more massive clumps. The first possibility is investigated in Figure \ref{fig:nclumps}, which shows that more massive clouds have more clumps associated with them. They also have more COHRS dendrogram leaves, which correspond to local maxima in the CO(3-2) intensity, and serve as confirmation of the trend in number of Hi-GAL clumps.
We fit the data in these plots to a power-law function using orthogonal distance regression. The reported slope uncertainties were determined by bootstrap resampling the fit residuals. This method was taken for all regressions in this work.
We find power law slopes of $\beta=0.54\pm0.03$ and $\beta=0.71\pm0.04$ relating cloud mass to number of clumps and number of dendrogram leaves, respectively. The $1\sigma$ dispersion of data about their regression is 0.3 dex for both data sets. Note that the large error bars make these fits statistically consistent with unity.

To investigate the second possibility, whether more massive clouds host more massive clumps, Figure \ref{fig:Mclump} shows the mean clump mass for each cloud plotted against the cloud mass. This demonstrates that more massive clouds also have more massive clumps, on average. This tendency could mean that the clump IMF is not universal, and instead favors more massive clumps in more massive clouds.
We find a power-law slope of $\beta=0.48\pm0.03$ relating mean clump mass to cloud mass and a $1\sigma$ dispersion about the regression of 0.4 dex. Note that we eliminated the intercept term for this fit, which is the reason for the smaller slope uncertainties here, in comparison with the fits from Figure \ref{fig:nclumps}. This slope is less than unity, and thus clump masses increase more slowly than cloud masses. The sum of the two effects (more clumps and more massive clumps) produces the universal clump mass fraction that we observed earlier. 

\begin{figure}[!htbp]
\begin{center}
    \includegraphics[width=0.6\textwidth]{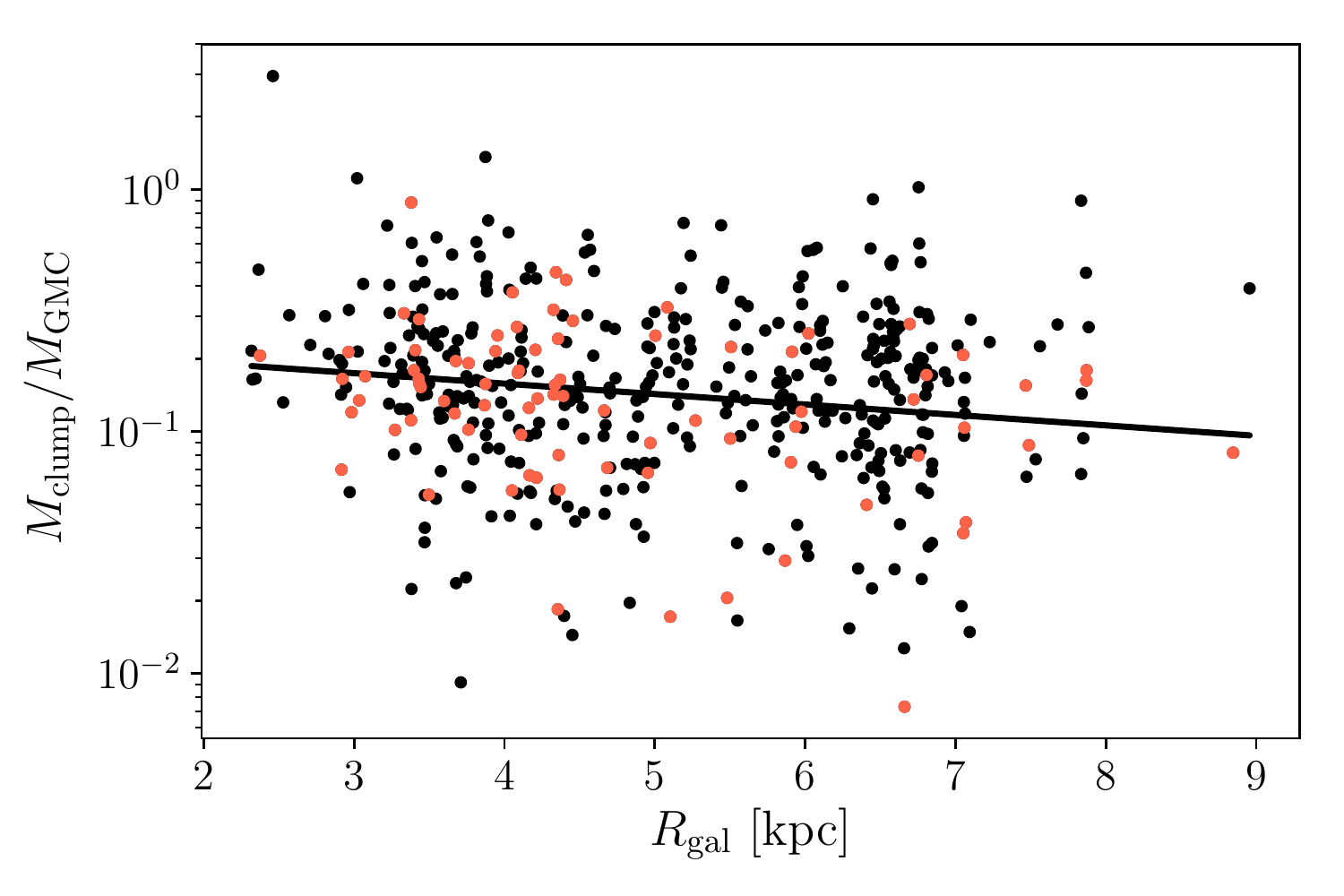}
\caption{Clump mass fraction as a function of Galactocentric radius. Red points indicate clouds with virial parameters $\alpha_\mathrm{vir}<2$. The regression shown as a black line has a slope of $\beta =-0.09\pm0.02$ and a $1\sigma$ dispersion about the regression of 0.4 dex.}
\label{fig:CFE_Rgal}
\end{center}
\end{figure}

Finally, we investigate the effects of environment on clump mass fraction, and find a very slight trend towards decreasing clump mass fractions with increasing distance from the Galactic Center (Figure \ref{fig:CFE_Rgal}) A linear fit to $\log{(M_\mathrm{clump}/M_\mathrm{GMC})}$ vs. $R_\mathrm{gal}$ yields a slope of $\beta = -0.09\pm0.02$ and a $1\sigma$ dispersion about the regression of 0.2 dex. 
The enhancements in cloud counts around 4 kpc and 6.5 kpc are likely due to presence of the molecular ring and a spiral arm. These locations do not appear to be associated with increased clump mass fractions.
This is similar to the findings of \citet{Ragan16}, concerning the fraction of Hi-GAL objects which were determined to be star-forming. They found that the star-forming fraction (the fraction of Hi-GAL sources with a 70 $\mu$m counterpart) decreased weakly with Galactocentric radius, and was unaffected by the presence of spiral arms. Likewise, \citet{Eden12} found that clump mass function and CFE are unaffected by Galactic environment (i.e. spiral arms), when looking at the Scutum tangent region.

\section{Star Formation Rates}
\label{sec:sfrs}
Some fraction of the mass found in clumps goes on to collapse further and form stars. We examine the rate at which clumps embedded in our clouds are forming stars, by way of 70 $\mu$m Hi-GAL emission. We also investigate which of a cloud's properties influence its star formation rate.
The median star formation rate in our sample of clouds is $1.1\times10^{-5}~\mathrm{M}_\odot~\mathrm{yr}^{-1}$. The rms is nearly an order of magnitude, with the mean SFR in log-space being $\langle \log_{10}(\mathrm{SFR}) \rangle = -5.0\pm0.9~\log_{10}(\mathrm{M}_\odot~\mathrm{yr}^{-1})$. The total SFR for our cloud sample is $0.03~\mathrm{M}_\odot~\mathrm{yr}^{-1}$. When scaled to the whole Galactic Plane --- by dividing the SFR of our sample by the fraction of mass in the COHRS catalog which was in our sample, and by the fraction of the Galactic Plane which COHRS observed --- this results in a SFR of $0.2~\mathrm{M}_\odot~\mathrm{yr}^{-1}$ for the Milky Way. This is low compared to more precise studies of this number \citep[e.g.,][]{Robitaille10,Chomiuk11}, but within an order of magnitude. At least part of this difference due to our inclusion only of star formation associated spatially and morphologically with the molecular clouds in our sample.  There remains a significant amount of star formation that will not be linked to catalogued molecular gas features.

The distribution of cloud SFRs is seen in Figure \ref{fig:sfr_hist}. There is no significant correlation with clump mass fraction, as seen in Figure \ref{fig:cfe_sfr} (left). Thus, even if a cloud puts more of its mass in clumps, it does not necessarily form stars at a greater rate. In addition, we can examine effects on mass-normalized quantities: depletion times. We define cloud and clump depletion times as
\begin{align}
    t_\mathrm{dep,GMC} &= M_\mathrm{GMC} / \mathrm{SFR} \label{eq:tdepgmc} \\
    t_\mathrm{dep,clump} &= M_\mathrm{clump} / \mathrm{SFR} \label{eq:tdepclump},
\end{align}
where SFR is the star formation rate measured for the cloud in both cases. Investigating these normalized quantities, we see no correlation between star formation rate and clump mass fraction. On the other hand, a positive correlation, partially driven by clouds with extreme clump mass fractions, is seen between $t_\mathrm{dep,clump}$ and clump mass fraction (Figure \ref{fig:cfe_sfr}, center). This
implies that on average either clump mass is higher in clouds with higher clump mass fractions or the star formation rate in clumps is smaller in clouds with greater clump mass fractions.  Figure \ref{fig:cfe_sfr} (right) shows that it is the former:  clump masses tend to be larger in clouds with greater clump mass fractions.   Figure \ref{fig:cfe_sfr} also shows that clumps with lower virial parameters tend to have higher star formation rates (left) and that more massive clumps tend to have smaller virial parameters (right).  While the scatter is very large,  taken together these weak correlations are consistent with  an evolutionary trend in which clumps and clump mass fractions grow within clouds over time, become increasingly gravitationally unstable, and have correspondingly increased star formation rates.

\begin{figure}[!htbp]
\begin{center}
	\includegraphics[width=0.6\textwidth]{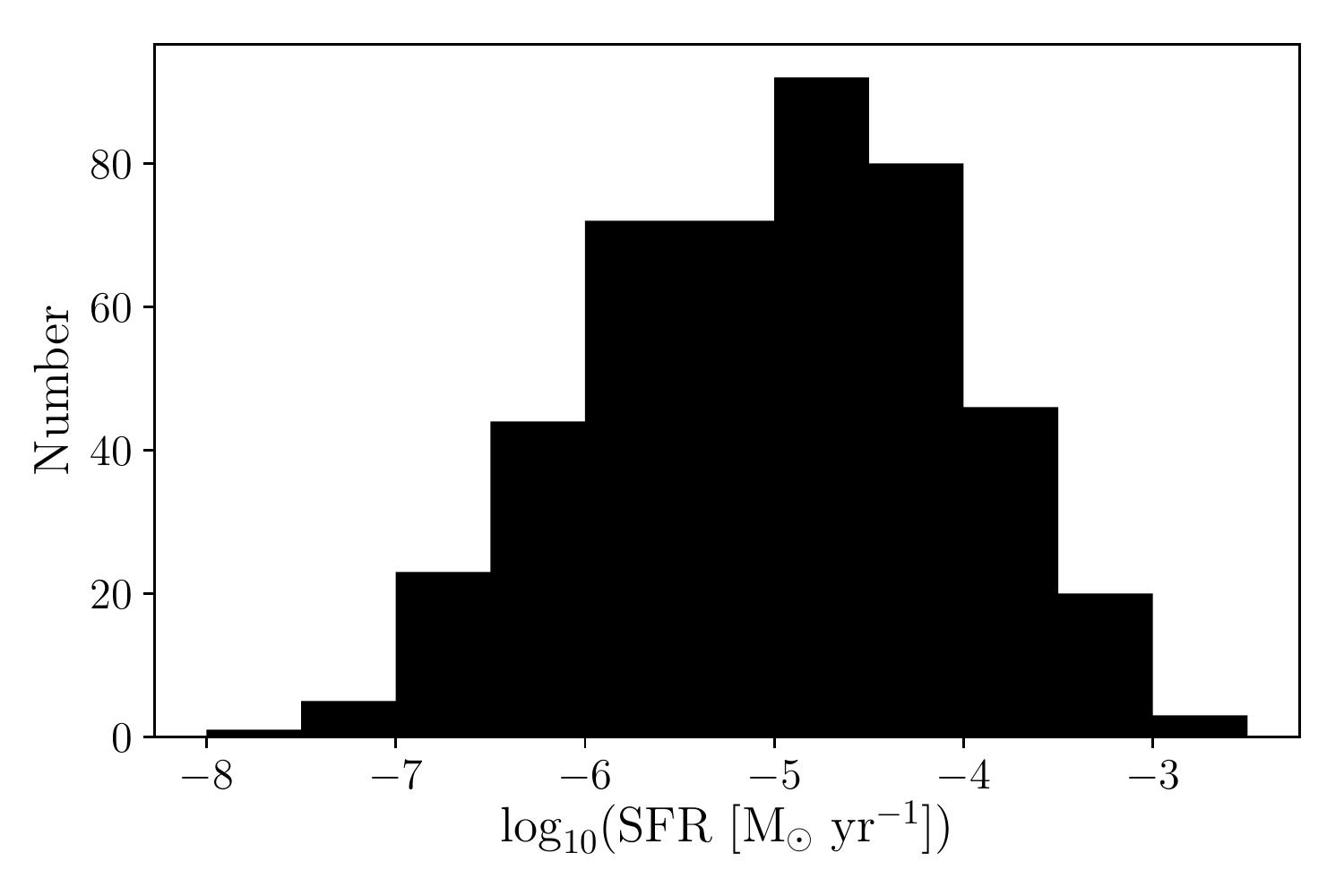}
\caption{Distribution of cloud star formation rates derived from Hi-GAL 70 $\mu$m emission. The mean and standard deviation in log-space are $-5.0\pm0.9~\log_{10}(\mathrm{M}_\odot~\mathrm{yr}^{-1})$.}
\label{fig:sfr_hist}
\end{center}
\end{figure}

\begin{figure}[!htbp]
\begin{center}
	\includegraphics[width=0.32\textwidth]{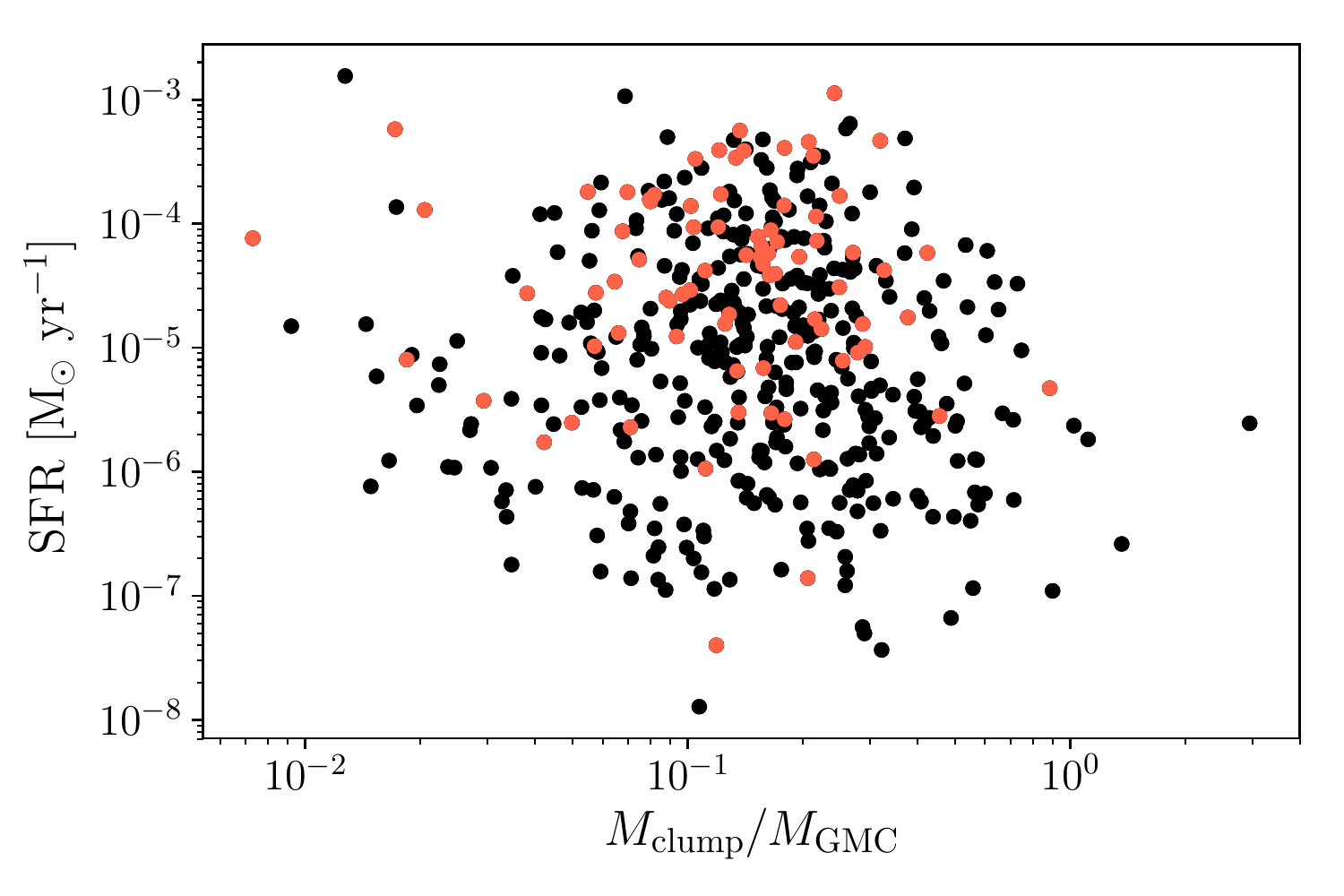}
	\includegraphics[width=0.32\textwidth]{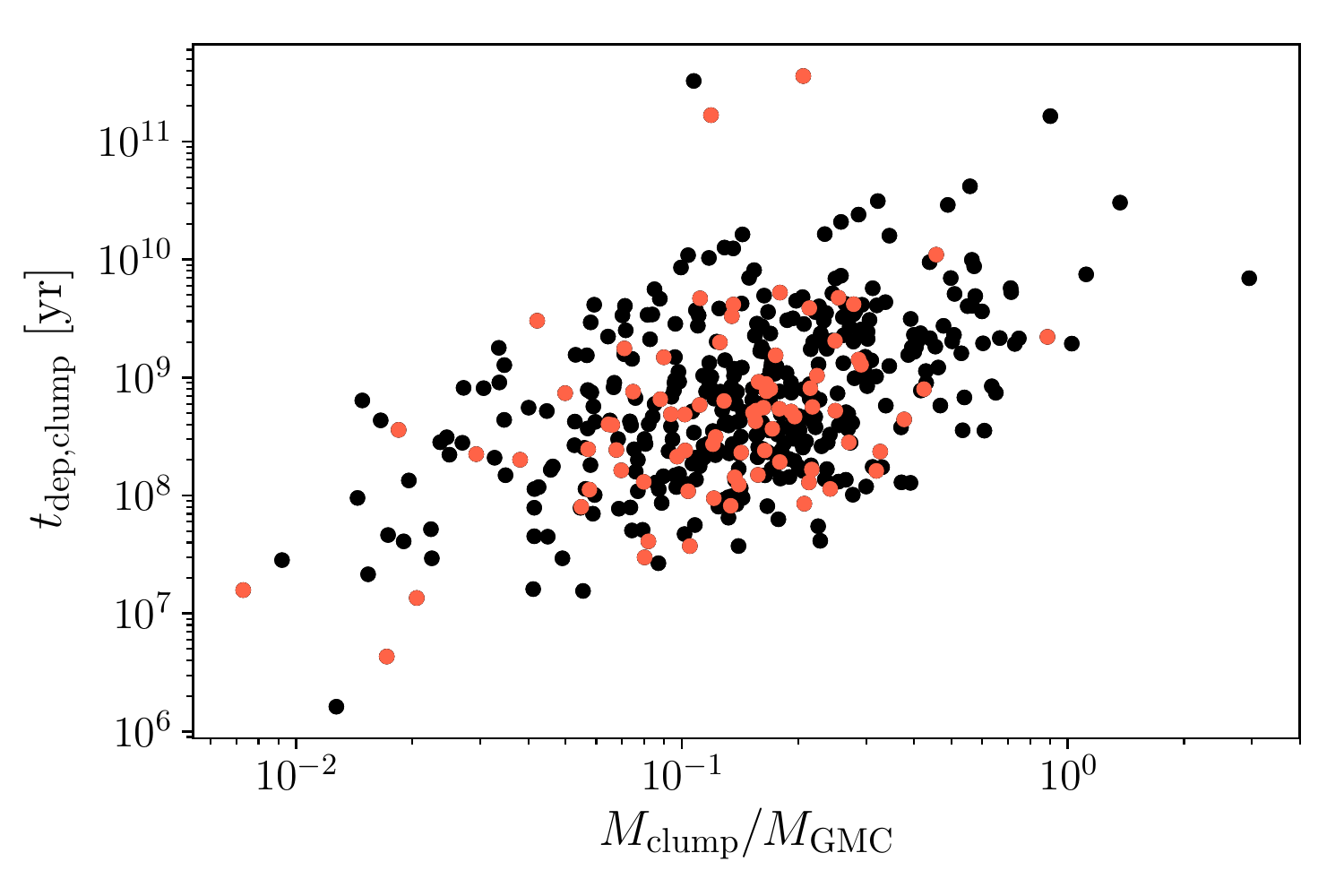}	
	\includegraphics[width=0.32\textwidth]{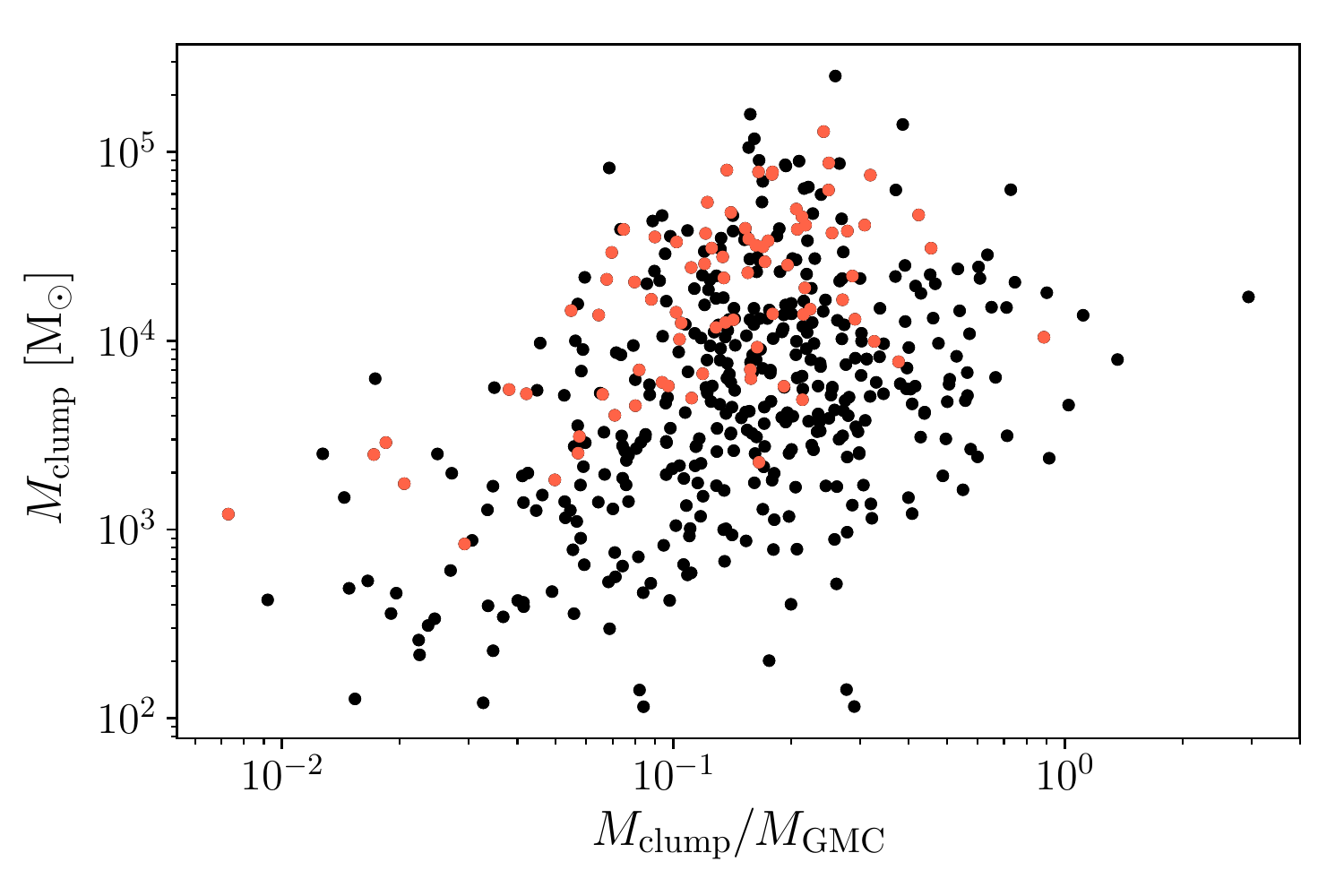}
\caption{{\it Left}: Star formation rate derived from Hi-GAL 70 $\mu$m emission as a function of clump mass fraction. {\it Center}: Clump mass depletion time (see Equation \ref{eq:tdepclump}) as a function of clump mass fraction.  {\it Right}: Clouds' total mass in clumps as a function of clump mass fraction. Red points indicate clouds with virial parameters $\alpha_\mathrm{vir}<2$.}
\label{fig:cfe_sfr}
\end{center}
\end{figure}

\begin{figure}[!htbp]
\begin{center}
	\includegraphics[height=0.8\textheight]{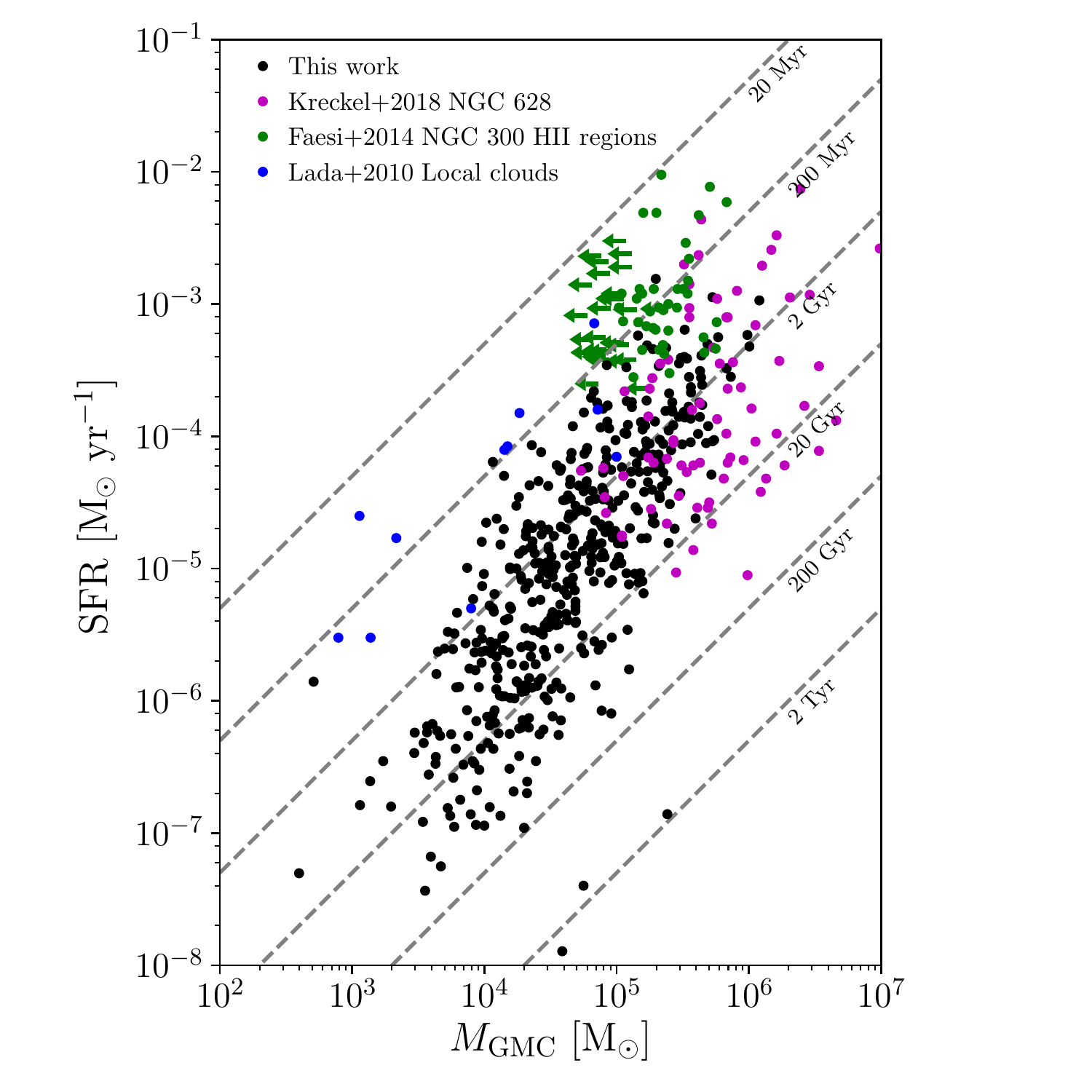}
\caption{Star formation rates, as derived from Hi-GAL 70 $\mu$m emission, as a function of cloud molecular gas masses. Clouds from this work are shown in black. Also included are NGC 628 HII regions matched to GMCs (magenta) from \citet{Kreckel18}, NGC 300 HII regions (green) as analyzed by \citet{Faesi14}, and local Milky Way clouds (blue) from the work of \citet{Lada10}. The dashed diagonal lines indicate mass depletion times.}
\label{fig:sfr_Mgmc}
\end{center}
\end{figure}

Figure \ref{fig:sfr_Mgmc} shows that more massive clouds host more star formation.  This behavior has been seen before in other studies, including: the Gould Belt clouds in the Solar Neighborhood \citep{Lada10}, {\sc Hii} regions in the nearby galaxies NGC 300 \citep{Faesi14}, and NGC 628 \citep{Kreckel18}, among other work.  We place these other data on Figure \ref{fig:sfr_Mgmc}, alongside our own.  The work of \citet{Lada10} argued that the local molecular clouds followed a linear correlation between cloud mass and star formation rate, arguing that star formation was well described by a constant volume density threshold, so that a fixed fraction of cloud mass would form stars over time.  The measurements for the Local Clouds are ``clean'' relying on counting of Young Stellar Objects for the star formation rate and using dust extinction for measuring the mass distributions.  The data are thus free from many of the biases introduced by the excitation based tracers used farther afield. The {\sc Hii} regions studied in NGC 300 are consistent with the extrapolation of the trend seen in local clouds, but the data in NGC 628 as well as our own study have relatively lower star formation rates per unit cloud mass.  Our data, in particular, also seem to show lower star formation rates at lower cloud masses when compared to the local molecular clouds.  This finding is not surprising in one context: the regions studied in \citet{Lada10} and \citet{Faesi14} are all focused on know star formation region whereas the regions studied in \citet{Kreckel18} and this work focus on the molecular clouds.  The different focus on different stages of star formation (GMCs vs star forming regions) can lead to a divergence in measurement \citep{Schruba10, Onodera10}.  The star formation rates in our current study are relatively coarse estimates compared to \citet{Lada10}, but the number of objects in our study is significantly larger and shows a non-linear scaling with cloud mass. Overall, this work shows that the constant-gas-fraction arguments for star formation in molecular clouds cannot apply to the GMC scale throughout the Milky Way. If this were the case, we would see a linear scaling of GMC mass and SFR, at the same levels as \citet{Lada10}. That we, along with \citet{Kreckel18}, do not, indicates that there are other factors at play.

In contrast, a uniform star formation efficiency for molecular clumps is consistent with our data, as explored in Figure \ref{fig:sfr_trends}. Panel (a) shows a tight correlation between clump mass and star formation rate which is consistent with the linear slope and short depletion times found in the \citet{Lada10} study.  Panel (b) shows that the depletion time of a population of clumps in a cloud is consistent as a function of the mean clump mass in the cloud.  This shows that clouds hosting higher mass clumps do not show different star formation properties. Panel (c) illustrates the weak decrease in depletion time for GMCs as a function of cloud mass, which as also reflected in Figure \ref{fig:sfr_Mgmc}.  In general, we find that higher mass clouds show star formation increases greater than linearly, which result in decreases in cloud depletion times. Panel (d) illustrates that clouds with more mass in clumps show a progressively tighter convergence to a single value of $t_\mathrm{dep,clump}$. This tightening is likely due to averaging of SFRs over more clumps as $M_\mathrm{clump}$ increases. With star formation happening in a greater number of clumps, fluctuations become insignificant in the mean, leading to a smaller spread in cloud-wide SFRs for clouds with more mass in clumps. Clump masses in panels (a) and (d) are more tightly correlated with SFR than cloud masses in Figure \ref{fig:sfr_Mgmc} and panel (c), particularly at the high-mass ends. This is in agreement with the work of \citet{Vutisalchavakul16}, who found that dense molecular gas is a better predictor of SFR than total molecular gas, and that the relationship is consistent with linear.

When comparing the vertical scales in panels (c) and (d), we see that the mass depletion time for clouds is greater than that for clumps. This provides a positive consistency check on our model of clumps as substructures of clouds. The depletion time in clumps is approximately 5 times shorter than in clouds. The difference in free-fall times is comparable to this, at 4 times shorter in clumps than in clouds (Figure \ref{fig:nn}).

\begin{figure}[!htbp]
\begin{center}
	\includegraphics[width=0.49\textwidth]{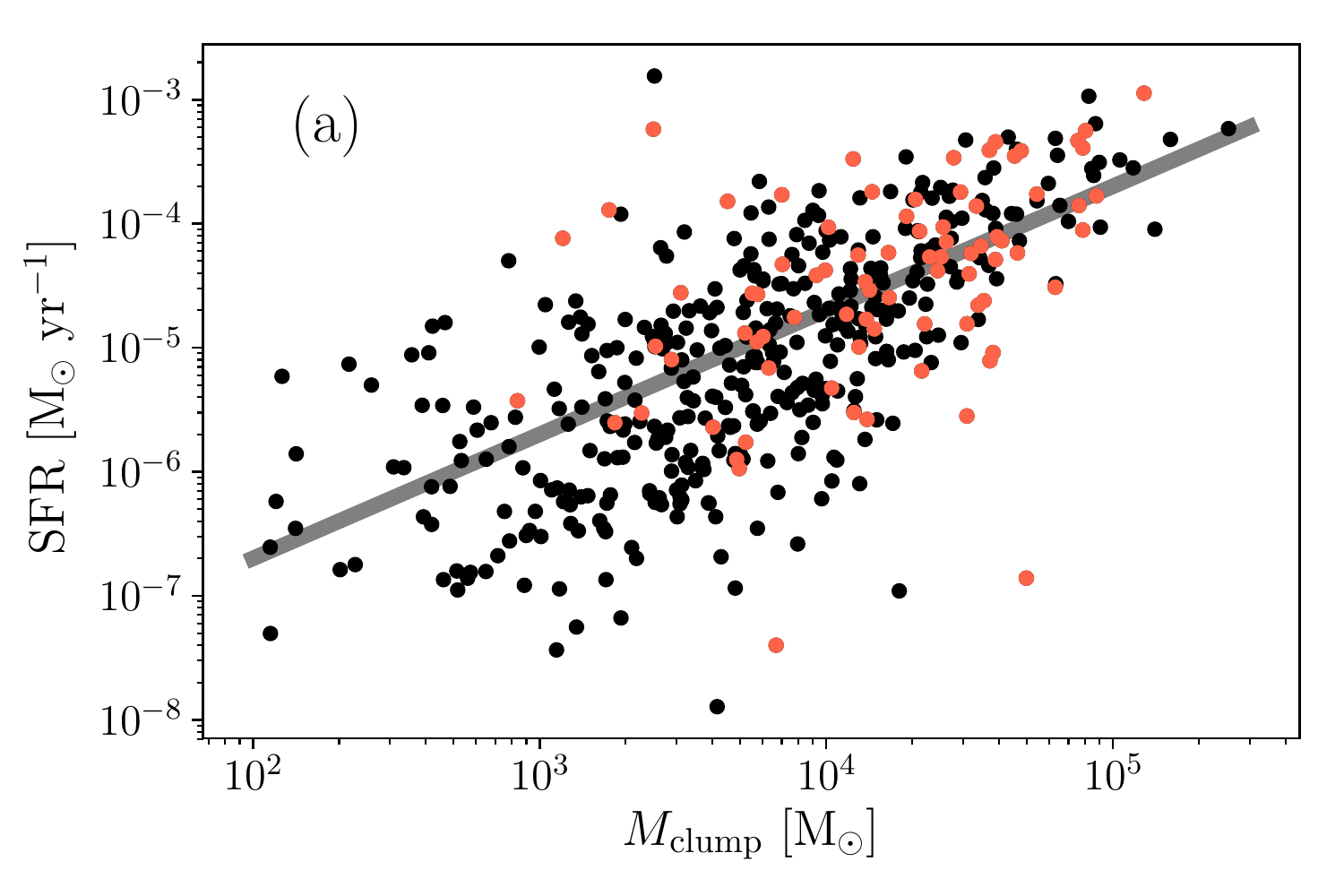}
	\includegraphics[width=0.49\textwidth]{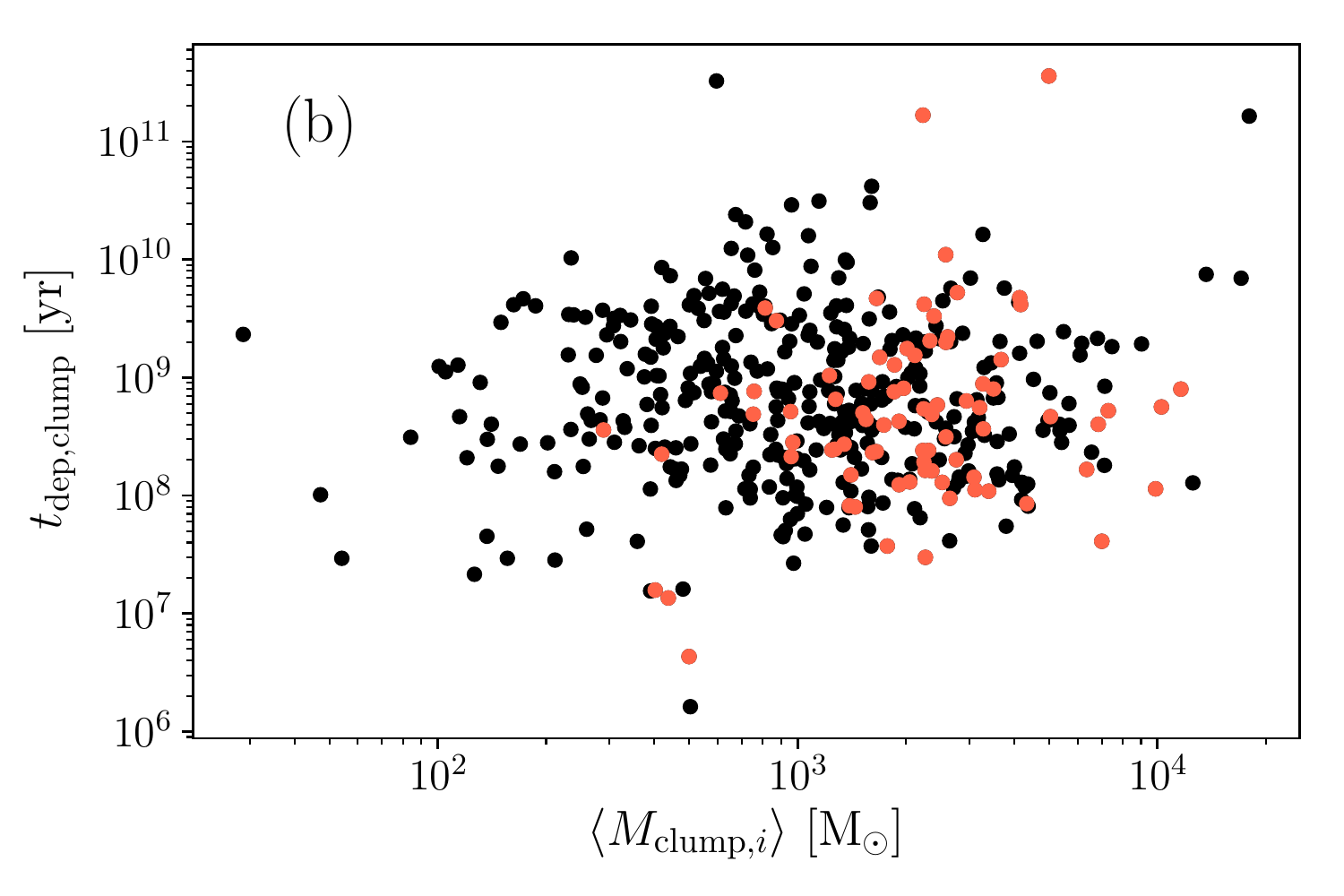}
	\includegraphics[width=0.49\textwidth]{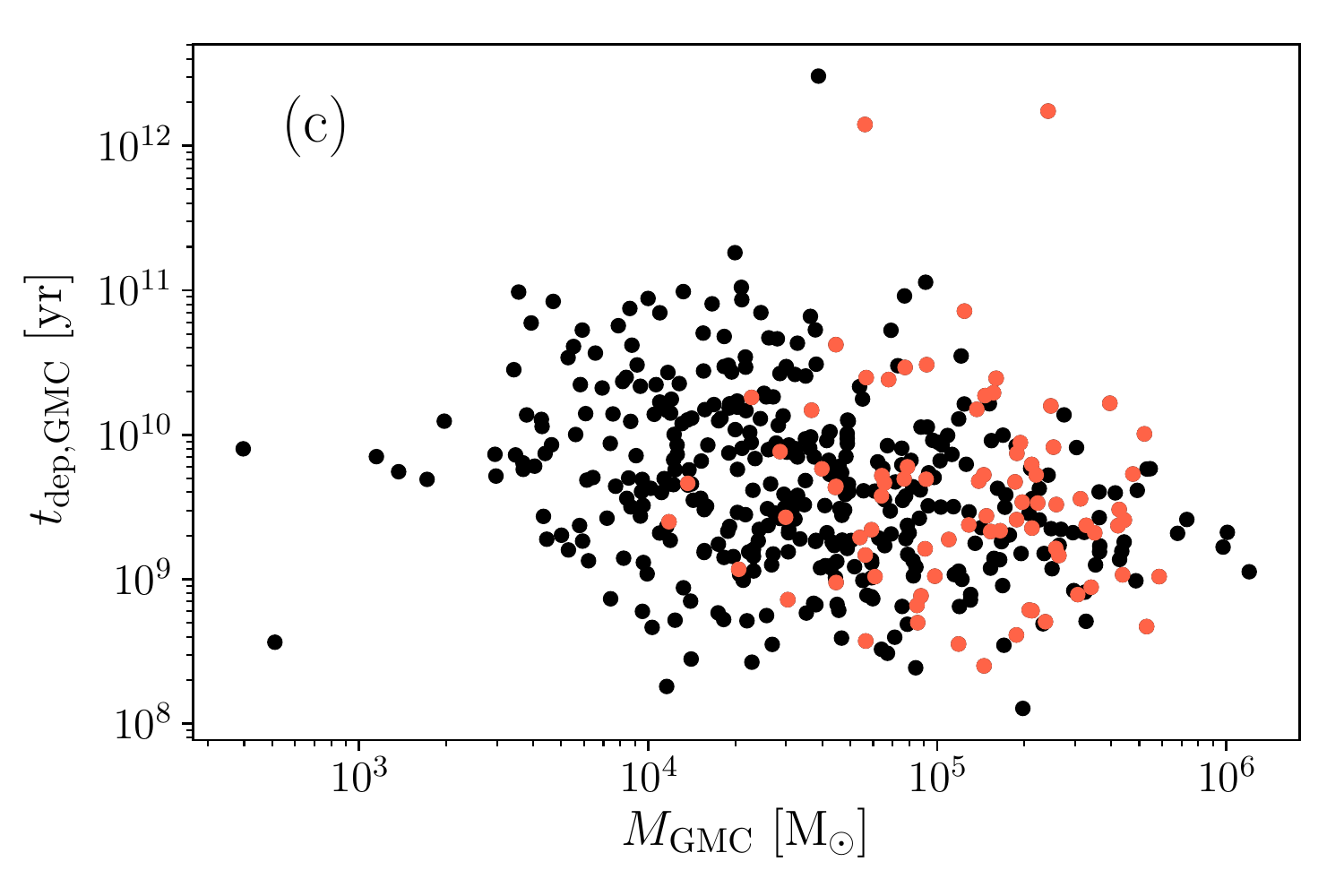}
	\includegraphics[width=0.49\textwidth]{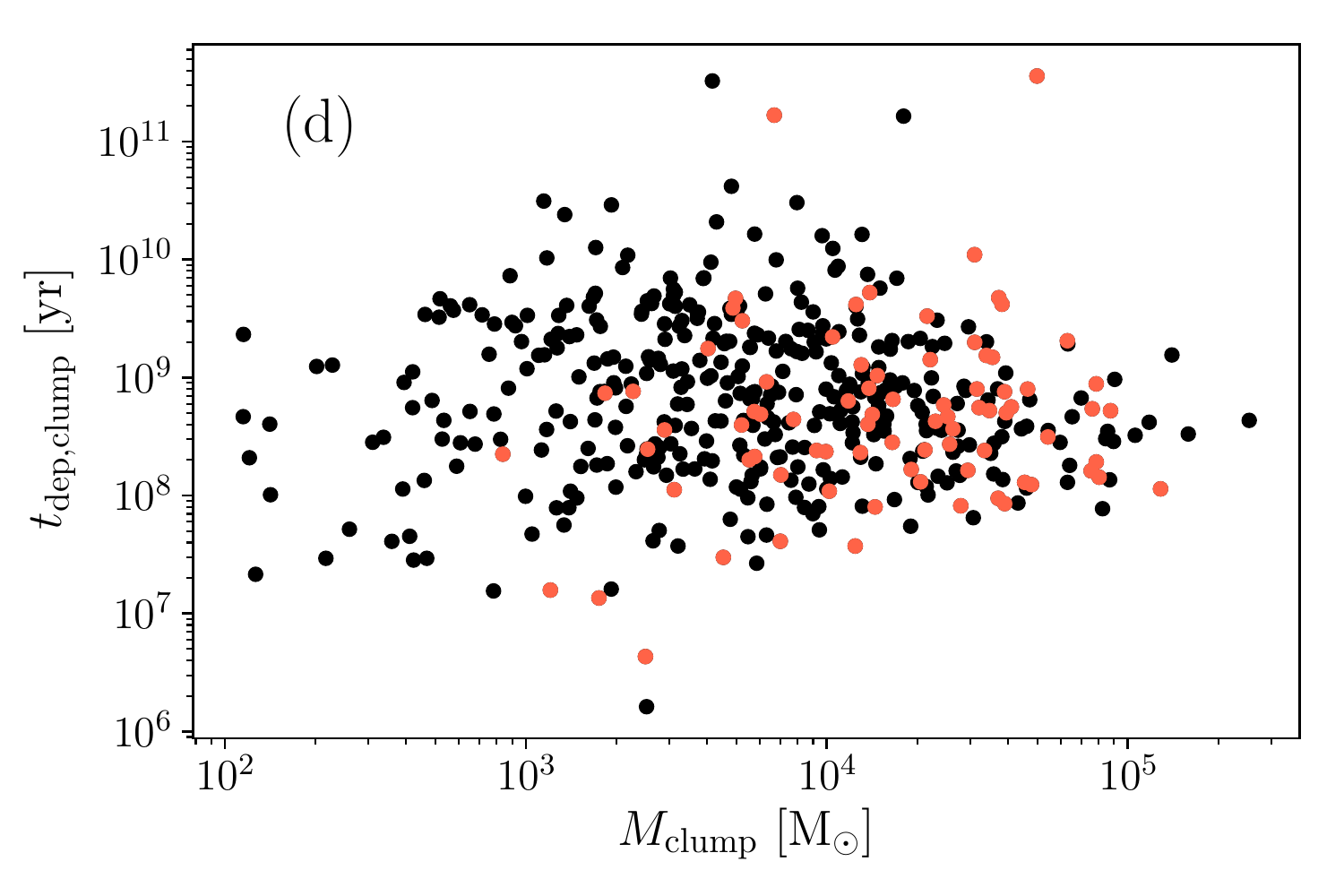}
\caption{Star formation rates derived from Hi-GAL 70 $\mu$m emission, compared with cloud and clump masses. (a): SFR as a function of clump mass. The grey line represents unity slope. (b) Clump mass depletion time, against mean clump mass. (c) and (d): Mass depletion times for clouds and clumps as a function of cloud and clump masses. As in Figure \ref{fig:cfeprops}, red points indicate clouds with virial parameters $\alpha_\mathrm{vir}<2$.}
\label{fig:sfr_trends}
\end{center}
\end{figure}

\begin{figure}[!htbp]
\begin{center}
	\includegraphics[width=0.49\textwidth]{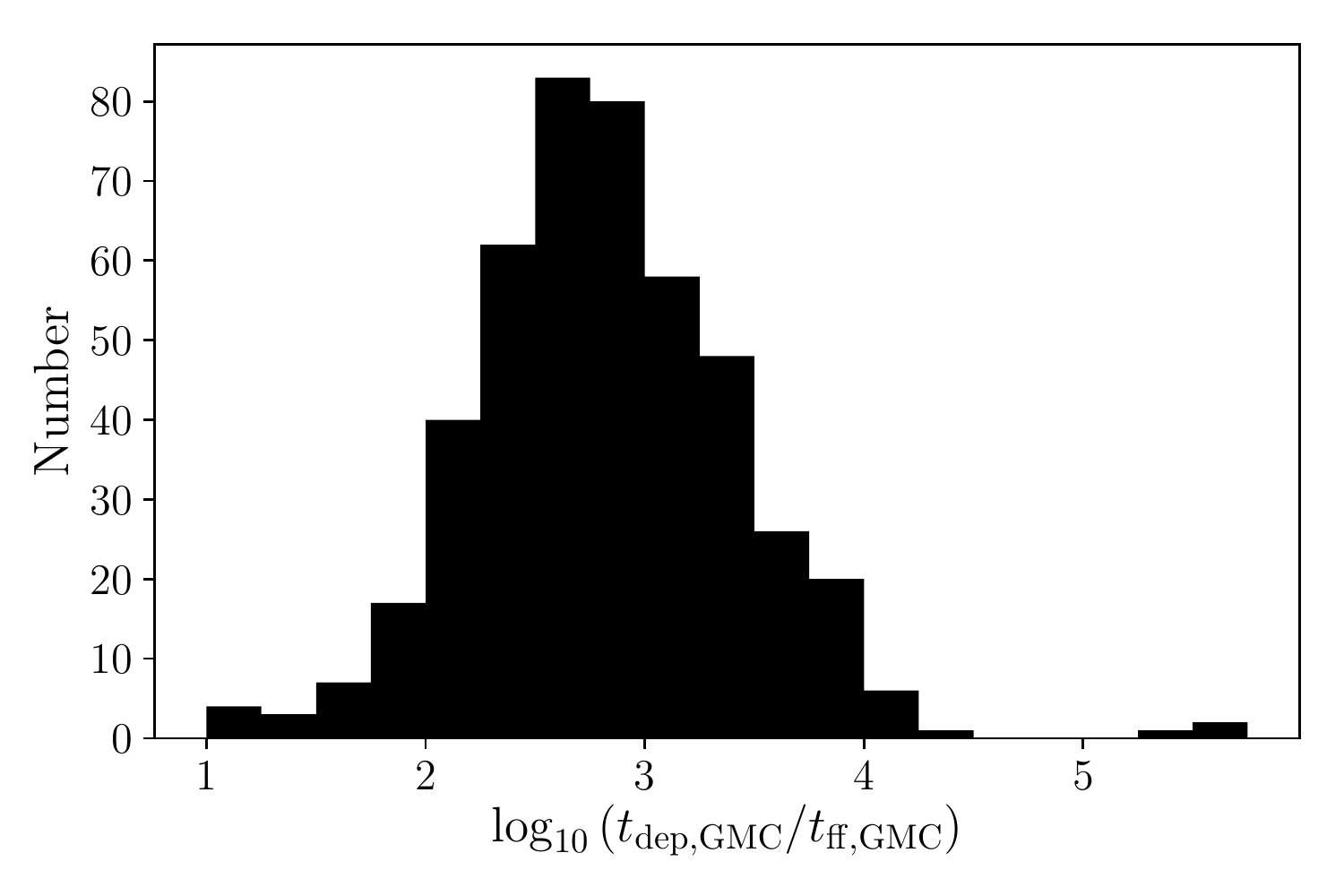}
	\includegraphics[width=0.49\textwidth]{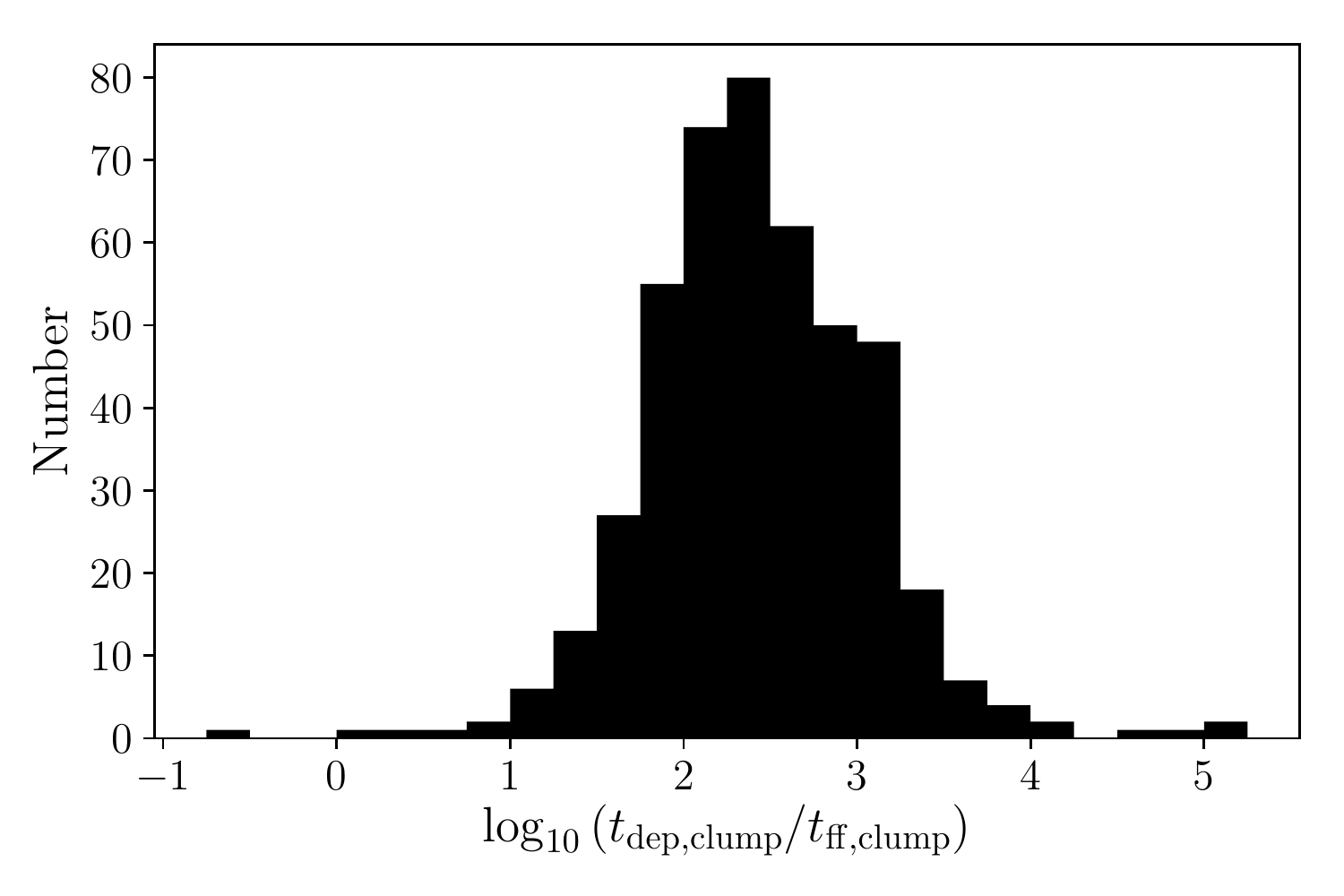}
\caption{{\it Left}: The distribution of log$_{10}(t_\mathrm{dep}/t_\mathrm{ff})$ for clouds. The mean value is $2.81\pm0.61$. {\it Right}: Same as above, but for clumps. The mean value is $2.43\pm0.65$.}
\label{fig:tdep_tff}
\end{center}
\end{figure}

In Figure \ref{fig:tdep_tff} we show the distributions of the ratios of the depletion times to the free-fall times for both clouds and clumps. For GMCs, we find the equivalent of $\epsilon_{\mathrm{ff,GMC}} = 0.15\%$, which is lower than the characteristic 0.7\% found by \citet{Utomo18} for nearby galaxies.  However, the nature of the \citet{Utomo18} measurement includes star formation over kpc scales and compares it to GMC-scale free-fall times.  Thus, star formation not directly associated with the molecular gas is included in their measurement but excluded in our measurement.  Based on the scaling of star formation rate of gas in the catalogue to that expected for the Milky Way as a whole (Section \ref{sec:sfrs}), this could easily account for the discrepancy.  More interesting is the significant difference between $\epsilon_{\mathrm{ff,clump}}=0.37\%$ for the clumps and the lower value for the clouds.  This implies there is a change in the conditions of the molecular gas that is relevant to setting the star formation rate beyond just density in clumps vs.~clouds.

To test the density thresholds used by theoretical numerical simulations, we plot SFR against number densities across entire clouds, and the number densities in their constituent clumps (Figure \ref{fig:sfr_n}). We find SFR is weakly correlated with number densities in clumps, but when the densities are instead calculated over whole clouds, this correlation is no longer present. However, when depletion times are investigated in place of star formation rates, this correlation disappears (Figure \ref{fig:tdep_n}). Thus the correlation seen in Figure \ref{fig:sfr_n} is likely merely a consequence of more massive clouds having greater clump number densities, as seen in Figure \ref{fig:n_Mgmc}. Thus, while it seems valid to convert a fixed fraction of cloud mass into stars once densities indicate a GMC has formed in a simulation, as per Figure \ref{fig:sfr_trends}, a  dispersion consistent with observations should be included also.

\begin{figure}[!htbp]
\begin{center}
	\includegraphics[width=0.49\textwidth]{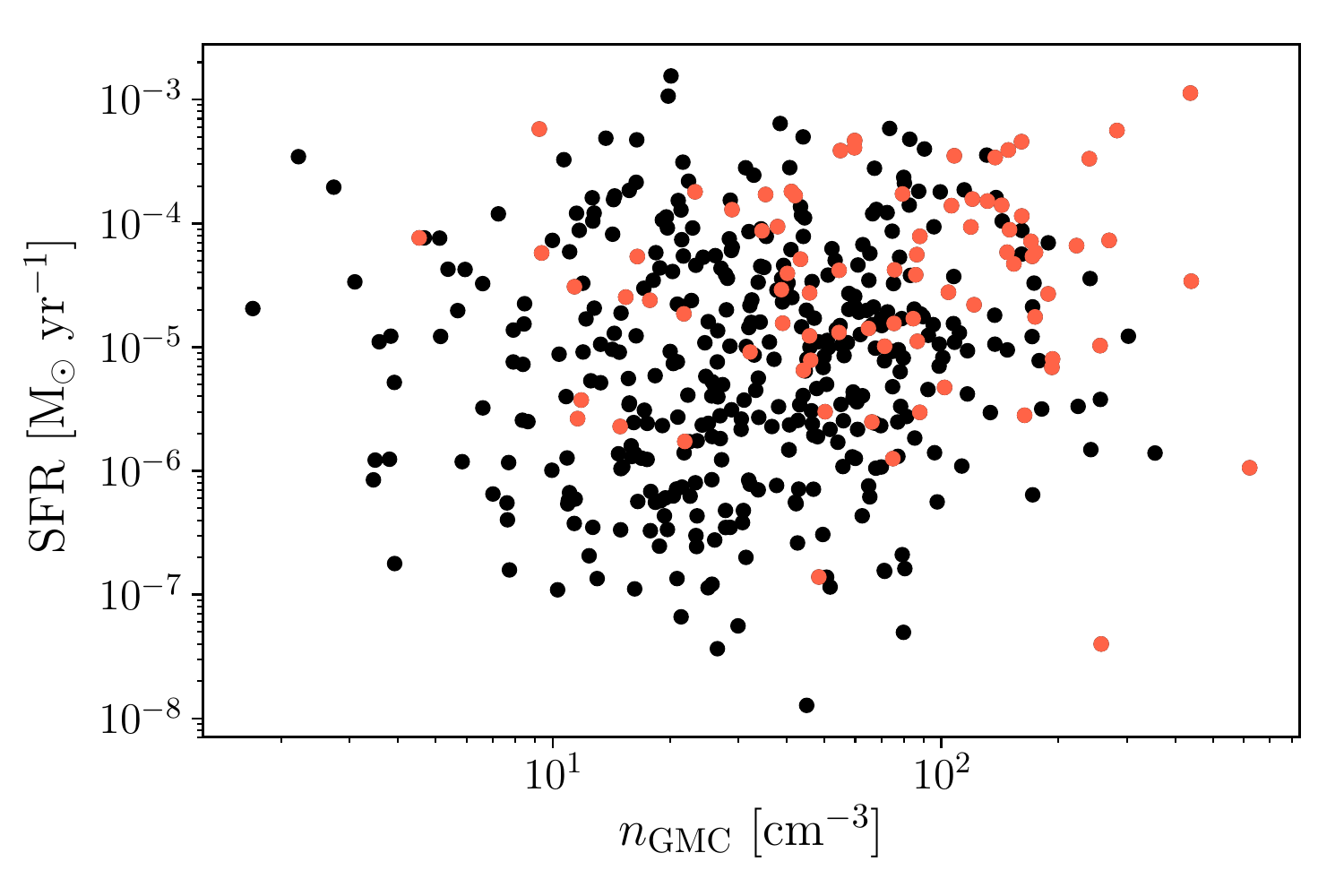}
	\includegraphics[width=0.49\textwidth]{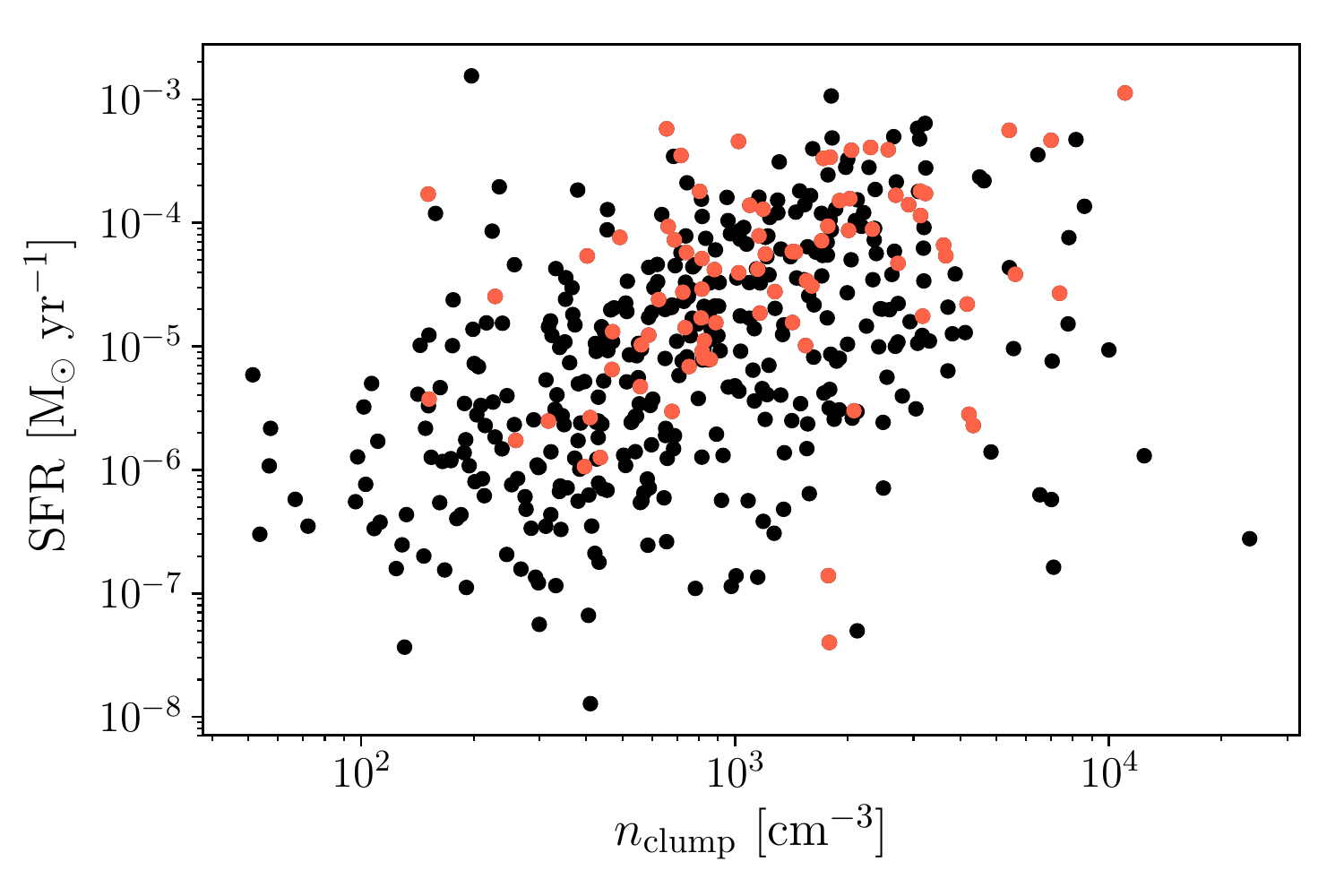}
\caption{{\it Left}: 70 $\mu$m star formation rates against the average number density across the entire cloud. {\it Right}: Same as above, but for average number density for clumps within each cloud.  Red points indicate clouds with virial parameters $\alpha_\mathrm{vir}<2$.}
\label{fig:sfr_n}
\end{center}
\end{figure}

\begin{figure}[!htbp]
\begin{center}
	\includegraphics[width=0.6\textwidth]{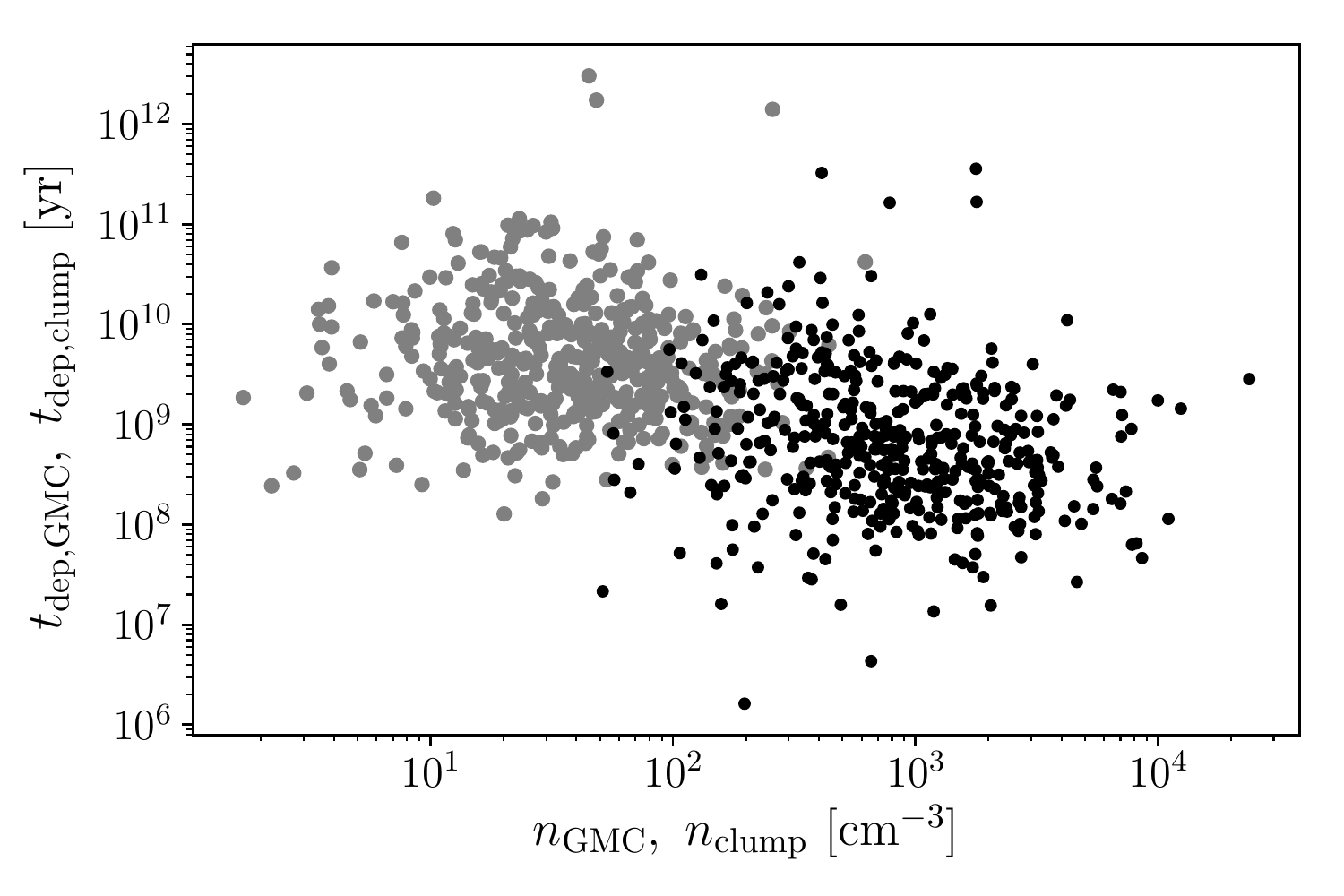}
\caption{Mass depletion time as a function of number density. Cloud values are shown in grey, and mean clump values are shown in black.}
\label{fig:tdep_n}
\end{center}
\end{figure}

\section{SFR Densities compared to the Kennicutt-Schmidt Law}
\label{s:ks}

The Kennicutt-Schmidt (KS) law relates the SFR surface density ($\Sigma_\mathrm{SFR}$) to the surface mass density of gas ($\Sigma_\mathrm{gas}$) through,
\begin{equation}
	\Sigma_\mathrm{SFR} = A \left(\frac{\Sigma_\mathrm{gas}}{1~\mathrm{M}_\odot~\mathrm{pc}^{-2}}\right)^N,
\label{eq:ks}
\end{equation}
where $A = (2.5\pm0.7)\times10^{-10}$ M$_\odot$ yr$^{-1}$ pc$^{-2}$ and $N=1.4 \pm 0.15$ is the power-law exponent \citep{Kennicutt1998}. In Figure \ref{fig:ks}, we compare our cloud data to the KS law (shown as a thick grey line). Our data fit the law remarkably well, although with a significant degree of dispersion. There are a few clouds with abnormally low $\Sigma_\mathrm{SFR}$ values. We find no other abnormalities in these clouds, and they fall within the overall scatter of star formation rates calculated using alternate calibrations from 24 $\mu$m emission \citep{Rieke09} and total infrared emission \citep{Hao11,Murphy11}.

\begin{figure}[!htbp]
\begin{center}
	\includegraphics[width=0.6\textwidth]{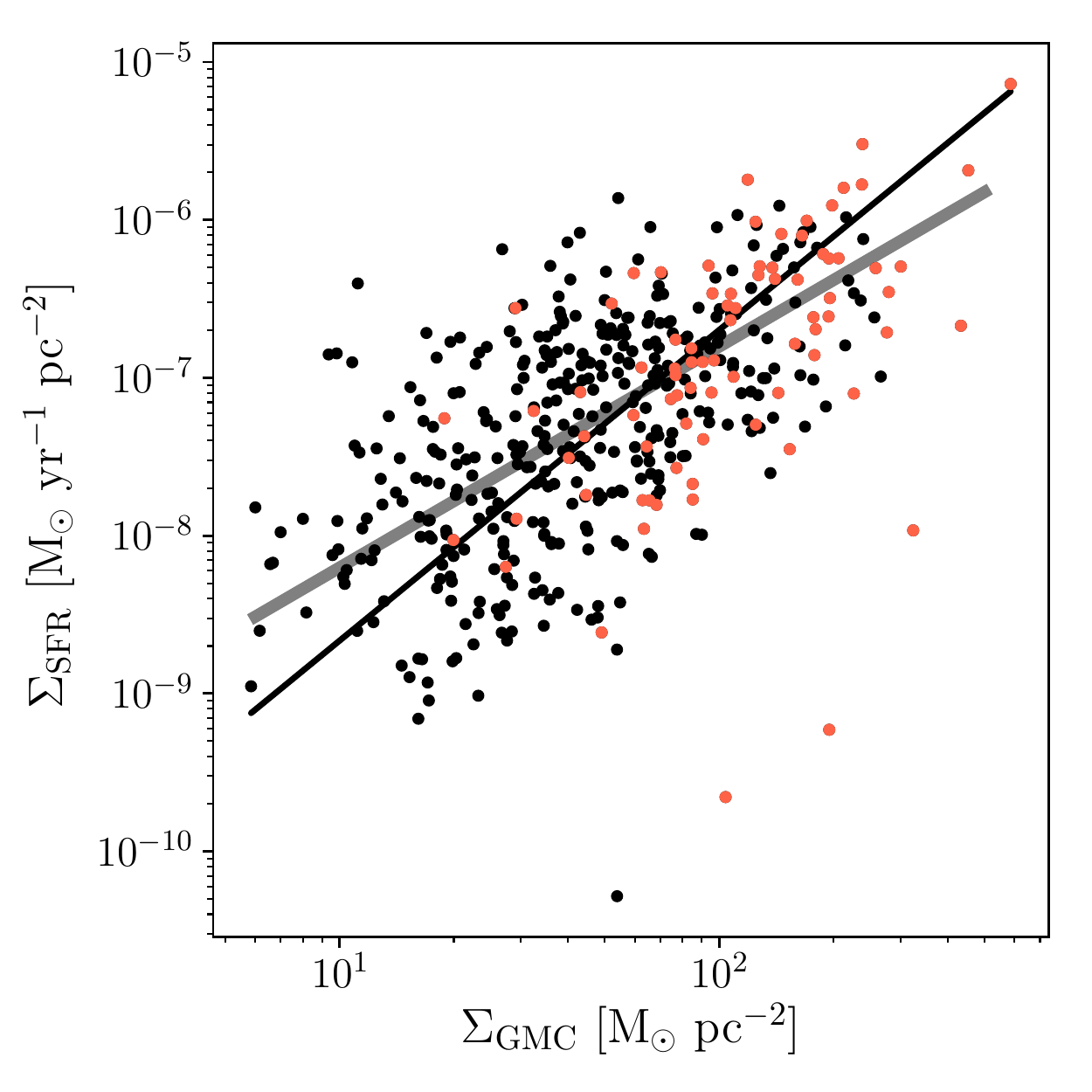}
\caption{The Kennicutt-Schmidt law, compared to our star formation rate surface densities. The KS law (Equation \ref{eq:ks}) is shown as a thick grey line. Thin black line represents the power-law fit to the data, and has a slope of $\beta = 2.0\pm0.1$.  Red points indicate clouds with virial parameters $\alpha_\mathrm{vir}<2$.}
\label{fig:ks}
\end{center}
\end{figure}

We fit Equation \eqref{eq:ks} to these data and find parameters of $A = (2.3\pm1.1)\times10^{-11}$ M$_\odot$ yr$^{-1}$ pc$^{-2}$ and $N=2.0 \pm 0.1$. The $1\sigma$ dispersion about this regression is 0.6 dex. Our fit slope is steeper than KS. However, there are caveats to applying the KS law here. First, \citet{Kennicutt1998} derived this law from disk-averaged values of galaxies, not on a cloud-by-cloud basis. \citet{Schruba10} and \citet{Onodera10} showed that on GMC scales the scatter about the KS law increases so greatly that the relationship essentially breaks down. While we still see evidence for this relationship, we also see a large degree of scatter about that relationship. Second, the KS law was derived using total gas mass (H{\sc i} + H$_2$), whereas we are measuring only molecular gas. Using kpc-scale analysis based on lines-of-sight, \citet{Leroy13} studied this relationship with molecular gas, and found a slope of $N=1.0\pm0.2$. 
Finally, all of the above references anchor the star formation rates to H$\alpha$ emission tracer \citep[][supplemented with UV and 24 $\mu$m emission]{Onodera10,Leroy13}. Compared with our 70 $\mu$m emission tracer, H$\alpha$ is more sensitive to star formation at the upper end of the IMF and to newer star formation. Furthermore, while 70 $\mu$m emission depends on dust attenuation in order to trace star formation, H$\alpha$ is hampered by it.  Finally, our measurement of star formation rate is based on spatial and morphological coincidence but may associate star formation with molecular clouds over a long path length, averaging multiple star forming regions along the line of sight.  All these systematic effects could easily account for the differences in the observed KS law index.

\section{Virial Parameters}
\label{s:vir}

In the previous sections, we focused on the scalings of clump and cloud properties and the resulting star formation properties. However, in Figures \ref{fig:cfes}, \ref{fig:nclumps}, \ref{fig:Mclump}, \ref{fig:CFE_Rgal}, \ref{fig:cfe_sfr}, \ref{fig:sfr_trends}, \ref{fig:sfr_n}, and \ref{fig:ks}, we have indicated those clouds with $\alpha_\mathrm{vir}<2$ to determine if they have distinct star formation properties compared to those with $\alpha_\mathrm{vir}>2$.

\begin{figure}[!htbp]
\begin{center}
	\includegraphics[width=0.6\textwidth]{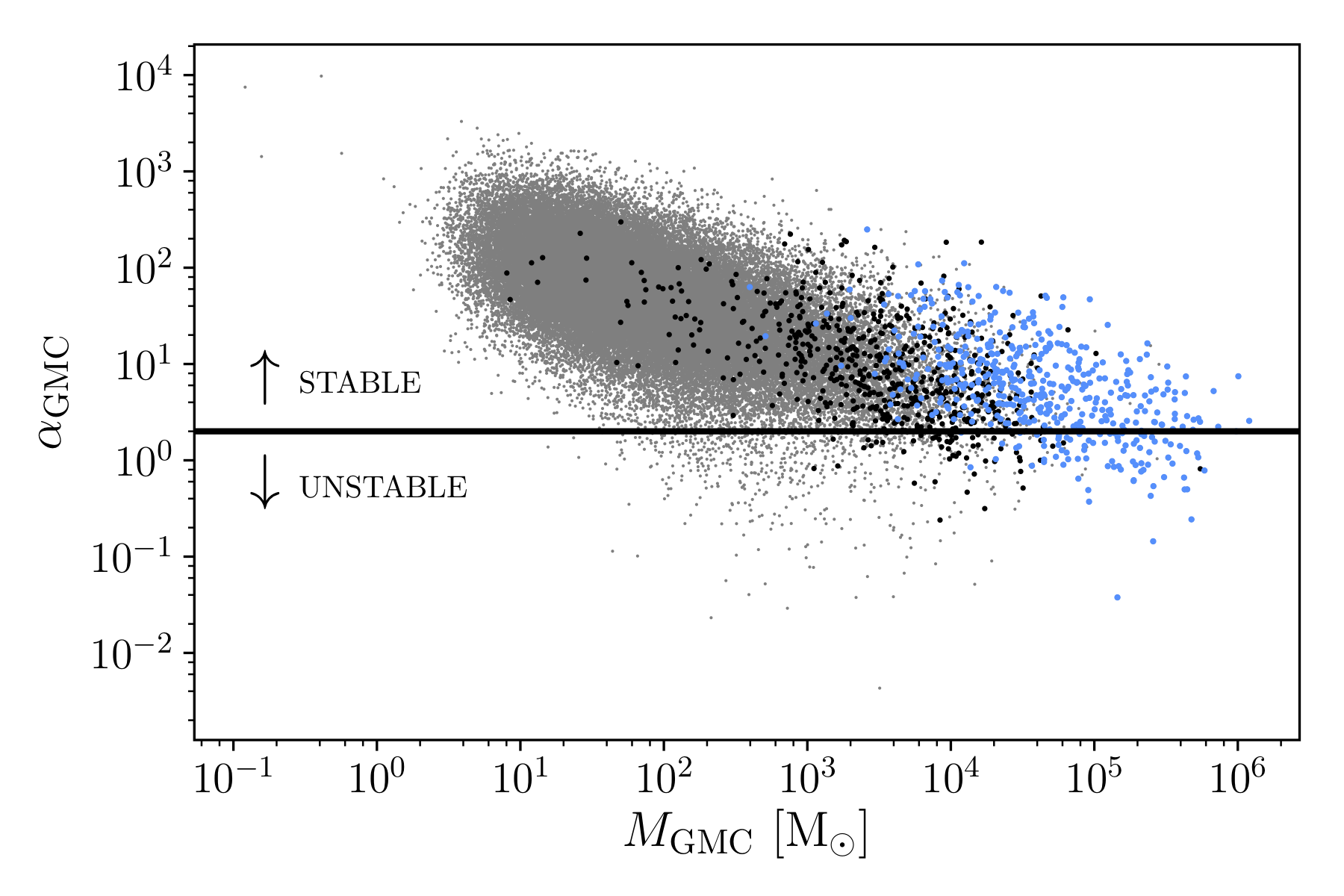}
\caption{Virial parameters and masses of all clouds in the COHRS catalog (grey), as well as those which were matched to Hi-GAL clumps, but did not meet our $3'$ cloud angular size requirement (black), and those used in our analysis (blue). The black line denotes $\alpha=2$, the nominal lower bound on stable clouds.}
\label{fig:all_a}
\end{center}
\end{figure}

Overall, there are not significant differences between the low- and high-virial parameter clouds in terms of their star formation or clump formation properties.  Most of the variation can be understood in terms of the variations of the GMC population itself.  In Figure \ref{fig:all_a}, we show the virial parameter distribution for all the molecular clouds in \citet{Colombo18} and the well-resolved subset of the clouds included in our sample (blue) and those that matched to Hi-GAL clumps but did not have a sufficiently large angular size to be included in the sample (black).  This figure shows that high mass clouds typically have lower virial parameters.

In the plots above, those clouds which have $\alpha<2$ are colored red. In Section \ref{ss:cfeprops} we noted that despite evidence elsewhere for star formation in clouds with masses of a few $10^3~\mathrm{M}_\odot$, none of our relatively low mass clouds have $\alpha_\mathrm{vir}<2$. However, this is not a matter of their lack of existence. Rather, they likely did not make our sample restriction to well-resolved clouds.  Figure \ref{fig:all_a} shows that potentially unstable clouds exist down to $10~\mathrm{M}_\odot$, and some with a few $10^3~\mathrm{M}_\odot$ were matched to Hi-GAL clumps but did not make our sample cut. While the COHRS catalog was created to identify clouds, it also contains a multitude of small objects, too small to contain clumps. 80\% of COHRS objects are smaller than 30'' and 100 M$_\odot$, and are therefore too small to be considered clouds for the purposes of this study. It is these objects which make up the grey points in Figure \ref{fig:all_a}.

The vast majority of COHRS objects not containing clumps are stable against gravitational collapse. This is expected, since clouds undergoing collapse should have denser regions, which we would detect as clumps. Furthermore, only a small portion of the clouds in the catalog have small virial parameters, and of those, we identify clumps in a fraction enhanced above the overall detection rate. Also, the mean $\alpha$ value for clouds not matched to clumps is roughly an order of magnitude greater than the mean for those clouds containing clumps, suggesting that those without clumps are more stable against collapse. It should also be noted that many of these clouds with high virial parameters are too small for us to identify any substructure due to our high-pass filter.

Of those clouds containing clumps, both virialized and non-virialized clouds exist throughout the full range of clump mass fractions (e.g. Figure \ref{fig:cfeprops}), cloud- and clump-mass-normalized SFRs (Figure \ref{fig:sfr_trends}), and SFR surface densities (Figure \ref{fig:ks}). For each of these quantities, the mean value for the virialized population is not distinguishable from that of the non-virialized population, except inasmuch as higher mass clouds tend to have lower virial parameters. The dispersion caused by observational uncertainties causes scatter in the derived $\alpha$s, and we find no evidence that the virial parameter is a robust discriminator for substructure formation or star formation.

This absence of evidence should not be taken as an indication that gravitational effects are uninmportant.  The simple virial parameter can be a poor diagnostic of internal energies on small scales \citep{Beaumont13}.  On clump scales, \citet{Traficante18} also found that the virial parameter is a poor descriptor of stability. They found infalling motions in both virialized and non-virialized clumps, and the distributions of virial parameters for populations at various evolutionary stages were indistinguishable. Overall, the virial theorem is a blunt instrument, and the effects of self-gravitation are too weak to overcome the observational limitations of our diagnostic.

\section{Velocity Distributions and Spiral Arms}

We examine cloud and clump velocities to investigate the relative distributions of clouds and their clumps, as well as the sources of support in virialized clouds. Figure \ref{fig:vel} shows the velocity distributions for the clouds and clumps therein. The left panel shows the velocity disperion of each cloud. The mean $\sigma_v$ is $3.9\pm2.0$ km s$^{-1}$. The sound speed in the ISM can be calculated as
\begin{equation}
	c_s = \sqrt{\frac{kT}{\mu m_\mathrm{H}}},
\end{equation}
where $k$ is the Boltzmann constant, $\mu=2.4$ is the mean molecular mass in the molecular ISM, and $m_\mathrm{H}$ is the mass of a hydrogen atom. This results in a sound speed of $0.18$ to $0.32~\mathrm{km}~\mathrm{s}^{-1}$ for gas temperatures of 10 to 30 K. Comparing this to the distribution of cloud line widths, we see that velocities in all of our clouds are supersonic. The clouds which are unstable according to the virial theorem (shown in red) have a mean $\sigma_v$ of $2.6\pm0.9$ km s$^{-1}$, lower than that of the whole sample, as expected since $\alpha_\mathrm{vir}\propto \sigma_v^2$.

\begin{figure}[!htbp]
\begin{center}
	\includegraphics[height=0.25\textheight]{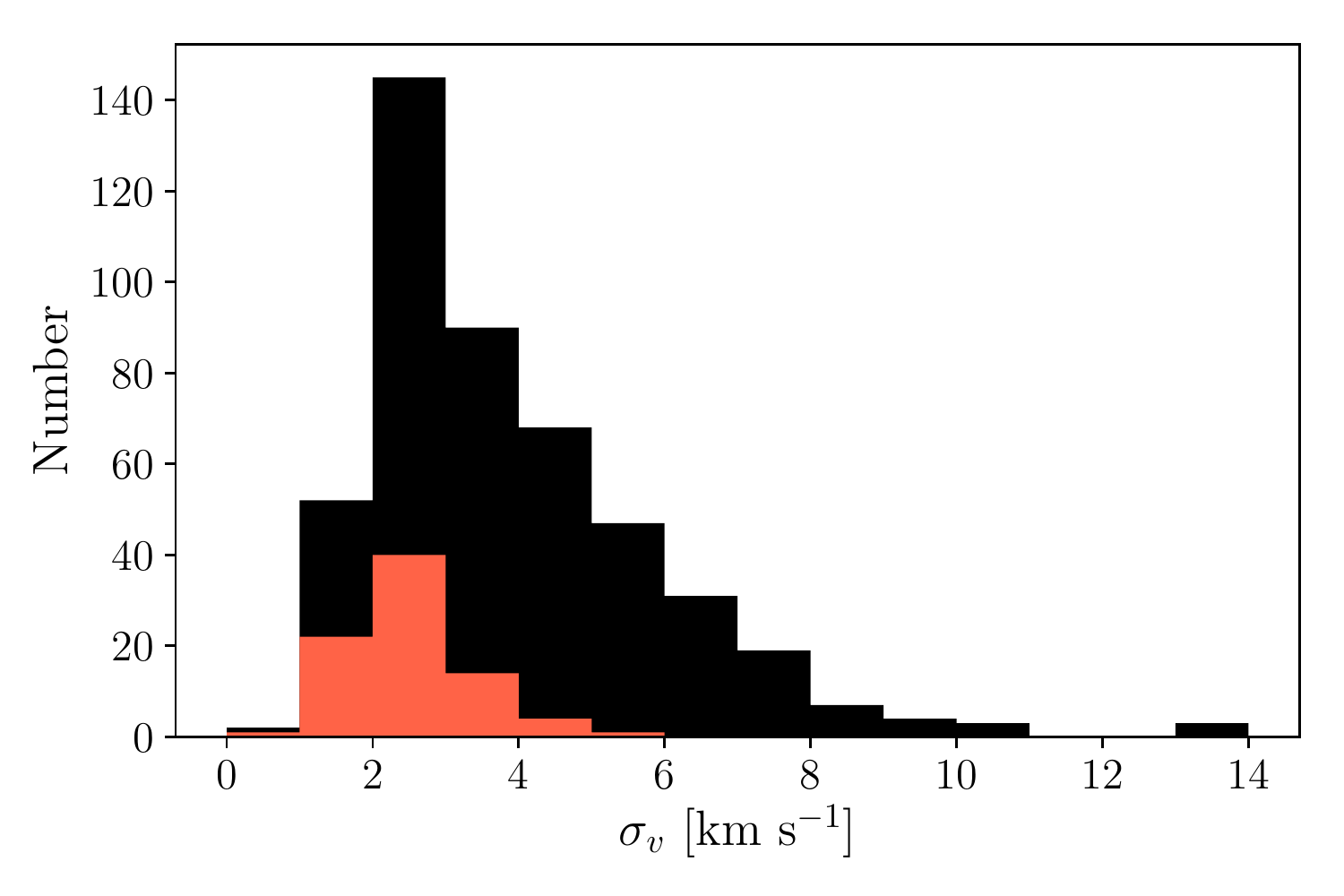}
	\includegraphics[height=0.25\textheight]{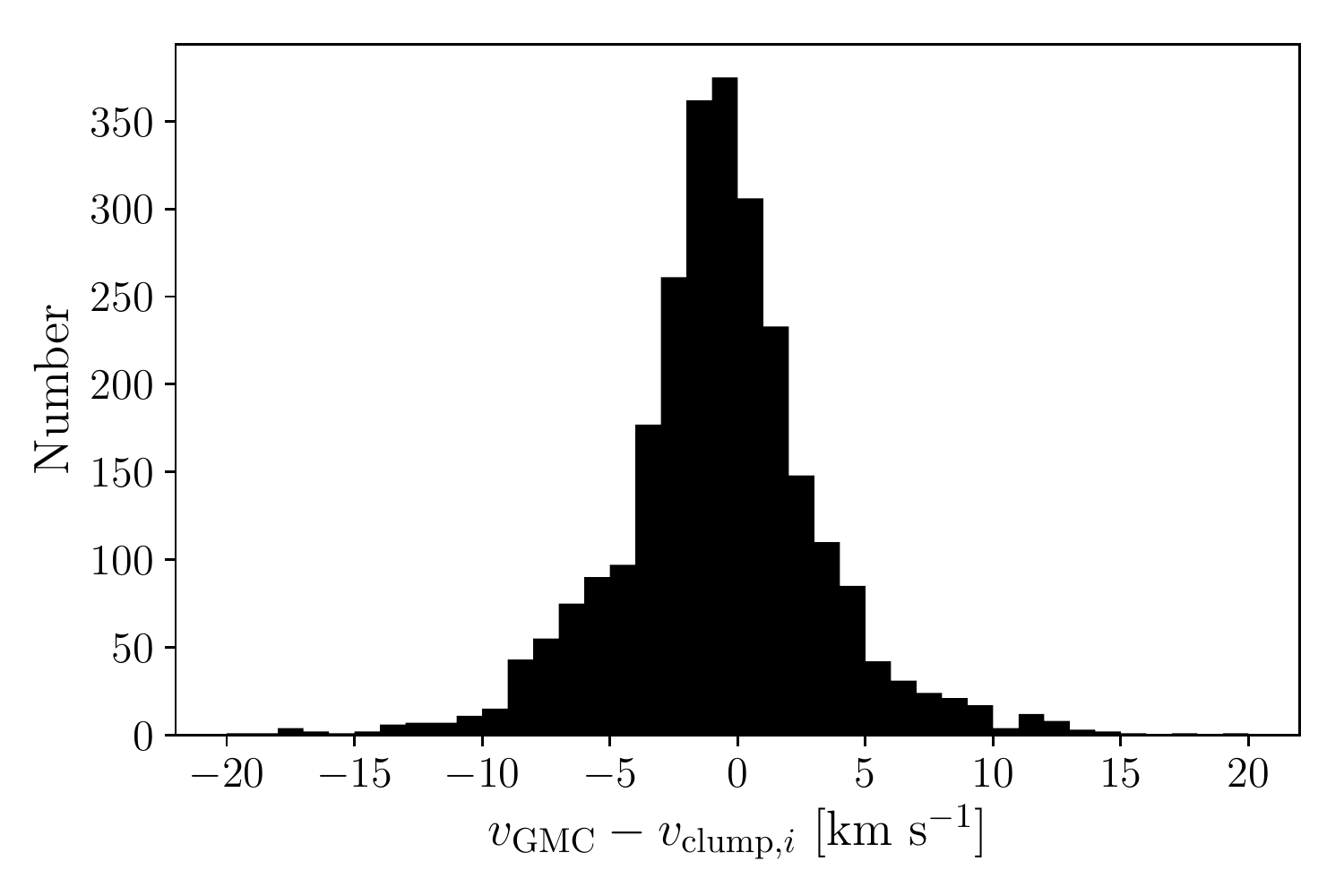}
	\includegraphics[height=0.25\textheight]{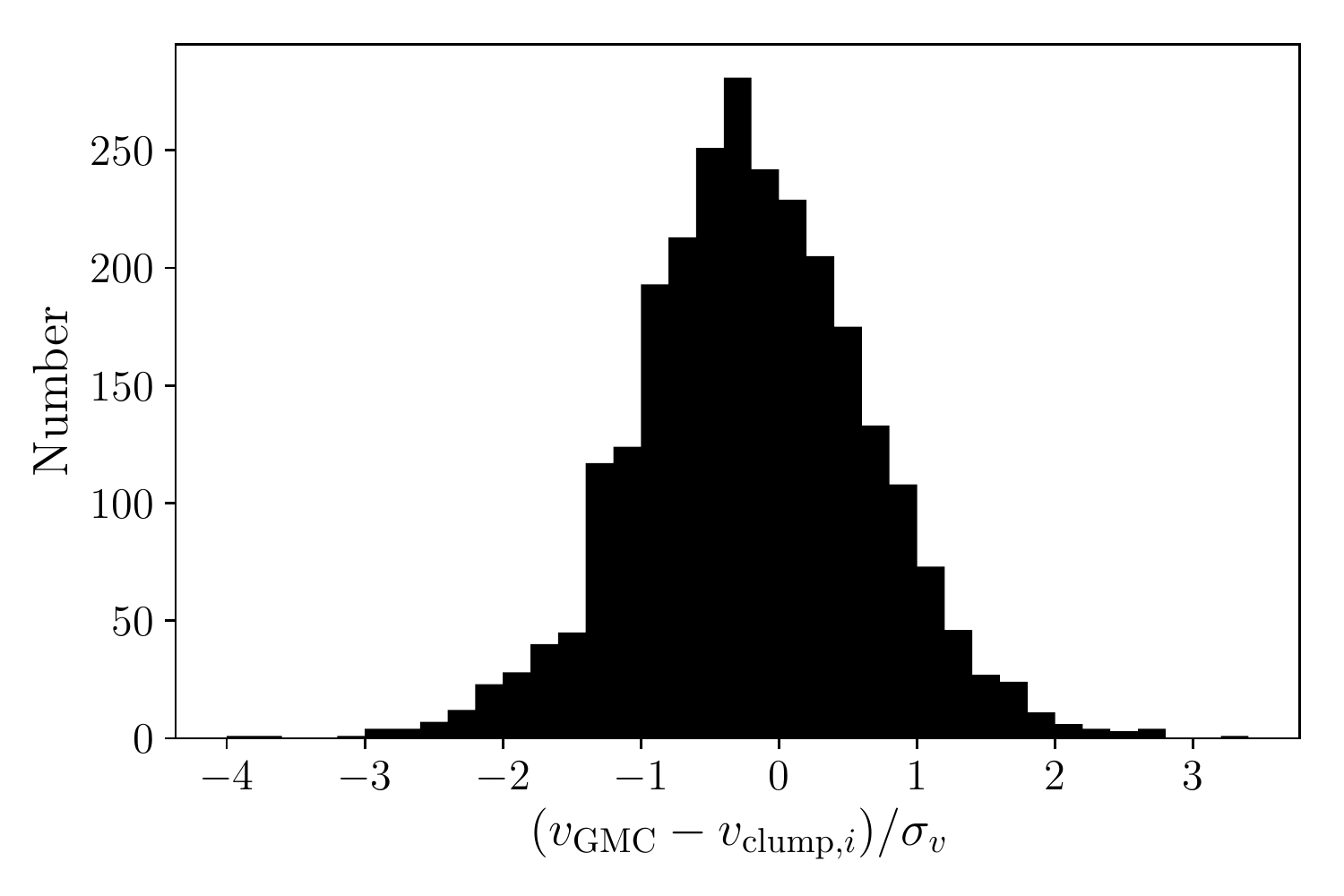}
\caption{{\it Left}: Distribution of the line widths, $\sigma_v$, for clouds. Red shows only those clouds with virial parameters $\alpha_\mathrm{vir}<2$. {\it Right}: Distribution of the differences between cloud centroid velocity and velocities of each of its clumps. {\it Lower}: Same as right panel, but normalized by cloud line width.}
\label{fig:vel}
\end{center}
\end{figure}

The right panel in Figure \ref{fig:vel} shows the difference between the cloud's $v_\mathrm{LSR}$ and that of each of its constituent clumps. This is skewed slightly towards greater clump velocities, with a mean difference of $-0.73\pm0.08$ km s$^{-1}$. The offset is present both when each clump is considered individually (as shown) and when the average clump velocity for each cloud is compared to the cloud velocity ($v_\mathrm{GMC}-\langle v_{\mathrm{clump},i}\rangle$). This is within the 1 km s$^{-1}$ velocity resolution of COHRS, but nevertheless could be a valid trend. A similar distribution shape was found by \citet{Battisti14}, who did an analogous analysis using GRS for clouds and BGPS for clumps. Note we have some overlap in velocity sources with this study, which could contribute to the similarity in distribution shape.

In the lower panel, we show these velocity differences normalized by the cloud line width. We find a smooth normal distribution with a mean and standard deviation of $-0.2$ and 0.8, respectively. Negative values refer to greater line-of-sight velocities in clumps, which correspond to the inner edges of spiral arms for clouds on the near side of the Galactic Center (when observing in the fourth Galactic quadrant), where the majority of our clouds are located. This is in line with \citet{Vallee14}, who found that $^{13}$CO is found preferentially on the inner edge of spiral arms, as whereas $^{12}$CO defines the mid-lane. Furthermore, 24 $\mu$m and 60 $\mu$m emission is found even closer to the inner edges \citep[according to][]{Vallee14}. This suggest that denser gas is found toward the inner edges of spiral arms, in line with our velocity offsets.

\section{Conclusions}

We have paired our catalog of Hi-GAL dust-identified molecular cloud clumps with the CO-identified molecular cloud catalog of \citet{Colombo18} in the Galactic longitude range $10^\circ < \ell < 56^\circ$. This took the form of matching clouds to their constituent clumps. Our final sample of matched objects consists of 3,674 clumps matched to 473 clouds. We draw a number of conclusions from the properties of these nested structures.

\begin{enumerate}


\item Clumps are the high density substructures of clouds, with mean densities of $60~\mathrm{cm}^{-3}$ for clouds and $1700~\mathrm{cm}^{-3}$ for clumps (Figure \ref{fig:nn}).

\item Averaged across our entire sample, a cloud's mass is not correlated with its average number density. However, more massive clouds tend to host denser clumps (Figure \ref{fig:n_Mgmc}).

\item The mean clump mass fraction is $\langle M_{\mathrm{clump}}/M_\mathrm{GMC} \rangle = 0.20^{+0.13}_{-0.10}$ (Figure \ref{fig:cfehists}). We find (at best) very weak negative correlations of clump mass fraction with cloud mass and cloud surface mass density, although scatter in the mean properties allows for undiscovered correlations. We find that, on average, clump formation efficiency is independent of the virial parameter of the cloud, and the cloud's line width (Figure \ref{fig:cfeprops}).

\item More massive clouds contain both more clumps, and more massive clumps (there is a dearth of high-mass clumps in low-mass clouds). Both the number and mass of clumps grow at rates more shallow than linear with cloud mass. The power-law slopes for mean clump mass and number of clumps as functions of cloud mass are $\beta=0.48\pm0.03$ and $\beta=0.54\pm0.03$, respectively (Figure \ref{fig:nclumps} and \ref{fig:Mclump}).

\item The mean clump mass fraction shows a weak decrease with Galactocentric radius. While the amount of gas present in a cloud likely has the larger effect on clump masses and star formation rates, the environment in which the gas is found may have some influence (Figure \ref{fig:CFE_Rgal}).


\item Within the observational scatter, there is no clear dependence of star formation rates times on cloud substructure as measured through the clump mass fraction. There is, however, a positive correlation between star formation depletion time in clumps and clump mass fraction (Figure \ref{fig:cfe_sfr}), and clump masses tend to be larger in clouds with greater clump mass fractions.
The scatter is large; however,  the weak correlations are consistent with an evolutionary trend in which clumps and clump mass fractions grow within clouds over time, become increasingly gravitationally unstable, and have correspondingly increased star formation rates. 

\item More massive clouds host not only more and larger clumps than less massive clouds, but also more star formation (Figure \ref{fig:sfr_trends}).  A cloud's star formation rate is dependent predominately on its mass or equivalently, its total clump mass

\item The star formation efficiency per free-fall time for GMCs is $\epsilon_{\mathrm{ff, GMC}} =0.15\%$ but the for clumps it is significantly larger: $\epsilon_{\mathrm{ff, clump}}=0.37\%$ (Figure \ref{fig:tdep_tff}).  
We find that the star formation rate of clouds is independent on their average number density, but a positive scaling is seen between the star formation rate and the clump density (Figure \ref{fig:sfr_n}, black points). However, this correlation is not present in the mass-normalized star formation measure, $t_\mathrm{dep}$ (Figure \ref{fig:tdep_n}), leading to the conclusion that the original scaling was due to the trend towards denser clumps in more massive clouds (Figure \ref{fig:n_Mgmc}).

\item The star formation rate surface densities of the clouds in the sample show a steeper slope ($N=2.0\pm0.1$) than the Kennicutt-Schmidt law ($N=1.4\pm0.15$) (Figure \ref{fig:ks}). This is in tension with extragalactic studies, which show the KS law breaking down at GMC scales.  The difference is likely due to the use of different star formation tracers and size scales probed by the observations.

\item The virial parameter of a cloud is a poor predictor of clump and star formation. Both virialized and non-virialized clouds exist throughout the observed ranges of clump mass fractions, specific SFRs, and SFR surface densities. For each of these quantities, the difference in mean values for the $\alpha>2$ and $\alpha<2$ populations are statistically insignificant (compare the red data for $\alpha<2$ shown throughout the figures to the black data for $\alpha >2$).

\item Clumps are distributed fairly uniformly within the velocity distributions of their host clouds. However, there is an offset towards higher line-of-sight velocities in clumps, potentially corresponding to denser structures being found preferentially toward the inner edges of spiral arms (Figure \ref{fig:vel}).

\end{enumerate}

\acknowledgements

SPIRE has been developed by a consortium of institutes led by Cardiff University (UK) and including Univ. Lethbridge (Canada); NAOC (China); CEA, LAM (France); IFSI, Univ. Padua (Italy); IAC (Spain); Stockholm Observatory (Sweden); Imperial College London, RAL, UCL-MSSL, UKATC, Univ. Sussex (UK); and Caltech, JPL, NHSC, Univ. Colorado (USA). This development has been supported by national funding agencies: CSA (Canada); NAOC (China); CEA, CNES, CNRS (France); ASI (Italy); MCINN (Spain); SNSB (Sweden); STFC (UK); and NASA (USA).

ER acknowledges the support of the Natural Sciences and Engineering Research Council of Canada (NSERC), funding reference number RGPIN-2017-03987.

This work made use of the {\sc astropy} \citep{astropy} and {\sc matplotlib} \citep{matplotlib} software packages and we are grateful for the continued efforts of the open source community at improving our capabilities in science.

Part of this work based on archival data, software or online services provided by the ASI SCIENCE DATA CENTER (ASDC).

\textit{Facility}: Herschel (SPIRE)

\bibliographystyle{apj}
\bibliography{Bibliography}

\begin{thebibliography}{}
\expandafter\ifx\csname natexlab\endcsname\relax\def\natexlab#1{#1}\fi

\bibitem[{{Astropy Collaboration} {et~al.}(2013){Astropy Collaboration},
  {Robitaille}, {Tollerud}, {Greenfield}, {Droettboom}, {Bray}, {Aldcroft},
  {Davis}, {Ginsburg}, {Price-Whelan}, {Kerzendorf}, {Conley}, {Crighton},
  {Barbary}, {Muna}, {Ferguson}, {Grollier}, {Parikh}, {Nair}, {Unther},
  {Deil}, {Woillez}, {Conseil}, {Kramer}, {Turner}, {Singer}, {Fox}, {Weaver},
  {Zabalza}, {Edwards}, {Azalee Bostroem}, {Burke}, {Casey}, {Crawford},
  {Dencheva}, {Ely}, {Jenness}, {Labrie}, {Lim}, {Pierfederici}, {Pontzen},
  {Ptak}, {Refsdal}, {Servillat}, \& {Streicher}}]{astropy}
{Astropy Collaboration}, {Robitaille}, T.~P., {Tollerud}, E.~J., {et~al.} 2013,
  \aap, 558, A33

\bibitem[{{Barnes} {et~al.}(2017){Barnes}, {Longmore}, {Battersby}, {Bally},
  {Kruijssen}, {Henshaw}, \& {Walker}}]{Barnes17}
{Barnes}, A.~T., {Longmore}, S.~N., {Battersby}, C., {et~al.} 2017, \mnras,
  469, 2263

\bibitem[{{Battisti} \& {Heyer}(2014)}]{Battisti14}
{Battisti}, A.~J., \& {Heyer}, M.~H. 2014, \apj, 780, 173

\bibitem[{{Beaumont} {et~al.}(2013){Beaumont}, {Offner}, {Shetty}, {Glover}, \&
  {Goodman}}]{Beaumont13}
{Beaumont}, C.~N., {Offner}, S. S.~R., {Shetty}, R., {Glover}, S. C.~O., \&
  {Goodman}, A.~A. 2013, \apj, 777, 173

\bibitem[{{Bergin} \& {Tafalla}(2007)}]{Bergin2007}
{Bergin}, E.~A., \& {Tafalla}, M. 2007, \araa, 45, 339

\bibitem[{{Bigiel} {et~al.}(2016){Bigiel}, {Leroy}, {Jim{\'e}nez-Donaire},
  {Pety}, {Usero}, {Cormier}, {Bolatto}, {Garcia-Burillo}, {Colombo},
  {Gonz{\'a}lez-Garc{\'\i}a}, {Hughes}, {Kepley}, {Kramer}, {Sandstrom},
  {Schinnerer}, {Schruba}, {Schuster}, {Tomicic}, \& {Zschaechner}}]{Bigiel16}
{Bigiel}, F., {Leroy}, A.~K., {Jim{\'e}nez-Donaire}, M.~J., {et~al.} 2016,
  \apj, 822, L26

\bibitem[{{Bolatto} {et~al.}(2013){Bolatto}, {Wolfire}, \& {Leroy}}]{Bolatto13}
{Bolatto}, A.~D., {Wolfire}, M., \& {Leroy}, A.~K. 2013, \araa, 51, 207

\bibitem[{{Calzetti} {et~al.}(2010){Calzetti}, {Wu}, {Hong}, {Kennicutt},
  {Lee}, {Dale}, {Engelbracht}, {van Zee}, {Draine}, {Hao}, {Gordon},
  {Moustakas}, {Murphy}, {Regan}, {Begum}, {Block}, {Dalcanton}, {Funes}, {Gil
  de Paz}, {Johnson}, {Sakai}, {Skillman}, {Walter}, {Weisz}, {Williams}, \&
  {Wu}}]{Calzetti10}
{Calzetti}, D., {Wu}, S.-Y., {Hong}, S., {et~al.} 2010, \apj, 714, 1256

\bibitem[{{Chomiuk} \& {Povich}(2011)}]{Chomiuk11}
{Chomiuk}, L., \& {Povich}, M.~S. 2011, \aj, 142, 197

\bibitem[{{Colombo} {et~al.}(2015){Colombo}, {Rosolowsky}, {Ginsburg},
  {Duarte-Cabral}, \& {Hughes}}]{Colombo15}
{Colombo}, D., {Rosolowsky}, E., {Ginsburg}, A., {Duarte-Cabral}, A., \&
  {Hughes}, A. 2015, \mnras, 454, 2067

\bibitem[{{Colombo} {et~al.}(2019){Colombo}, {Rosolowsky}, {Duarte-Cabral},
  {Ginsburg}, {Glenn}, {Zetterlund}, {Hernand ez}, {Dempsey}, \&
  {Currie}}]{Colombo18}
{Colombo}, D., {Rosolowsky}, E., {Duarte-Cabral}, A., {et~al.} 2019, \mnras,
  483, 4291

\bibitem[{{Dempsey} {et~al.}(2013){Dempsey}, {Thomas}, \& {Currie}}]{Dempsey13}
{Dempsey}, J.~T., {Thomas}, H.~S., \& {Currie}, M.~J. 2013, \apjs, 209, 8

\bibitem[{{Dobbs} {et~al.}(2011){Dobbs}, {Burkert}, \& {Pringle}}]{Dobbs11}
{Dobbs}, C.~L., {Burkert}, A., \& {Pringle}, J.~E. 2011, \mnras, 417, 1318

\bibitem[{{Draine}(2003)}]{Draine03}
{Draine}, B.~T. 2003, \araa, 41, 241

\bibitem[{{Draine} \& {Li}(2007)}]{Draine07}
{Draine}, B.~T., \& {Li}, A. 2007, \apj, 657, 810

\bibitem[{{Dunham} {et~al.}(2011){Dunham}, {Rosolowsky}, {Evans}, {Cyganowski},
  \& {Urquhart}}]{BGPS7}
{Dunham}, M.~K., {Rosolowsky}, E., {Evans}, II, N.~J., {Cyganowski}, C., \&
  {Urquhart}, J.~S. 2011, \apj, 741, 110

\bibitem[{{Eden} {et~al.}(2012){Eden}, {Moore}, {Plume}, \& {Morgan}}]{Eden12}
{Eden}, D.~J., {Moore}, T.~J.~T., {Plume}, R., \& {Morgan}, L.~K. 2012, \mnras,
  422, 3178

\bibitem[{{Elia} {et~al.}(2017){Elia}, {Molinari}, {Schisano}, {Pestalozzi},
  {Pezzuto}, {Merello}, {Noriega-Crespo}, {Moore}, {Russeil}, {Mottram},
  {Paladini}, {Strafella}, {Benedettini}, {Bernard}, {Di Giorgio}, {Eden},
  {Fukui}, {Plume}, {Bally}, {Martin}, {Ragan}, {Jaffa}, {Motte}, {Olmi},
  {Schneider}, {Testi}, {Wyrowski}, {Zavagno}, {Calzoletti}, {Faustini},
  {Natoli}, {Palmeirim}, {Piacentini}, {Piazzo}, {Pilbratt}, {Polychroni},
  {Baldeschi}, {Beltr{\'a}n}, {Billot}, {Cambr{\'e}sy}, {Cesaroni},
  {Garc{\'{\i}}a-Lario}, {Hoare}, {Huang}, {Joncas}, {Liu}, {Maiolo}, {Marsh},
  {Maruccia}, {M{\`e}ge}, {Peretto}, {Rygl}, {Schilke}, {Thompson},
  {Traficante}, {Umana}, {Veneziani}, {Ward-Thompson}, {Whitworth}, {Arab},
  {Bandieramonte}, {Becciani}, {Brescia}, {Buemi}, {Bufano}, {Butora},
  {Cavuoti}, {Costa}, {Fiorellino}, {Hajnal}, {Hayakawa}, {Kacsuk}, {Leto}, {Li
  Causi}, {Marchili}, {Martinavarro-Armengol}, {Mercurio}, {Molinaro},
  {Riccio}, {Sano}, {Sciacca}, {Tachihara}, {Torii}, {Trigilio}, {Vitello}, \&
  {Yamamoto}}]{Elia17}
{Elia}, D., {Molinari}, S., {Schisano}, E., {et~al.} 2017, \mnras, 471, 100

\bibitem[{{Ellsworth-Bowers} {et~al.}(2015){Ellsworth-Bowers}, {Rosolowsky},
  {Glenn}, {Ginsburg}, {Evans}, {Battersby}, {Shirley}, \& {Svoboda}}]{BGPS12}
{Ellsworth-Bowers}, T.~P., {Rosolowsky}, E., {Glenn}, J., {et~al.} 2015, \apj,
  799, 29

\bibitem[{{Ellsworth-Bowers} {et~al.}(2013){Ellsworth-Bowers}, {Glenn},
  {Rosolowsky}, {Mairs}, {Evans}, {Battersby}, {Ginsburg}, {Shirley}, \&
  {Bally}}]{BGPS8}
{Ellsworth-Bowers}, T.~P., {Glenn}, J., {Rosolowsky}, E., {et~al.} 2013, \apj,
  770, 39

\bibitem[{{Enoch} {et~al.}(2008){Enoch}, {Evans}, {Sargent}, {Glenn},
  {Rosolowsky}, \& {Myers}}]{Enoch08}
{Enoch}, M.~L., {Evans}, II, N.~J., {Sargent}, A.~I., {et~al.} 2008, \apj, 684,
  1240

\bibitem[{{Enoch} {et~al.}(2007){Enoch}, {Glenn}, {Evans}, {Sargent}, {Young},
  \& {Huard}}]{Enoch07}
{Enoch}, M.~L., {Glenn}, J., {Evans}, II, N.~J., {et~al.} 2007, \apj, 666, 982

\bibitem[{{Evans} {et~al.}(2014){Evans}, {Heiderman}, \&
  {Vutisalchavakul}}]{Evans14}
{Evans}, Neal~J., I., {Heiderman}, A., \& {Vutisalchavakul}, N. 2014, \apj,
  782, 114

\bibitem[{{Faesi} {et~al.}(2014){Faesi}, {Lada}, {Forbrich}, {Menten}, \&
  {Bouy}}]{Faesi14}
{Faesi}, C.~M., {Lada}, C.~J., {Forbrich}, J., {Menten}, K.~M., \& {Bouy}, H.
  2014, \apj, 789, 81

\bibitem[{{Federrath} {et~al.}(2016){Federrath}, {Rathborne}, {Longmore},
  {Kruijssen}, {Bally}, {Contreras}, {Crocker}, {Garay}, {Jackson}, {Testi}, \&
  {Walsh}}]{Federrath16}
{Federrath}, C., {Rathborne}, J.~M., {Longmore}, S.~N., {et~al.} 2016, \apj,
  832, 143

\bibitem[{{Fujimoto} {et~al.}(2014){Fujimoto}, {Tasker}, {Wakayama}, \&
  {Habe}}]{Fujimoto14}
{Fujimoto}, Y., {Tasker}, E.~J., {Wakayama}, M., \& {Habe}, A. 2014, \mnras,
  439, 936

\bibitem[{{Gao} \& {Solomon}(2004)}]{Gao04}
{Gao}, Y., \& {Solomon}, P.~M. 2004, \apj, 606, 271

\bibitem[{{Griffin} {et~al.}(2010){Griffin}, {Abergel}, {Abreu}, {Ade},
  {Andr{\'e}}, {Augueres}, {Babbedge}, {Bae}, {Baillie}, {Baluteau}, {Barlow},
  {Bendo}, {Benielli}, {Bock}, {Bonhomme}, {Brisbin}, {Brockley-Blatt},
  {Caldwell}, {Cara}, {Castro-Rodriguez}, {Cerulli}, {Chanial}, {Chen},
  {Clark}, {Clements}, {Clerc}, {Coker}, {Communal}, {Conversi}, {Cox},
  {Crumb}, {Cunningham}, {Daly}, {Davis}, {de Antoni}, {Delderfield}, {Devin},
  {di Giorgio}, {Didschuns}, {Dohlen}, {Donati}, {Dowell}, {Dowell}, {Duband},
  {Dumaye}, {Emery}, {Ferlet}, {Ferrand}, {Fontignie}, {Fox}, {Franceschini},
  {Frerking}, {Fulton}, {Garcia}, {Gastaud}, {Gear}, {Glenn}, {Goizel},
  {Griffin}, {Grundy}, {Guest}, {Guillemet}, {Hargrave}, {Harwit}, {Hastings},
  {Hatziminaoglou}, {Herman}, {Hinde}, {Hristov}, {Huang}, {Imhof}, {Isaak},
  {Israelsson}, {Ivison}, {Jennings}, {Kiernan}, {King}, {Lange}, {Latter},
  {Laurent}, {Laurent}, {Leeks}, {Lellouch}, {Levenson}, {Li}, {Li},
  {Lilienthal}, {Lim}, {Liu}, {Lu}, {Madden}, {Mainetti}, {Marliani}, {McKay},
  {Mercier}, {Molinari}, {Morris}, {Moseley}, {Mulder}, {Mur}, {Naylor},
  {Nguyen}, {O'Halloran}, {Oliver}, {Olofsson}, {Olofsson}, {Orfei}, {Page},
  {Pain}, {Panuzzo}, {Papageorgiou}, {Parks}, {Parr-Burman}, {Pearce},
  {Pearson}, {P{\'e}rez-Fournon}, {Pinsard}, {Pisano}, {Podosek}, {Pohlen},
  {Polehampton}, {Pouliquen}, {Rigopoulou}, {Rizzo}, {Roseboom}, {Roussel},
  {Rowan-Robinson}, {Rownd}, {Saraceno}, {Sauvage}, {Savage}, {Savini},
  {Sawyer}, {Scharmberg}, {Schmitt}, {Schneider}, {Schulz}, {Schwartz},
  {Shafer}, {Shupe}, {Sibthorpe}, {Sidher}, {Smith}, {Smith}, {Smith},
  {Spencer}, {Stobie}, {Sudiwala}, {Sukhatme}, {Surace}, {Stevens}, {Swinyard},
  {Trichas}, {Tourette}, {Triou}, {Tseng}, {Tucker}, {Turner}, {Vaccari},
  {Valtchanov}, {Vigroux}, {Virique}, {Voellmer}, {Walker}, {Ward}, {Waskett},
  {Weilert}, {Wesson}, {White}, {Whitehouse}, {Wilson}, {Winter}, {Woodcraft},
  {Wright}, {Xu}, {Zavagno}, {Zemcov}, {Zhang}, \& {Zonca}}]{Griffin2010}
{Griffin}, M.~J., {Abergel}, A., {Abreu}, A., {et~al.} 2010, \aap, 518, L3

\bibitem[{{Guedes} {et~al.}(2011){Guedes}, {Callegari}, {Madau}, \&
  {Mayer}}]{Guedes11}
{Guedes}, J., {Callegari}, S., {Madau}, P., \& {Mayer}, L. 2011, \apj, 742, 76

\bibitem[{{Hao} {et~al.}(2011){Hao}, {Kennicutt}, {Johnson}, {Calzetti},
  {Dale}, \& {Moustakas}}]{Hao11}
{Hao}, C.-N., {Kennicutt}, R.~C., {Johnson}, B.~D., {et~al.} 2011, \apj, 741,
  124

\bibitem[{Hunter(2007)}]{matplotlib}
Hunter, J.~D. 2007, Computing In Science \& Engineering, 9, 90

\bibitem[{{Jackson} {et~al.}(2006){Jackson}, {Rathborne}, {Shah}, {Simon},
  {Bania}, {Clemens}, {Chambers}, {Johnson}, {Dormody}, {Lavoie}, \&
  {Heyer}}]{Jackson06}
{Jackson}, J.~M., {Rathborne}, J.~M., {Shah}, R.~Y., {et~al.} 2006, \apjs, 163,
  145

\bibitem[{{Kennicutt} \& {Evans}(2012)}]{Kennicutt12}
{Kennicutt}, R.~C., \& {Evans}, N.~J. 2012, \araa, 50, 531

\bibitem[{{Kennicutt}(1998)}]{Kennicutt1998}
{Kennicutt}, Jr., R.~C. 1998, \apj, 498, 541

\bibitem[{{Kreckel} {et~al.}(2018){Kreckel}, {Faesi}, {Kruijssen}, {Schruba},
  {Groves}, {Leroy}, {Bigiel}, {Blanc}, {Chevance}, {Herrera}, {Hughes},
  {McElroy}, {Pety}, {Querejeta}, {Rosolowsky}, {Schinnerer}, {Sun}, {Usero},
  \& {Utomo}}]{Kreckel18}
{Kreckel}, K., {Faesi}, C., {Kruijssen}, J.~M.~D., {et~al.} 2018, \apj, 863,
  L21

\bibitem[{{Krumholz} {et~al.}(2009){Krumholz}, {McKee}, \&
  {Tumlinson}}]{Krumholz09}
{Krumholz}, M.~R., {McKee}, C.~F., \& {Tumlinson}, J. 2009, \apj, 699, 850

\bibitem[{{Lada} {et~al.}(2012){Lada}, {Forbrich}, {Lombardi}, \&
  {Alves}}]{Lada12}
{Lada}, C.~J., {Forbrich}, J., {Lombardi}, M., \& {Alves}, J.~F. 2012, \apj,
  745, 190

\bibitem[{{Lada} {et~al.}(2010){Lada}, {Lombardi}, \& {Alves}}]{Lada10}
{Lada}, C.~J., {Lombardi}, M., \& {Alves}, J.~F. 2010, \apj, 724, 687

\bibitem[{{Lee} {et~al.}(2018){Lee}, {Leroy}, {Bolatto}, {Glover},
  {Indebetouw}, {Sandstrom}, \& {Schruba}}]{Lee18}
{Lee}, C., {Leroy}, A.~K., {Bolatto}, A.~D., {et~al.} 2018, \mnras, 474, 4672

\bibitem[{{Leroy} {et~al.}(2013){Leroy}, {Walter}, {Sandstrom}, {Schruba},
  {Munoz-Mateos}, {Bigiel}, {Bolatto}, {Brinks}, {de Blok}, {Meidt}, {Rix},
  {Rosolowsky}, {Schinnerer}, {Schuster}, \& {Usero}}]{Leroy13}
{Leroy}, A.~K., {Walter}, F., {Sandstrom}, K., {et~al.} 2013, \aj, 146, 19

\bibitem[{{Longmore} {et~al.}(2013){Longmore}, {Bally}, {Testi}, {Purcell},
  {Walsh}, {Bressert}, {Pestalozzi}, {Molinari}, {Ott}, {Cortese}, {Battersby},
  {Murray}, {Lee}, {Kruijssen}, {Schisano}, \& {Elia}}]{Longmore13}
{Longmore}, S.~N., {Bally}, J., {Testi}, L., {et~al.} 2013, \mnras, 429, 987

\bibitem[{{McKee} \& {Ostriker}(2007)}]{McKee07}
{McKee}, C.~F., \& {Ostriker}, E.~C. 2007, \araa, 45, 565

\bibitem[{{Molinari} {et~al.}(2010){Molinari}, {Swinyard}, {Bally}, {Barlow},
  {Bernard}, {Martin}, {Moore}, {Noriega-Crespo}, {Plume}, {Testi}, {Zavagno},
  {Abergel}, {Ali}, {Andr{\'e}}, {Baluteau}, {Benedettini}, {Bern{\'e}},
  {Billot}, {Blommaert}, {Bontemps}, {Boulanger}, {Brand}, {Brunt}, {Burton},
  {Campeggio}, {Carey}, {Caselli}, {Cesaroni}, {Cernicharo}, {Chakrabarti},
  {Chrysostomou}, {Codella}, {Cohen}, {Compiegne}, {Davis}, {de Bernardis}, {de
  Gasperis}, {Di Francesco}, {di Giorgio}, {Elia}, {Faustini}, {Fischera},
  {Fukui}, {Fuller}, {Ganga}, {Garcia-Lario}, {Giard}, {Giardino}, {Glenn},
  {Goldsmith}, {Griffin}, {Hoare}, {Huang}, {Jiang}, {Joblin}, {Joncas},
  {Juvela}, {Kirk}, {Lagache}, {Li}, {Lim}, {Lord}, {Lucas}, {Maiolo},
  {Marengo}, {Marshall}, {Masi}, {Massi}, {Matsuura}, {Meny}, {Minier},
  {Miville-Desch{\^e}nes}, {Montier}, {Motte}, {M{\"u}ller}, {Natoli}, {Neves},
  {Olmi}, {Paladini}, {Paradis}, {Pestalozzi}, {Pezzuto}, {Piacentini},
  {Pomar{\`e}s}, {Popescu}, {Reach}, {Richer}, {Ristorcelli}, {Roy}, {Royer},
  {Russeil}, {Saraceno}, {Sauvage}, {Schilke}, {Schneider-Bontemps},
  {Schuller}, {Schultz}, {Shepherd}, {Sibthorpe}, {Smith}, {Smith},
  {Spinoglio}, {Stamatellos}, {Strafella}, {Stringfellow}, {Sturm}, {Taylor},
  {Thompson}, {Tuffs}, {Umana}, {Valenziano}, {Vavrek}, {Viti}, {Waelkens},
  {Ward-Thompson}, {White}, {Wyrowski}, {Yorke}, \& {Zhang}}]{Molinari2010}
{Molinari}, S., {Swinyard}, B., {Bally}, J., {et~al.} 2010, \pasp, 122, 314

\bibitem[{{Molinari} {et~al.}(2016){Molinari}, {Schisano}, {Elia},
  {Pestalozzi}, {Traficante}, {Pezzuto}, {Swinyard}, {Noriega-Crespo}, {Bally},
  {Moore}, {Plume}, {Zavagno}, {di Giorgio}, {Liu}, {Pilbratt}, {Mottram},
  {Russeil}, {Piazzo}, {Veneziani}, {Benedettini}, {Calzoletti}, {Faustini},
  {Natoli}, {Piacentini}, {Merello}, {Palmese}, {Del Grande}, {Polychroni},
  {Rygl}, {Polenta}, {Barlow}, {Bernard}, {Martin}, {Testi}, {Ali},
  {Andr{\'e}}, {Beltr{\'a}n}, {Billot}, {Carey}, {Cesaroni}, {Compi{\`e}gne},
  {Eden}, {Fukui}, {Garcia-Lario}, {Hoare}, {Huang}, {Joncas}, {Lim}, {Lord},
  {Martinavarro-Armengol}, {Motte}, {Paladini}, {Paradis}, {Peretto},
  {Robitaille}, {Schilke}, {Schneider}, {Schulz}, {Sibthorpe}, {Strafella},
  {Thompson}, {Umana}, {Ward-Thompson}, \& {Wyrowski}}]{Molinari16b}
{Molinari}, S., {Schisano}, E., {Elia}, D., {et~al.} 2016, \aap, 591, A149

\bibitem[{{Murphy} {et~al.}(2011){Murphy}, {Condon}, {Schinnerer}, {Kennicutt},
  {Calzetti}, {Armus}, {Helou}, {Turner}, {Aniano}, {Beir{\~a}o}, {Bolatto},
  {Brandl}, {Croxall}, {Dale}, {Donovan Meyer}, {Draine}, {Engelbracht},
  {Hunt}, {Hao}, {Koda}, {Roussel}, {Skibba}, \& {Smith}}]{Murphy11}
{Murphy}, E.~J., {Condon}, J.~J., {Schinnerer}, E., {et~al.} 2011, \apj, 737,
  67

\bibitem[{{Onodera} {et~al.}(2010){Onodera}, {Kuno}, {Tosaki}, {Kohno},
  {Nakanishi}, {Sawada}, {Muraoka}, {Komugi}, {Miura}, {Kaneko}, {Hirota}, \&
  {Kawabe}}]{Onodera10}
{Onodera}, S., {Kuno}, N., {Tosaki}, T., {et~al.} 2010, \apjl, 722, L127

\bibitem[{{Ossenkopf} \& {Henning}(1994)}]{OH94}
{Ossenkopf}, V., \& {Henning}, T. 1994, \aap, 291, 943

\bibitem[{{Padoan} {et~al.}(1997){Padoan}, {Nordlund}, \& {Jones}}]{Padoan97}
{Padoan}, P., {Nordlund}, A., \& {Jones}, B.~J.~T. 1997, \mnras, 288, 145

\bibitem[{{Poglitsch} {et~al.}(2010){Poglitsch}, {Waelkens}, {Geis},
  {Feuchtgruber}, {Vandenbussche}, {Rodriguez}, {Krause}, {Renotte}, {van
  Hoof}, {Saraceno}, {Cepa}, {Kerschbaum}, {Agn{\`e}se}, {Ali}, {Altieri},
  {Andreani}, {Augueres}, {Balog}, {Barl}, {Bauer}, {Belbachir}, {Benedettini},
  {Billot}, {Boulade}, {Bischof}, {Blommaert}, {Callut}, {Cara}, {Cerulli},
  {Cesarsky}, {Contursi}, {Creten}, {De Meester}, {Doublier}, {Doumayrou},
  {Duband}, {Exter}, {Genzel}, {Gillis}, {Gr{\"o}zinger}, {Henning},
  {Herreros}, {Huygen}, {Inguscio}, {Jakob}, {Jamar}, {Jean}, {de Jong},
  {Katterloher}, {Kiss}, {Klaas}, {Lemke}, {Lutz}, {Madden}, {Marquet},
  {Martignac}, {Mazy}, {Merken}, {Montfort}, {Morbidelli}, {M{\"u}ller},
  {Nielbock}, {Okumura}, {Orfei}, {Ottensamer}, {Pezzuto}, {Popesso},
  {Putzeys}, {Regibo}, {Reveret}, {Royer}, {Sauvage}, {Schreiber}, {Stegmaier},
  {Schmitt}, {Schubert}, {Sturm}, {Thiel}, {Tofani}, {Vavrek}, {Wetzstein},
  {Wieprecht}, \& {Wiezorrek}}]{Poglitsch2010}
{Poglitsch}, A., {Waelkens}, C., {Geis}, N., {et~al.} 2010, \aap, 518, L2

\bibitem[{{Ragan} {et~al.}(2016){Ragan}, {Moore}, {Eden}, {Hoare}, {Elia}, \&
  {Molinari}}]{Ragan16}
{Ragan}, S.~E., {Moore}, T.~J.~T., {Eden}, D.~J., {et~al.} 2016, \mnras, 462,
  3123

\bibitem[{{Rathborne} {et~al.}(2014){Rathborne}, {Longmore}, {Jackson},
  {Kruijssen}, {Alves}, {Bally}, {Bastian}, {Contreras}, {Foster}, {Garay},
  {Testi}, \& {Walsh}}]{Rathborne14}
{Rathborne}, J.~M., {Longmore}, S.~N., {Jackson}, J.~M., {et~al.} 2014, \apj,
  795, L25

\bibitem[{{Rieke} {et~al.}(2009){Rieke}, {Alonso-Herrero}, {Weiner},
  {P{\'e}rez-Gonz{\'a}lez}, {Blaylock}, {Donley}, \& {Marcillac}}]{Rieke09}
{Rieke}, G.~H., {Alonso-Herrero}, A., {Weiner}, B.~J., {et~al.} 2009, \apj,
  692, 556

\bibitem[{{Robitaille} \& {Whitney}(2010)}]{Robitaille10}
{Robitaille}, T.~P., \& {Whitney}, B.~A. 2010, \apjl, 710, L11

\bibitem[{{Rosolowsky} \& {Leroy}(2006)}]{Rosolowsky06}
{Rosolowsky}, E., \& {Leroy}, A. 2006, \pasp, 118, 590

\bibitem[{{Rosolowsky} {et~al.}(2010){Rosolowsky}, {Dunham}, {Ginsburg},
  {Bradley}, {Aguirre}, {Bally}, {Battersby}, {Cyganowski}, {Dowell},
  {Drosback}, {Evans}, {Glenn}, {Harvey}, {Stringfellow}, {Walawender}, \&
  {Williams}}]{BGPS2}
{Rosolowsky}, E., {Dunham}, M.~K., {Ginsburg}, A., {et~al.} 2010, \apjs, 188,
  123

\bibitem[{{Sanders} {et~al.}(1986){Sanders}, {Clemens}, {Scoville}, \&
  {Solomon}}]{sanders86}
{Sanders}, D.~B., {Clemens}, D.~P., {Scoville}, N.~Z., \& {Solomon}, P.~M.
  1986, \apjs, 60, 1

\bibitem[{{Schruba} {et~al.}(2010){Schruba}, {Leroy}, {Walter}, {Sandstrom}, \&
  {Rosolowsky}}]{Schruba10}
{Schruba}, A., {Leroy}, A.~K., {Walter}, F., {Sandstrom}, K., \& {Rosolowsky},
  E. 2010, \apj, 722, 1699

\bibitem[{{Scoville} \& {Solomon}(1975)}]{Scoville75}
{Scoville}, N.~Z., \& {Solomon}, P.~M. 1975, \apjl, 199, L105

\bibitem[{{Shirley} {et~al.}(2013){Shirley}, {Ellsworth-Bowers}, {Svoboda},
  {Schlingman}, {Ginsburg}, {Rosolowsky}, {Gerner}, {Mairs}, {Battersby},
  {Stringfellow}, {Dunham}, {Glenn}, \& {Bally}}]{Shirley13}
{Shirley}, Y.~L., {Ellsworth-Bowers}, T.~P., {Svoboda}, B., {et~al.} 2013,
  \apjs, 209, 2

\bibitem[{{Solomon} {et~al.}(1987){Solomon}, {Rivolo}, {Barrett}, \&
  {Yahil}}]{Solomon87}
{Solomon}, P.~M., {Rivolo}, A.~R., {Barrett}, J., \& {Yahil}, A. 1987, \apj,
  319, 730

\bibitem[{{Svoboda} {et~al.}(2016){Svoboda}, {Shirley}, {Battersby},
  {Rosolowsky}, {Ginsburg}, {Ellsworth-Bowers}, {Pestalozzi}, {Dunham},
  {Evans}, {Bally}, \& {Glenn}}]{Svoboda16}
{Svoboda}, B.~E., {Shirley}, Y.~L., {Battersby}, C., {et~al.} 2016, \apj, 822,
  59

\bibitem[{{Traficante} {et~al.}(2018){Traficante}, {Duarte-Cabral}, {Elia},
  {Fuller}, {Merello}, {Molinari}, {Peretto}, {Schisano}, \& {Di
  Giorgio}}]{Traficante18}
{Traficante}, A., {Duarte-Cabral}, A., {Elia}, D., {et~al.} 2018, ArXiv
  e-prints, arXiv:1803.08929

\bibitem[{{Usero} {et~al.}(2015){Usero}, {Leroy}, {Walter}, {Schruba},
  {Garc{\'{\i}}a-Burillo}, {Sandstrom}, {Bigiel}, {Brinks}, {Kramer},
  {Rosolowsky}, {Schuster}, \& {de Blok}}]{Usero15}
{Usero}, A., {Leroy}, A.~K., {Walter}, F., {et~al.} 2015, \aj, 150, 115

\bibitem[{{Utomo} {et~al.}(2018){Utomo}, {Sun}, {Leroy}, {Kruijssen},
  {Schinnerer}, {Schruba}, {Bigiel}, {Blanc}, {Chevance}, {Emsellem},
  {Herrera}, {Hygate}, {Kreckel}, {Ostriker}, {Pety}, {Querejeta},
  {Rosolowsky}, {Sandstrom}, \& {Usero}}]{Utomo18}
{Utomo}, D., {Sun}, J., {Leroy}, A.~K., {et~al.} 2018, \apj, 861, L18

\bibitem[{{Vall{\'e}e}(2014)}]{Vallee14}
{Vall{\'e}e}, J.~P. 2014, \aj, 148, 5

\bibitem[{{V{\'a}zquez-Semadeni} {et~al.}(2011){V{\'a}zquez-Semadeni},
  {Banerjee}, {G{\'o}mez}, {Hennebelle}, {Duffin}, \& {Klessen}}]{Vazquez11}
{V{\'a}zquez-Semadeni}, E., {Banerjee}, R., {G{\'o}mez}, G.~C., {et~al.} 2011,
  \mnras, 414, 2511

\bibitem[{{Vutisalchavakul} {et~al.}(2016){Vutisalchavakul}, {Evans}, \&
  {Heyer}}]{Vutisalchavakul16}
{Vutisalchavakul}, N., {Evans}, II, N.~J., \& {Heyer}, M. 2016, \apj, 831, 73

\bibitem[{{Wolfire} {et~al.}(2003){Wolfire}, {McKee}, {Hollenbach}, \&
  {Tielens}}]{Wolfire03}
{Wolfire}, M.~G., {McKee}, C.~F., {Hollenbach}, D., \& {Tielens}, A.~G.~G.~M.
  2003, \apj, 587, 278

\bibitem[{{Wu} {et~al.}(2007){Wu}, {Evans}, {Gao}, {Solomon}, {Shirley}, \&
  {vanden Bout}}]{Wu07}
{Wu}, J., {Evans}, N., {Gao}, Y., {et~al.} 2007, in Astronomical Society of the
  Pacific Conference Series, Vol. 375, From Z-Machines to ALMA: (Sub)Millimeter
  Spectroscopy of Galaxies, ed. A.~J. {Baker}, J.~{Glenn}, A.~I. {Harris},
  J.~G. {Mangum}, \& M.~S. {Yun}, 291

\bibitem[{{Zetterlund} {et~al.}(2017){Zetterlund}, {Glenn}, \&
  {Rosolowsky}}]{Zetterlund17}
{Zetterlund}, E., {Glenn}, J., \& {Rosolowsky}, E. 2017, \apj, 835, 203

\bibitem[{{Zetterlund} {et~al.}(2018){Zetterlund}, {Glenn}, \&
  {Rosolowsky}}]{Zetterlund18}
---. 2018, \mnras, 480, 893

\end{thebibliography}

\end{document}